\documentclass[a4paper,fleqn,usenatbib]{mnras}

\usepackage{newtxtext,newtxmath}
\usepackage{subcaption}
\usepackage[T1]{fontenc}
\usepackage{ae,aecompl}

\usepackage{graphicx}	
\usepackage{amsmath}	
\usepackage{amssymb}	

\title[Accretion and drag force for SMBHs]{Bondi or not Bondi: the impact of resolution on accretion and drag force modelling for Supermassive Black Holes}

\author[R. S. Beckmann et al.]{R. S. Beckmann$^{1,2}$\thanks{Email: ricarda.beckmann@iap.fr}, A. Slyz$^{1}$, J. Devriendt$^{1,3}$
\\
$^{1}$ Sub-department of Astrophysics, University of Oxford, Keble Road, Oxford OX1 3RH, UK\\
$^{2}$ Institut d'Astrophysique de Paris, CNRS \& Sorbonne Universites, UMR 7095, 98bis Boulevard Arago, F-75014, Paris, France \\
$^{3}$ Université de Lyon, Université Lyon 1, ENS de Lyon, CNRS, Centre de Recherche Astrophysique de Lyon UMR5574, F-69230 Saint-Genis-Laval, France\\}

\date{Accepted XXX. Received YYY; in original form ZZZ}

\pubyear{2017}

\begin{document}
\label{firstpage}
\pagerange{\pageref{firstpage}--\pageref{lastpage}}
\maketitle

\begin{abstract}
Whilst in galaxy-size simulations, supermassive black holes (SMBH) are entirely handled by sub-grid algorithms, computational power now allows the accretion radius of such objects to be resolved in smaller scale simulations. In this paper, we investigate the impact of resolution on two commonly used SMBH sub-grid algorithms; the Bondi-Hoyle-Lyttleton (BHL) formula for accretion onto a point mass, and the related estimate of the drag force exerted onto a point mass by a gaseous medium. We find that when the accretion region around the black hole scales with resolution, and the BHL formula is evaluated using local mass-averaged quantities, the accretion algorithm smoothly transitions from the analytic BHL formula (at low resolution) to a supply limited accretion (SLA) scheme (at high resolution). However, when a similar procedure is employed to estimate the drag force it can lead to significant errors in its magnitude, and/or apply this force in the wrong direction in highly resolved simulations. At high Mach numbers and for small accretors, we also find evidence of the advective-acoustic instability operating in the adiabatic case, and of an instability developing around the wake's stagnation point in the quasi-isothermal case. Moreover, at very high resolution, and Mach numbers above $\mathcal{M}_\infty \geq 3$, the flow behind the accretion bow shock becomes entirely dominated by these instabilities. As a result, accretion rates onto the black hole drop by about an order of magnitude in the adiabatic case, compa to the analytic BHL formula.
\end{abstract}



%

\section{Introduction}
\label{sec:introduction}

One common problem in astrophysics is that the range of scales involved in a given problem can be huge, from atomic to galactic, and as such, it is extremely difficult to fully capture a physical process in a single simulation. This is especially true for processes that suffer from an 'inverse cascade' problem, where (unresolved) small scale behaviour influences the outcome on (resolved) large scales. For these reasons, we have to rely on sub-grid models, which aim to replicate the impact of unresolved, small scale behaviour on scales relevant to the simulation at hand, using only information available in the simulation. The success of such a sub-grid algorithm therefore depends both on the relevance of the physics it contains, and its ability to account for it over the widest possible range of scales. 

Super massive black holes (SMBH) in cosmological or idealised galaxy simulations are one such problem. The gas fuelling the black hole flows from the Mpc scales of the cosmic web, through the kpc scales of the galaxy down to the last stable orbit and event horizon of the black hole on $au$ scales. It is currently impossible to adequately track the gas from where it originated all the way to the black hole in a single simulation. Instead, in Adaptive Mesh Refinement (AMR) particle-grid codes such as RAMSES, SMBH are typically modelled as ``sink'' particles \citep{Krumholz2004,Dubois2010}, i.e. the black hole is a massive particle that moves over the grid, removing gas from a small accretion region centred on its current location \citep[see][for the equivalent in Smoothed Particle Hydrodynamics codes]{Bate1995,Springel2005}. Although there exist alternatives (see next paragraph), accretion usually proceeds at the Bondi-Hoyle-Lyttleton (BHL) rate. More specifically, the amount of mass accreted by the sink is calculated using the analytic formula of \citet{Bondi1944} and \citet{Hoyle1939} derived for a point mass accretor moving at constant velocity through an homogeneous background. In this approach, quantities measured ``at infinity'' should in principle be used to infer the mass accretion rate onto the sink (see Section \ref{subsec:accretion} for details).

Whilst a good starting point in situations with limited information, this approach has several notable shortcomings, particularly the question of where ``infinity'' is to be defined in a busy galaxy simulation. The analytic solution also does not account for the background gravitational potential of the host galaxy \citep{Korol2016}, the self-gravity of the gas, nor does it consider density or velocity gradients, angular momentum or gas effects such as pressure or instabilities (see \citet{Edgar2004} for a detailed review of BHL accretion). A variety of suggestions have been put forward to address some of these shortcomings. They fall broadly into two camps: (i) replace the BHL model with another that takes as input different large-scale properties of the host galaxy, such as gravitational torque \citep{Hopkins2011,Angles-Alcazar2015}, vorticity  \citep{Krumholz2004} or velocity dispersion of the bulge \citep{Hobbs2011}; or (ii) model unresolved physics on smaller scales via accretion disc schemes \citep{Power2011,Dubois2014a} or unresolved circularisation and viscous transfer considerations \citep{Debuhr2010a,Rosas-Guevara2015}. We refer the interested reader to \citet{Negri2016} for a comprehensive review of current accretion schemes.

The BHL analytic solution assumes a uniform density and velocity of the gas at infinity, a situation that most closely resembles simulations where all characteristic length-scales associated with the SMBH are much smaller than the local resolution, and the gas reservoir per cell is large compared to the accreted mass. Indeed in this situation, ``infinity'' can reasonably be understood to mean the gas cells surrounding the sink, as the black hole's gravitational potential is unresolved, and therefore has little impact on the local gas flow. This is particularly true in the absence of feedback, as the cold, dense phase of the interstellar medium (ISM) gas that is expected to feed the black hole, is also under-resolved in low resolution simulations. However, while this has led some authors to introduce a simple boost factor to the Bondi rate to compensate for the lack of resolution \citep{Booth2009}, \citet{Negri2016} recently showed that the interplay between local gas density, accretion rate and resolution is much more subtle, and becomes even more so in the presence of black hole feedback.

Computational advances now allow us to resolve a much wider range of scales in galaxy evolution simulations, from the Mpc scales of the cosmic web at large scales to the length scales of the black hole at small scales, fast approaching the physical scale (Schwarschild or Kerr radius) of the black hole itself. In this paper, we investigate how the popular BHL accretion algorithm, where local mass or volume weighted average gas properties are used to approximate quantities at infinity, behaves as the gravitational influence of the black hole becomes better resolved, and the local cell mass becomes comparable to the mass accreted per time resolution element. In that respect, the work presented here is similar to \citet{Ruffert1995,Ruffert1996}, who investigated the validity and convergence of the analytic BHL formula using a range of fixed spherical, inflow-only regions to represent the sink. However, our work focusses less on the validity of the BHL analytic solution itself --- even though we do discuss the observed instabilities and how they cause systematic deviations from the analytic solution --- than on testing the impact of representing the BH as a sink particle, as well as estimating local gas averages from partially or fully resolved density features within the accretion region of the sink. 

Our work also builds on \citet{Krumholz2004}, which first introduced the use of Lagrangian sink particles in grid codes, and \citet{Dubois2010}, who implemented it in RAMSES. More specifically, we carry out the most thorough exploration of the BHL parameter space performed to date using such a model, both in Mach number and resolution, and for two gas adiabatic indices. This allows us to test the sub-grid accretion algorithm behaviour in a variety of regimes were it has or will be used in galaxy simulations, including two features only briefly mentioned in \citet{Krumholz2004}: a change in accretion mode at sufficiently high resolution, and  accretion becoming inefficient when the resolution approaches the scale radius of the black hole. 

As the total mass accreted onto SMBH in galaxy simulations depends not only on the amount of gas removed, but also on the ability of the black hole to stay attached to local high density structures, we then look in detail at the sub-grid algorithm used to account for the dynamical friction exerted by the gaseous gravitational wake behind the sink particle. The particular simulations presented here were performed with RAMSES, but we believe that the conclusions we reach are widely applicable to sub-grid algorithms where the size of the accretion region, over which local gas properties are measured and from which gas is removed, decreases with increasing resolution. 

The structure of the paper is as follows: in Section \ref{sec:setup} we present the setup of the simulations, and explain the sink particle algorithm in RAMSES in detail. Section \ref{sec:accretion} investigates accretion for a range of Mach numbers and resolutions, including a detailed study of the Bondi problem (Section \ref{sec:bondi}) and the Hoyle-Lyttleton problem (Section \ref{sec:hoyle_lyttleton}). Section \ref{sec:fdrag} present results for the impact of Mach number and resolution on dynamical friction, both resolved on the grid and calculated analytically. Section \ref{subsec:discussion_instability} discusses the origin of instabilities observed in the flow. Conclusions can be found in Section \ref{sec:conclusion}.


\section{The simulations}
\label{sec:setup}

To study gas dynamics in the vicinity of the sink particle, we set up a series of 3D isolated boxes where the sink particle is embedded in a uniform gas flow. Each simulation is parametrised by a Mach number, $ \mathcal{M_\infty} = v_\infty / c_{s,\infty}$, where $c_{s,\infty}$ and $v_\infty$ are the sound speed and flow velocity respectively. As the sink is held fixed at the centre of the box, $v_\infty$ is both the absolute flow velocity, and the flow velocity relative to the sink. Resolution is measured as the number of cells, $N$, within the sink particle's scale radius $R_\infty^S$, so that $N = R_\infty^S/\Delta x_\mathrm{min}$, where $\Delta x_\mathrm{min}$ is the size of the smallest grid cell in the simulation.

This characteristic scale radius depends on $\mathcal{M_\infty}$ as follows:
\begin{equation}
	R_\infty^{S} = \begin{cases} 
	 GM_\mathrm{sink} / c_{s,\infty}^2 & \rm{if } \ \ \mathcal{M_\infty} \leq 1 \\
	 2GM_\mathrm{sink} / v_\infty^2 & \rm{if } \ \ \mathcal{M_\infty} >1 
	\end{cases}
	\label{eq:r_s}
\end{equation}
where $G$ is the gravitational constant and $M_\mathrm{sink}$ is the mass of the sink particle. Setting $G=c_{s,\infty}=\rho_\infty=1$ for all simulations reduces the number of parameters of the problem. Under this assumption, $M_\mathrm{sink} = R^S_\infty / L_\mathrm{box}$, so the mass of the sink determines the relative size of the local scale radius to the size of the box. In order to reduce edge effects, we set $L_\mathrm{box} = 1000 \times R^{S}_\infty$, which leads to $M_\mathrm{sink} = 0.001$. The final parameter is the gas pressure, which is set to $P=1/\gamma$ for all simulations. We perform simulations for two values of $\gamma$, an adiabatic one where $\gamma = 1.3334$, and a quasi-isothermal one, where $\gamma = 1.0001$.

Most work on BHL accretion in the literature parametrises resolution by $r^*/R^S$, where $r^*$ is the size of the accretor. However, with the sink particle algorithm used here, $r^*$ cannot be unambiguously defined, as we remove gas from a region spanned by a Gaussian kernel, whose width varies from $\Delta x_\mathrm{min}/4 <  r_\mathrm{kernel} < 2 \Delta x_\mathrm{min}$, depending on local conditions (see Section \ref{subsec:accretion} for details). We therefore use $N$ as the resolution parameter instead, but write $r^* \approx 2 \Delta x_\mathrm{min}$ i.e. $ r^* / R_\infty^S \approx 2 / N$ when comparing to previous work.

\subsection{Nomenclature}
In the remainder of the paper, quantities denoted with a $\infty$ subscript, $X_\infty$, are analytical values which parametrise the simulations presented here, defining both initial and boundary conditions.  Quantities denoted with a $\bullet$ subscript, $X_\bullet$, are calculated numerically on the fly, and are based on mass-averaged quantities within the accretion region of the sink. We explain in detail what this means in Section \ref{subsec:accretion}. Quantities with a $\diamond$ subscript, $X_\diamond$, are calculated from cell values across the entire simulation box. The term ``local'' refers to the immediate vicinity of the black hole, whereas the term ``global'' describes conditions at the boundaries of the simulation, far away from the sink particle. Time averaged values are denoted by triangular brackets, $\left< x \right>$.

The input parameters for each simulation are summarised in its name as follows: a simulation labelled m$\mathcal{M}_\infty$n$N$$x$ is adiabatic ($\gamma=1.3334$) when $x=a$, and quasi-isothermal ($\gamma=1.0001$) when $x=i$.  For example, a simulation called m10.0n4.8a has $\gamma=1.3334$, $\mathcal{M}_\infty=10$ and $N=4.8$.

Scalar quantities are denoted with italic lettering, $x$, whereas vector quantities are denoted in bold $\bold{x}$. {$r$ denotes a radial distance from the accretor, whereas $s$ denotes the distance from the accretor measured along the line axis of symmetry of the wake, with positive values measured downstream of the accretor, and negative values upstream.}

\subsection{Simulation setup}
\label{subsec:setup}

\begin{figure}
	\centering
	\includegraphics[height=0.8\columnwidth]{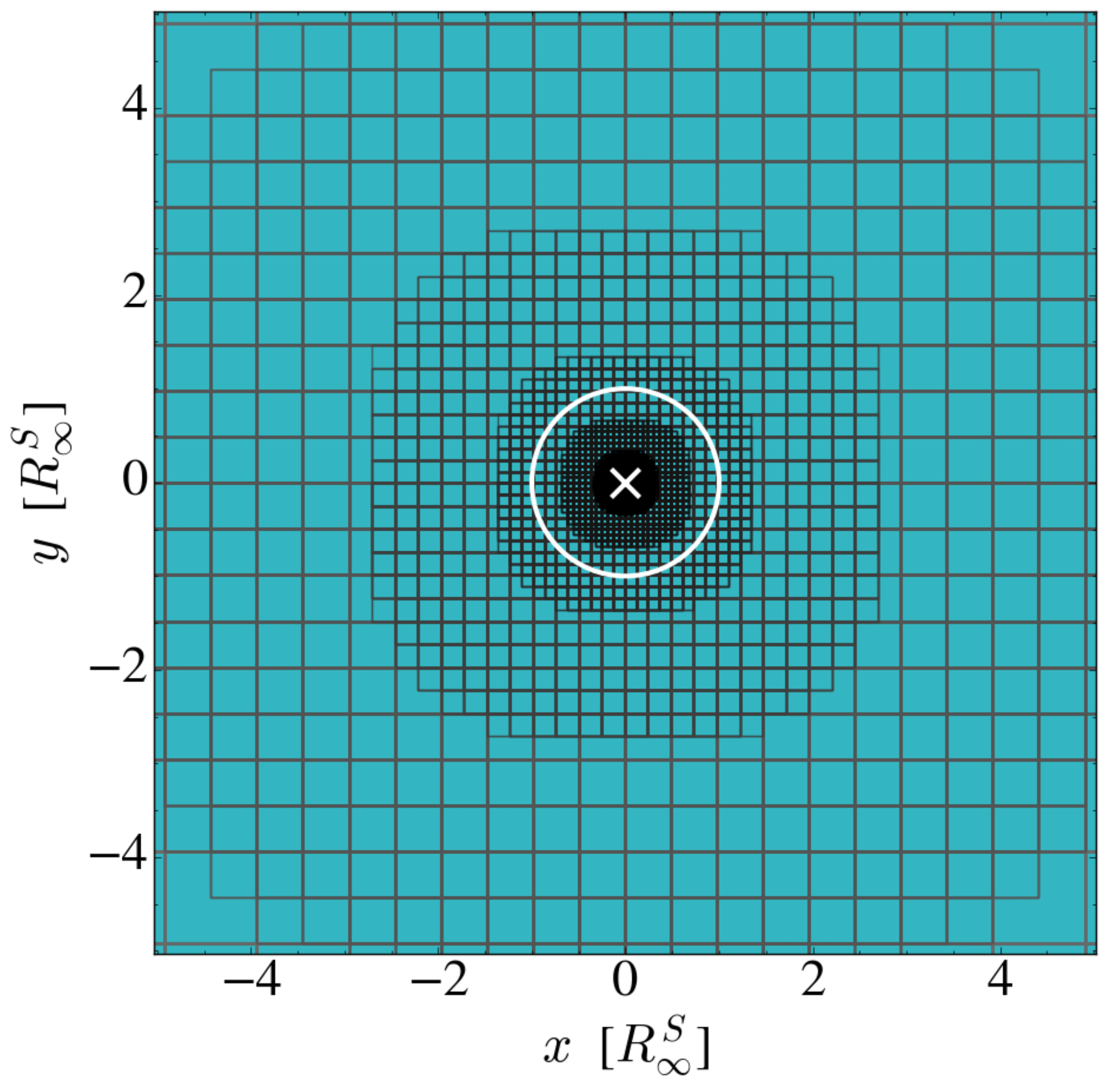}
	\includegraphics[height=0.8\columnwidth]{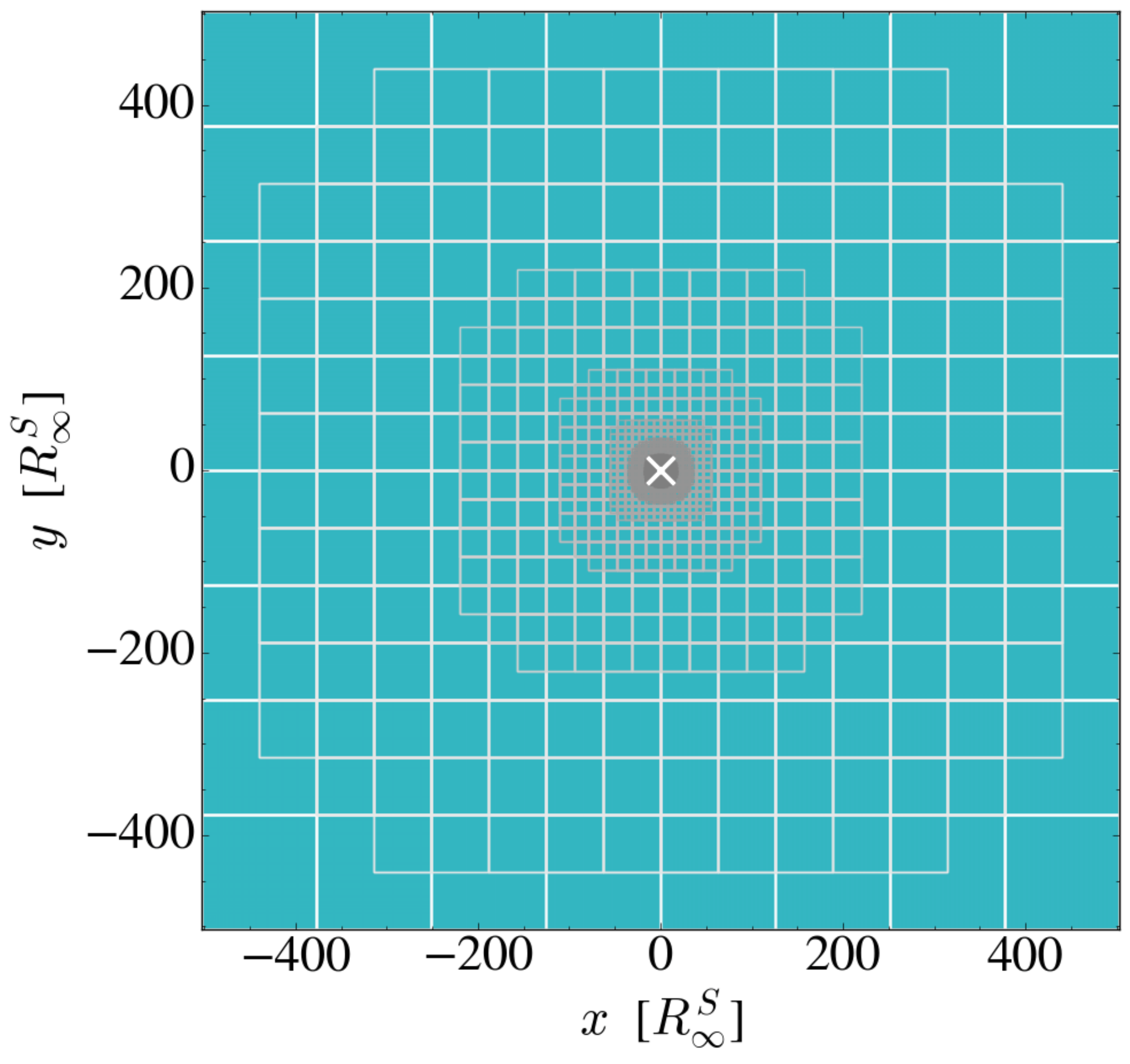}
	\caption{Grid structure showing the size of the box (top panel), and the nested grids in the vicinity of the sink (bottom panel), for a representative simulation (mach0.0n100a). The black hole position is indicated by a white cross, and the dashed line in the bottom plot denotes the Bondi radius.}
	\label{fig:grids}
\end{figure}

All simulations presented here are performed with the AMR code RAMSES \citep{Teyssier2002} and consist of a simple three-dimensional isolated cubic box, with a sink particle kept fixed at the centre. Gravity is prescribed by an analytic gravitational field for a point mass, which inherits the mass and location of the sink and a gravitational softening length equal to the smallest cell size, $\Delta x_\mathrm{min}$. There is no radiative cooling and the gas is not self-gravitating. 

Gas enters the simulation box diagonally, from the bottom left corner, to avoid issues associated with grid aligned flows\footnote{Such as odd-even decoupling \citep{Quirk1994} or the Carbuncle phenomenon at shock fronts \citep{Peery1988,Elling2009}}. All ghost regions (cells outside of the simulation domain) are set using a zero-gradient scheme, except in the $\mathcal{M}=0$ case where constant inflow boundary conditions are used instead to avoid edge effect propagation. Accretion proceeds via the usual RAMSES sink particle algorithm \citep{Krumholz2004,Dubois2010}. We use a linear reconstruction scheme for conservative variables, and a Courant factor of $0.36$ as higher order reconstructions and/or larger Courant factors lead to numerical artefacts.

To reach the required number of levels of refinement within a sufficiently large box, we employ a fixed nested grid strategy, as seen in Figure \ref{fig:grids}. In all cases, the black hole is surrounded by concentric shells of progressively lower refinement down to a root grid of $16^3$ cells, with the highest resolution shell determined by the chosen resolution value of $N$. Test of various nested grid configurations show that the results presented here are insensitive to the exact grid layout, so the radius of grids was doubled for each level to balance the size of refined regions with computational cost. For all simulations with $N\leq100$, all levels of refinement are present from the beginning of the simulation. Simulations with $N>100$ developed a shock feature during the initial settling phase, which led to a permanent breaking of the symmetry of the flow (see Appendix \ref{sec:initial_conditions}). This issue can be prevented by running simulations with $N>100$ with a lower initial resolution of $N \approx 60$ until $t \approx 3$, and refining to the desired level after this pre-conditioning period. To further avoid numerical effects when refining, only a single additional level of refinement is added per timestep until the correct $N$ is reached. In order to give all simulations enough time to damp initial condition transients, all time-averaged values in the paper are measured for $t>10$, where the time $t$ is given in units of scale radius crossing time:
\begin{equation}
  [t] = \frac{R_\infty^{S}}{(c_{s,\infty}+v_\infty) }
\label{eq:time}
\end{equation}

\subsection{The accretion algorithm}
\label{subsec:accretion}

\begin{figure}
	\centering
	\includegraphics[width=\columnwidth]{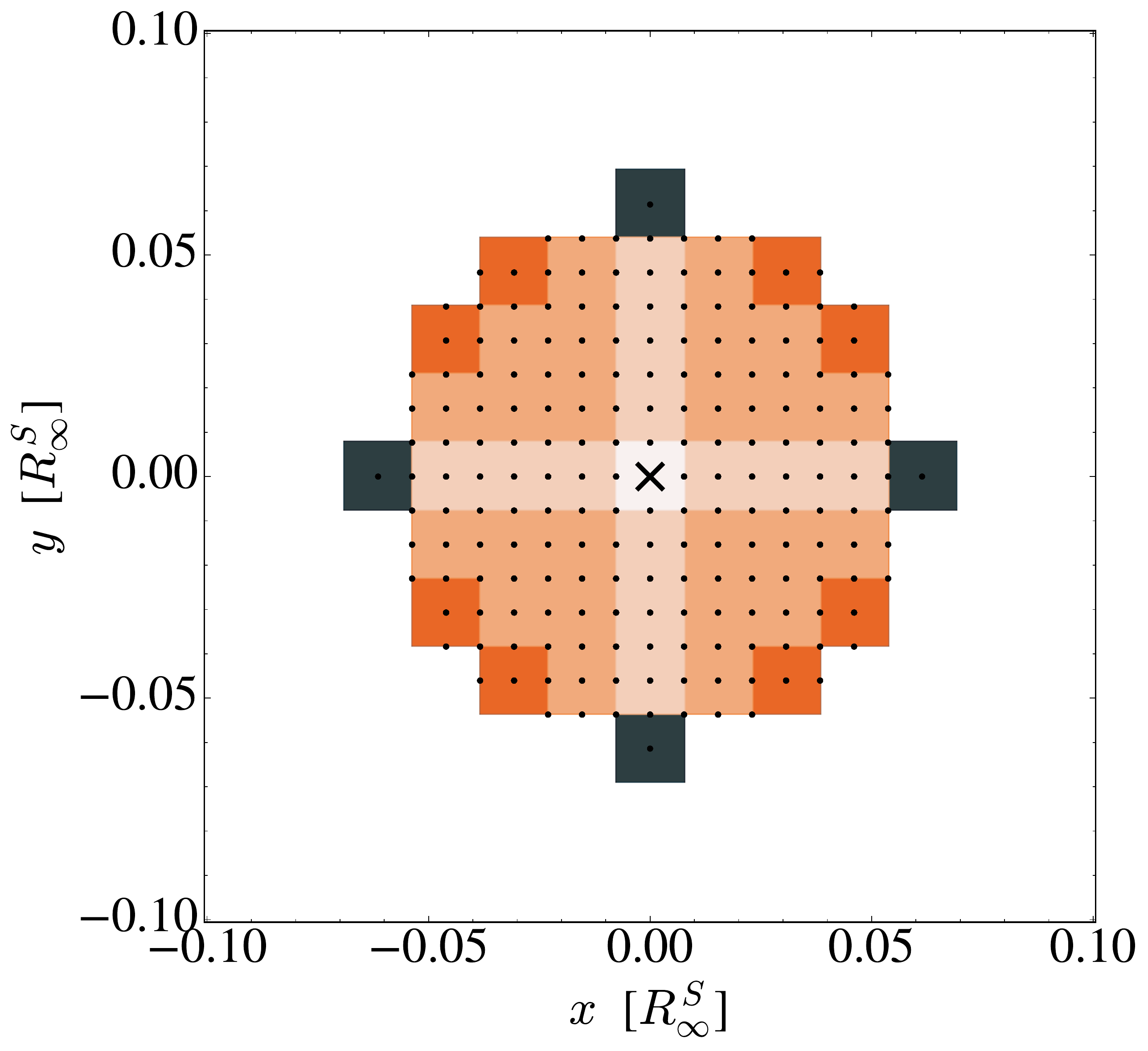}
	\caption{Cloud particles are scattered at regular distances from the sink, to calculate mass-averaged gas quantities of the local black hole environment for use in the Bondi-Hoyle accretion formula. The sink is denoted by a black cross and the cloud particles by black dots. Cells probed by the cloud particles are highlighted in colour, while cells outside the accretion region are shown in white. Colours reflect the number of cloud particles present in that cell.} 
	\label{fig:cloud_distribution}
\end{figure}

The standard sink particle accretion algorithm in RAMSES is that implemented by \citet{Dubois2010} from an algorithm originally proposed by \citet{Krumholz2004}, using the Bondi-Hoyle-Lyttleton interpolation formula \citep{Hoyle1939,Bondi1944,Edgar2004}, 
\begin{equation}
\dot{M}^\mathrm{BHL}= \frac{4 \pi G^2 M_\mathrm{sink}^2 \rho_\infty}{(c_{s,\infty}^2+v_\infty^2)^{3/2}}
\label{eq:bondi_hoyle}
\end{equation}
where the gas properties, such as the density $\rho_\infty$, the bulk velocity $v_\infty$ and the sound speed $c_{s,\infty}$ are non-local measures meant to be taken at infinity, or ``far from the gravitational influence of the point mass''. 

To calculate these non-local properties, RAMSES uses a cloud-particle system \citep{Dubois2010,Teyssier2011}, where a total of 2109 cloud particles are scattered around the sink with a spacing of  $\Delta_\mathrm{cloud}=\Delta x_\mathrm{loc}/2$, filling a sphere with radius $r_\mathrm{cloud} = 4 \times \Delta x_\mathrm{loc}$, where $\Delta x_\mathrm{loc}$ is the local cell width. For simulations more complex than the isolated case presented here, the sink is free to move across the grid, making the cloud particles invaluable in sampling local cells, whether the sink is positioned at the cell centre or not. However, in the simulations discussed in this paper, placing the sink at the centre of a cell, combined with the spacing of $\Delta x_\mathrm{loc} /2= \Delta x_\mathrm{min}/2 $ creates ambiguity in the assignation of parent cell to cloud particles, as some of these particles are located exactly on cell boundaries. RAMSES assigns ambiguous particles preferentially to the cell located downwards and to the left in the coordinate system used here, creating a preferential accretion direction. To avoid this issue, we move all cloud particles radially inward by distributing them with a separation of $\Delta x_\mathrm{loc} / 2.001$. This restores spherical symmetry as each cloud particle is unambiguously assigned to a host cell (See Figure \ref{fig:cloud_distribution} for a 2D slice through the cloud of particles).

Each cloud particle samples the properties of its host cell. The mass-weighted average gas properties used in the BHL formula are then calculated by summing over all cloud particles according to: 
\begin{equation}
	x_\bullet = \frac{\sum x_i m_i \alpha_i }{\sum \alpha_i \sum m_i} 
\end{equation}
where the contribution of each cloud particle $i$ is weighted by a Gaussian kernel,
\begin{equation}
\alpha_i = \exp \left( - \frac{r_{i}^2} { r_\mathrm{kernel}^2 } \right)
\label{eq:kernel}
\end{equation}
and $r_i$ is the cloud particle distance from the sink. The width of the kernel, $r_\mathrm{kernel}$, depends on the value of an interpolated scale radius, 
\begin{equation}
 r^\mathrm{BHL}_\mathrm{host}  = \frac{G M_\mathrm{sink}}{{v}_\mathrm{host}^2+ {c}^2_\mathrm{s,host}}, 
 \end{equation}
 where $v_\mathrm{host}$ and $c_\mathrm{s,host}$ are the relative velocity and sound speed of the sink host cell respectively. $r_\mathrm{kernel}$ is also limited by 
the local cell size, so that altogether:
 \begin{equation}
 	r_\mathrm{kernel} = \begin{cases}
		{ \Delta x_\mathrm{loc}  }/{4} & \mbox{ if } r^\mathrm{BHL}_\mathrm{host} < { \Delta x_\mathrm{loc}  }/{4} \\
		2 \Delta x_\mathrm{loc} & \mbox{ if } r^\mathrm{BHL}_\mathrm{host} > 2 \Delta x_\mathrm{loc} \\		r^\mathrm{BHL}_\mathrm{host} & \mbox{otherwise}.
	\end{cases}
\end{equation}
 
Gas mass is removed from local cells, and added to the sink particle, by looping over the cloud particles and removing mass from each host cell according to
\begin{equation}
\Delta m_{i}^\mathrm{cell} = \frac{\dot{M}_\bullet^\mathrm{BHL}\mathrm{d}t} { \sum \alpha_i}  \times \alpha_i
\label{eq:m_acc}
\end{equation}
where $\mathrm{d}t$ is the local timestep. $\dot{M}_\bullet^\mathrm{BHL}= \dot{M}^\mathrm{BHL}( \rho_\bullet, v_\bullet, c_{s,\bullet})$ is the accretion rate calculated on the fly according to the BHL formula in Equation \ref{eq:bondi_hoyle}, evaluated using local mass-weighted average quantities.

To avoid creating numerical instabilities by removing too much gas mass from a single cell, the total mass accreted per timestep is capped at 75 \% of the cloud particle host cell gas mass. This criteria is commonly employed in RAMSES, and ensures that no single cell is emptied in a single timestep when the sink particle moves across the grid. As we force the sink particle to remain in a specific host cell, we also introduce a density floor, $\rho_\mathrm{min}=10^{-5}$, below which accretion is not permitted. We emphasize that this extra parameter is specific to the set of simulations presented in this work rather than a standard RAMSES parameter. Our  results are insensitive to the choice of $\rho_\mathrm{min}$ as long as $\rho_\mathrm{min} \ll \rho_\infty$.  

\subsection{The drag force algorithm}
\label{subsec:fdrag_analytic}

One way in which the black hole interacts with gas in its vicinity is through dynamical friction. The relative velocity between the two creates an overdensity downstream of the black hole, and the enhanced gravitational pull that ensues acts as a drag force. This process transfers momentum from the sink to the gas --- contrary to accretion which transfers momentum from the gas to the black hole --- thus boosting the tendency of the sink to stay attached to overdense local gaseous structures. When this drag is unresolved, black holes often dynamically decouple from the gas, leading, in the worst cases, to their ejection from the galaxy disc. This has been a recurring problem in galaxy simulations including SMBHs \citep{Sijacki2007,Volonteri2016,Biernacki2017}.

RAMSES includes a sub-grid model for dynamical friction \citep{Dubois2013}, in which the drag force on the sink is estimated according to the analytic formula derived by \citet{Ostriker1998}:
\begin{equation}
	\bold{F}^{D}_\infty = - I \frac{4 \pi G^2 M_\mathrm{sink}^2 \rho_\infty}{v_\infty^2} \hat{\bold{v}}_\infty =  F^{D}_\infty \hat{\bold{v}}_\infty 
	\label{eq:fdrag}
\end{equation}
where $ \hat{\bold{v}}_\infty $ is the unit vector pointing in the direction of the relative velocity between the gas and the sink, and 
\begin{equation}
	I = \begin{cases}
		\frac{1}{2} \ln \left( \frac{ 1+ \mathcal{M}_\infty}{1-\mathcal{M}_\infty} \right) - \mathcal{M}_\infty  &  { \rm if }\ \ \mathcal{M}_\infty < 1\\
		\frac{1}{2} \ln \left( \mathcal{M}_\infty^2 -1 \right) + \ln (\Lambda)& { \rm if }\ \ \mathcal{M}_\infty > 1. \\
		\end{cases}
		\label{eq:Idrag}
\end{equation}	
In the previous equation, $\ln(\Lambda)=\ln \left( r_\mathrm{max}/r_\mathrm{min} \right) =3.2$ is the Coulomb logarithm, with the numerical value chosen according to \citet{Chapon2013}, $r_\mathrm{max} = (c_{s,\infty} + v_\infty) t $ is the maximum distance information has travelled downstream of the sink, and $r_\mathrm{min}$ is the softening length used to curtail the diverging density profile near the sink. We re-examine the choice for $\ln(\Lambda)$ in this paper, as the value from \citet{Chapon2013} is extracted from black hole merger simulations that do not include accretion.

In the simulations presented here, the sink particle is held at a fixed position, so the drag force sub-grid algorithm is not employed. However, we compare the analytic formula given by Equation \ref{eq:fdrag} to values measured directly from gravitational wakes of the sink particle when the latter is resolved. 
As with the BHL accretion rate, the analytic model (Equation \ref{eq:fdrag}) calculates the drag force from gas flow properties at infinity. However, the sub-grid model implemented in RAMSES evaluates the drag force on the sink based on the same local mass-weighted quantities used for BHL accretion,  $\rho_\bullet, \bold{v}_\bullet$ and $ \mathcal{M}_\bullet$.
 
In this paper, we thus distinguish between four different drag force calculations:
 \begin{enumerate}
 	\item $F^D_\infty$ is the drag force given by Equation \ref{eq:fdrag}, using the quantities at infinity.
	\item $F^D_\bullet=F^D(\rho_\bullet,v_\bullet,\mathcal{M}_\bullet)$ is the drag force given by Equation \ref{eq:fdrag}, using mass-weighted quantities from the accretion region, i.e. the drag force as estimated by the sub-grid model.
	\item $F^D_\diamond = \sum \frac{G M_\mathrm{sink} m_i}{r_i^2} \hat{\bf{r}}_i$ is the net gravitational force exerted on the sink by gas on the grid. It is found by summing over all cells $i$ in the box, where $m_i$ and $r_i$ are the cell mass and cell position vector relative to the sink respectively. As the setup is symmetric, this is equal to the total gravitational attraction of the wake downstream of the sink.
	\item $F^D_\mathrm{tot} = F^D_\bullet + F^D_\diamond$ is the total drag force acting on the sink when the sub-grid algorithm is active.
\end{enumerate}

\subsection{Plotting conventions}
{Streamlines annotated on slice plots follow the velocity field in the plane of the slice. They therefore do not represent three dimensional particle trajectories and are merely added to guide the eye as to the dominant flow patterns in the plot. For clarity, a minimum distance is enforced between streamlines. This discrete sampling can mean that some streamlines are interrupted. }
  

\section{Accretion}
\label{sec:accretion}

The BHL formula in Equation \ref{eq:bondi_hoyle} is based on two limiting cases. The pure Bondi problem, where   the sink is at rest relative to the uniform density background, i.e. $\mathcal{M}_\infty=0$, and the Hoyle-Lyttleton problem, where $\mathcal{M}_\infty \gg 1$. We investigate both cases in detail, in this section and the next, before exploring the parameter space further. 

\subsection{The Bondi problem}
\label{sec:bondi}

\begin{figure*}
	\begin{tabular}{cc}
	\includegraphics[height=0.8\columnwidth]{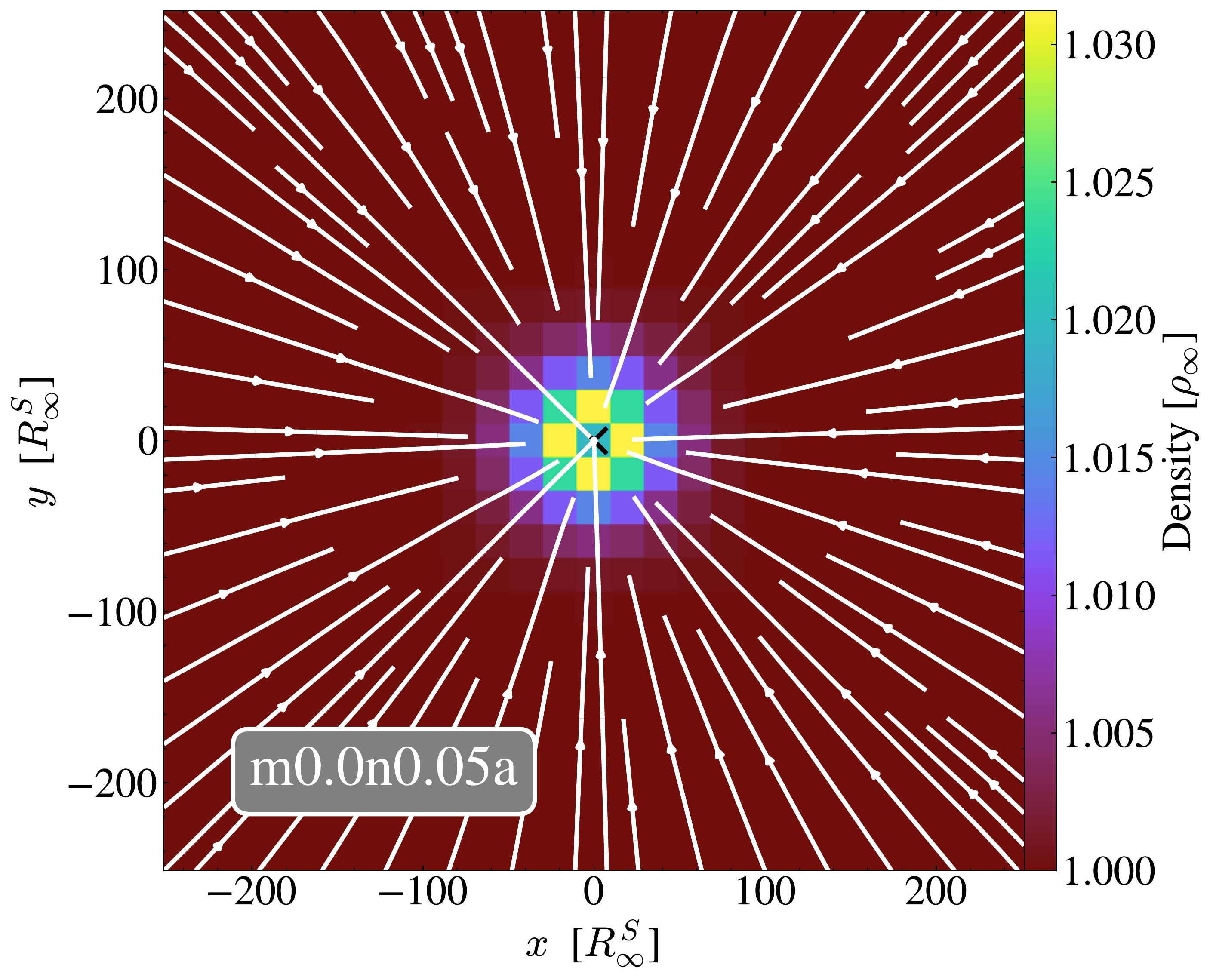} &
	\includegraphics[height=0.8\columnwidth]{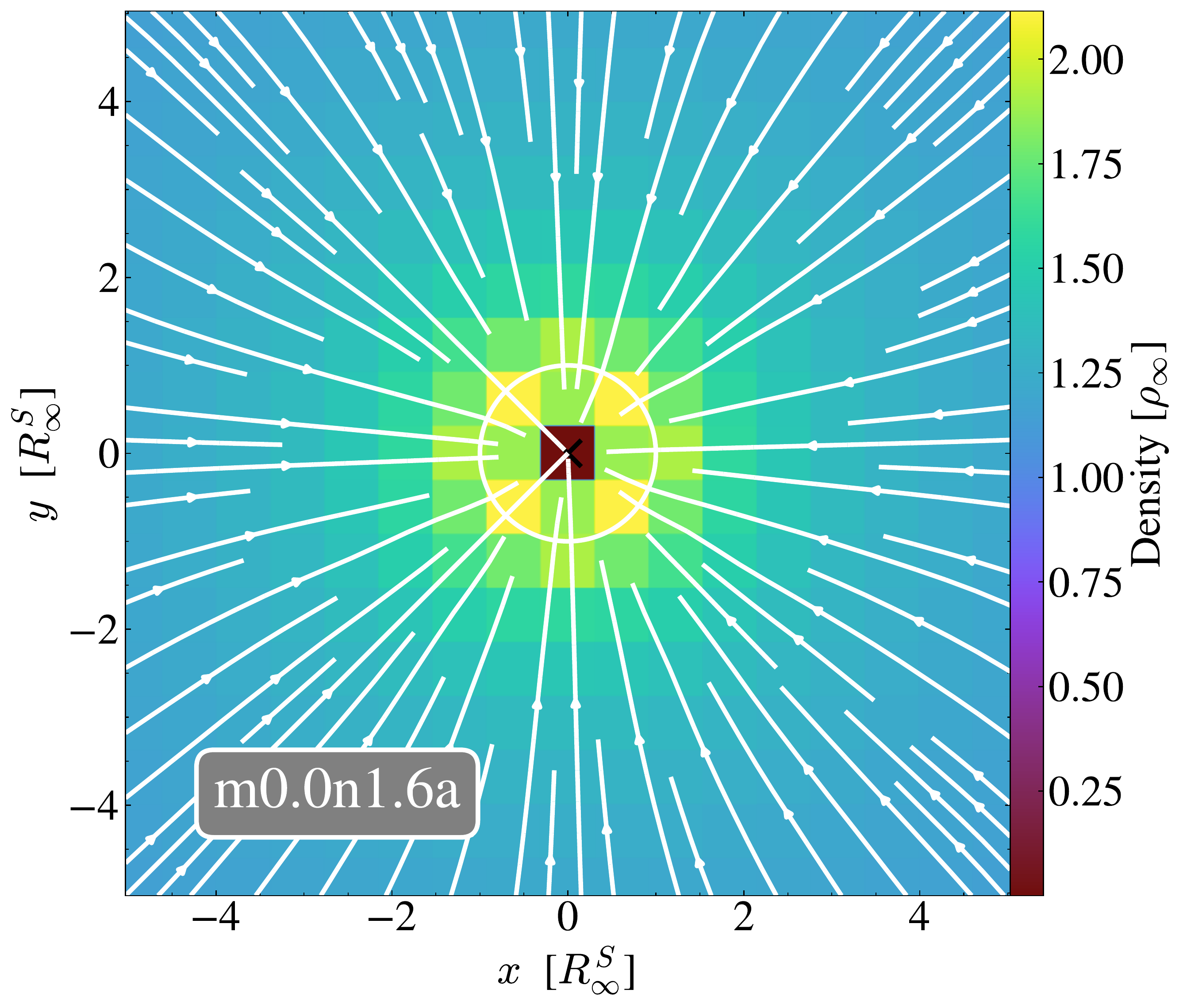} \\
	\includegraphics[height=0.8\columnwidth]{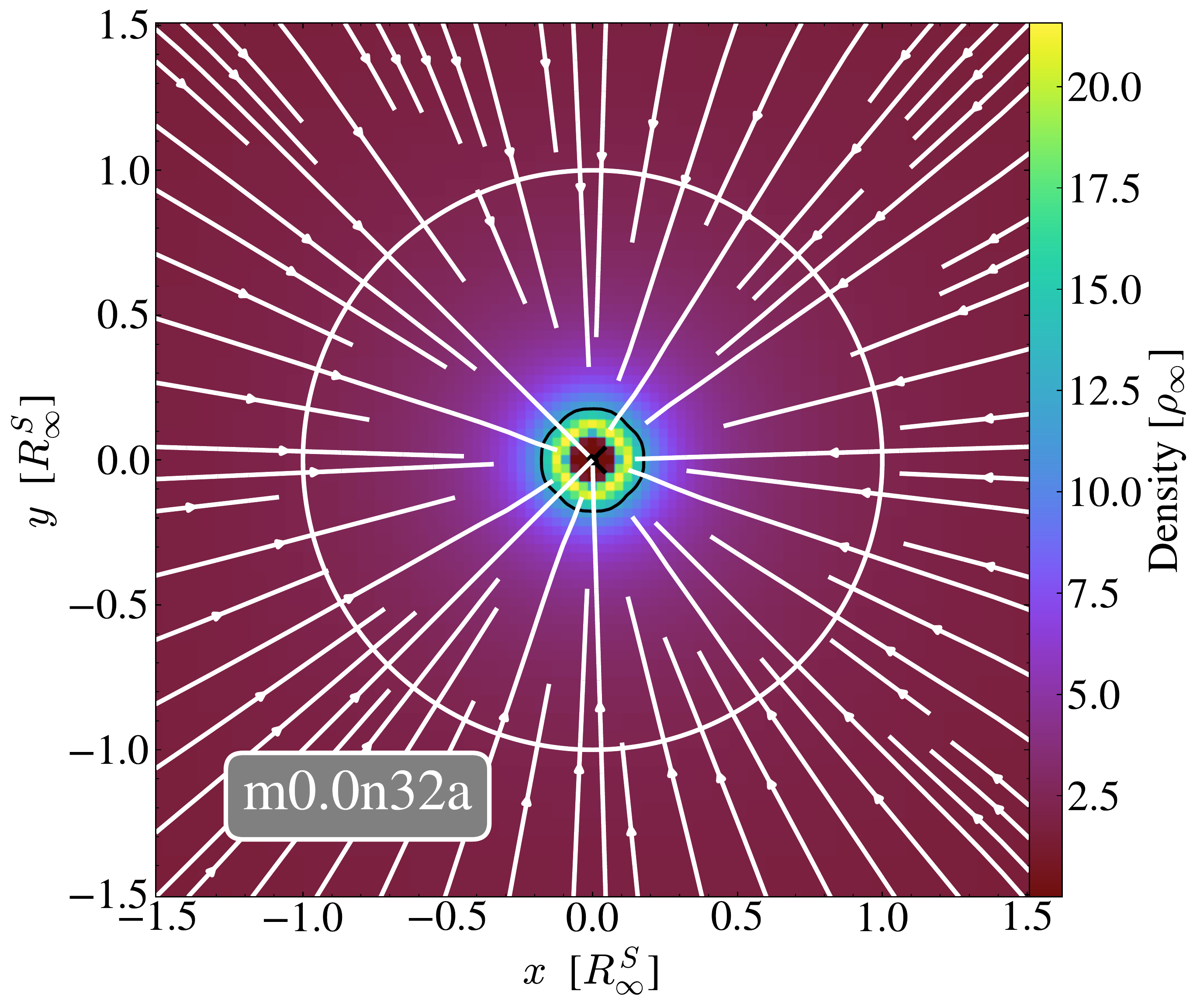} &
	\includegraphics[height=0.8\columnwidth]{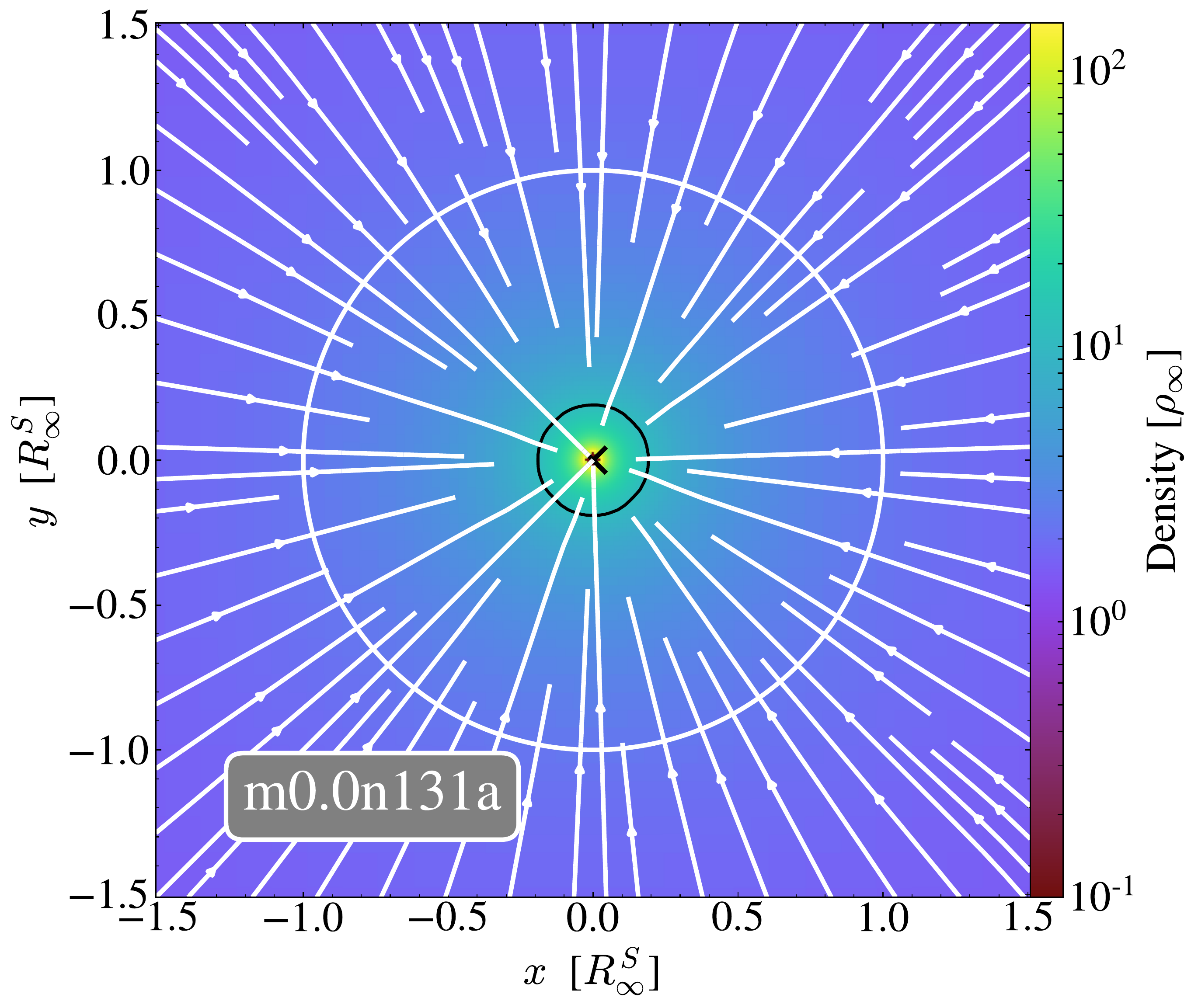} \\
	\end{tabular}
    	\caption{Slices through the central density peak surrounding the sink  for resolutions of $N=0.05,1.6,32$  and $131$ respectively. $R^S_\infty$ is denoted by a white circle. The sink is located in the centre of the box and denoted by a black cross. The black contour indicates the sonic surfaces in each slice, when these latter are resolved. Note that the top two panels are zoomed out in comparison to the bottom ones to show the relevant features at each resolution, hence the circles marking $R^S_\infty$ are much smaller in these panels. Each simulation is shown at at $t=25$.}
    \label{fig:m0_density}
\end{figure*}

\begin{figure}
	\includegraphics[width=\columnwidth]{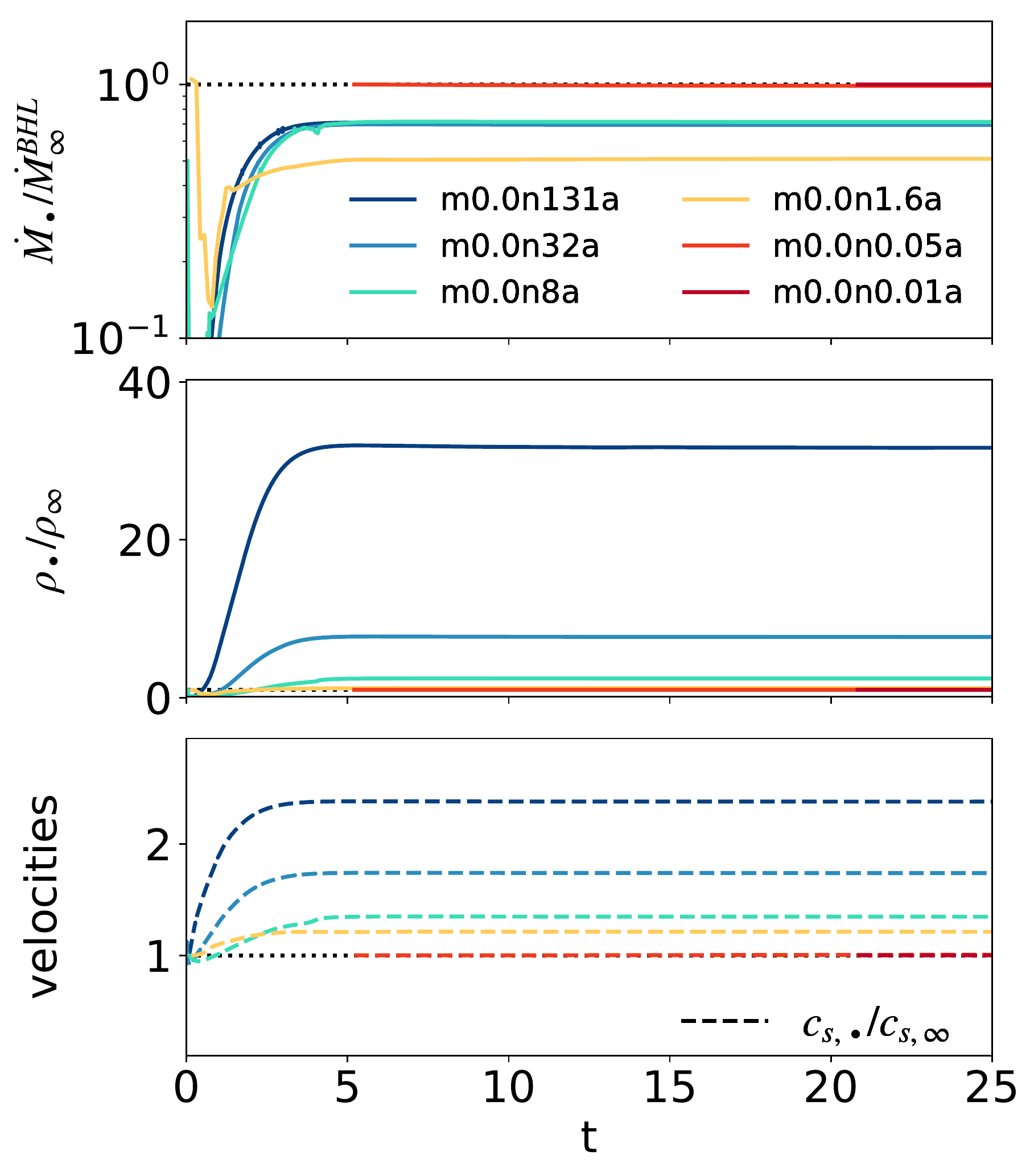}
    	\caption{Accretion rates and gas properties as sampled by the cloud particles, for a variety of resolutions, in the pure Bondi case, where $\mathcal{M}_\infty=0.0$. The bottom panel shows the sound speed only, as the relative velocity at infinity is zero, and can therefore not be used as a scaling factor.}
    \label{fig:m0_profiles}
\end{figure}

As expected from the analytic work by \citet{Bondi1944,Bondi1952}, Figure \ref{fig:m0_density} shows that the global flow pattern is symmetric and directed radially towards the sink even at very low resolutions, with the sink located at the density peak. Only the immediate sink environment is influenced by the resolution. There is an initial period during which the simulation is dominated by initial condition transients, which last until  $t \approx 5$ (see Figure \ref{fig:m0_profiles}). After the simulation has settled into a steady flow pattern, the accretion rate onto the sink, $\dot{M}_\bullet$, converges. Note that the actual accretion rate onto the sink, $\dot{M}_\bullet $ is not necessarily the same as the BHL accretion rate evaluated from the local gas properties $ \dot{M}^\mathrm{BHL}_\bullet$, as $\dot{M}$ is limited by the total gas mass present in the accretion region, $\sum ( \rho_i \Delta x_i^3 \alpha_i )$.

\begin{figure}
	\includegraphics[width=\columnwidth]{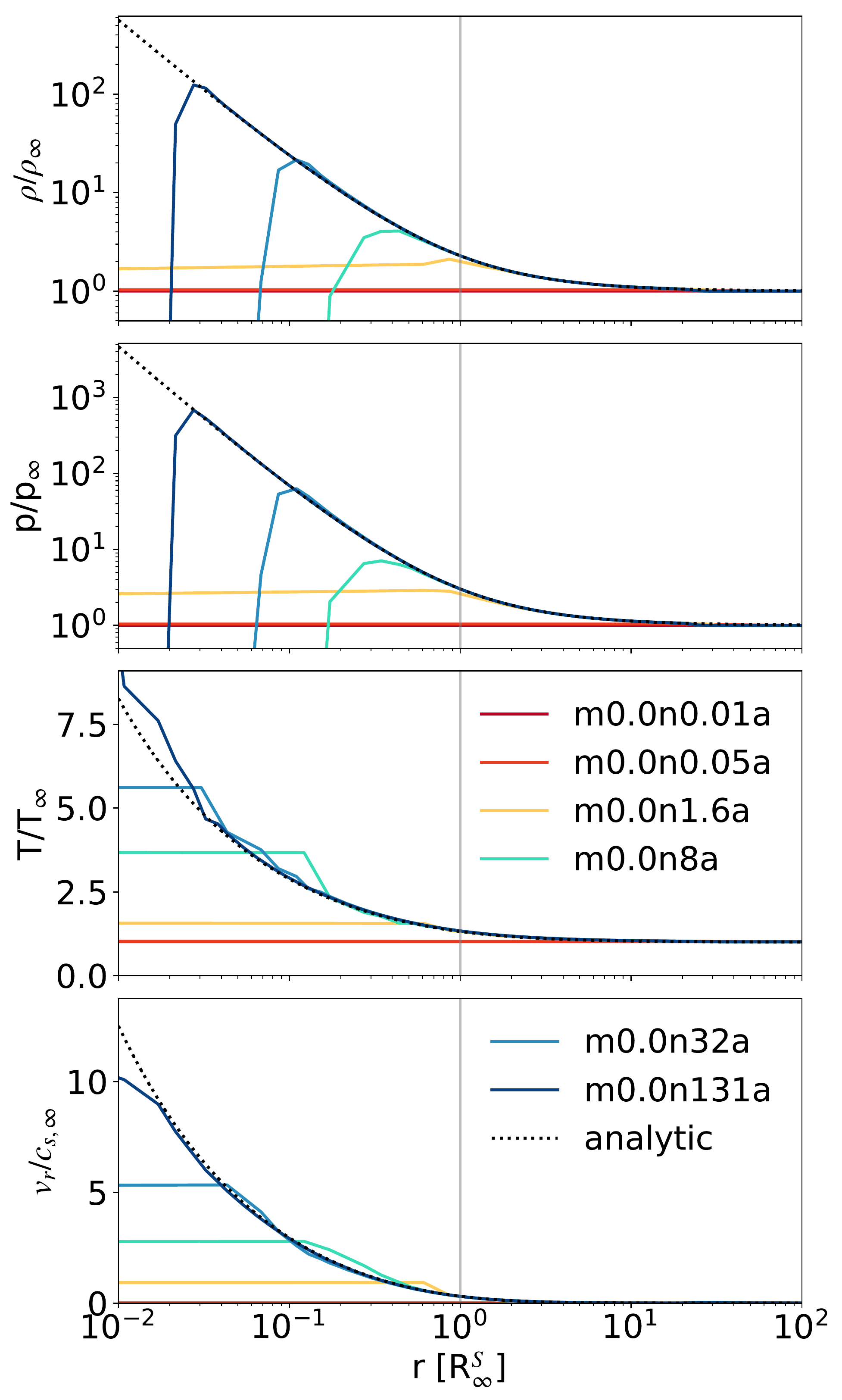}
	\caption{{Radial density, pressure, temperature and radial velocity profiles for the pure Bondi problem, at a range of resolutions, at time $t=20$. The dotted line denotes the analytic solution according to equation 5 in \citet{Bondi1952}.}}
	\label{fig:m0_radial}
\end{figure}

{Figure \ref{fig:m0_radial} demonstrates that for all simulations presented here, the radial density, pressure, temperature and velocity profiles provide excellent agreement with the analytic predictions by \citet{Bondi1952} at radii larger than the size of the accretion region, i.e. at $r > 2 \Delta x$. Within the accretion region, density and pressure drop sharply as mass is removed from the grid and added to the black hole, while temperatures can increase by up to a factor of 2 above the analytic value. As the analytic profiles diverge as $r \rightarrow 0$, some deviation within the central region is to be expected, particularly in the velocity field which has to go to zero at the centre of a spherically symmetric flow.}

The analytic solution for the BHL accretion rate assumes that gas properties are measured far from the influence of the sink, a situation that is best captured in the simulations presented here when the cell size is significantly larger than the sink's gravitational scale radius, i.e. $\Delta x \gg R^S_\infty$. Figure \ref{fig:m0_profiles} shows that for the unresolved case, here m0.0n0.05a, the accretion rate onto the black hole, as well as the local gas properties as calculated using the cloud particles, reflect the analytic solution closely. The local overdensity before accretion is very shallow, at $\rho \sim 1.035 \rho_\infty $ (see m0.0n0.05a Figure \ref{fig:m0_density}, top left), and is suppressed within the cell containing the sink once accretion starts.

The other extreme is the most highly resolved case probed here (m0.0n131a, bottom right panel of Figure \ref{fig:m0_density}), where an overdense peak develops within the Bondi radius. When the gas crosses the sonic radius, it transitions from subsonic to supersonic flow as it evolves towards a free fall solution before being accreted. The rising local density increases the BHL rate computed on the fly and the maximum amount of gas permitted is removed from the central cells at each time step. This means that the accretion algorithm effectively transitions from the BHL algorithm to a supply limited accretion scheme, where the accretion rate onto the sink is set by the gas inflow rate into the spherical accretion region of the sink, which has a radius of $ \approx 2  \Delta x_\mathrm{min}$. We refer to this as Supply Limited Accretion (SLA) throughout this paper. 

Note that at this resolution, the accretion rate onto the sink settles well below the analytic BHL rate, to $\dot{M}_\bullet/\dot{M}^\mathrm{BHL}_\infty \sim 0.78$, in agreement with results in \citet{Edgar2004}. Simulations with better force resolution probe the density profile on smaller scales, and therefore measure higher densities $\rho_\bullet$. As the contraction is adiabatic, higher densities have correspondingly higher sounds speed $c_{s,\bullet}$, as can be seen in Figure \ref{fig:m0_profiles}. For this reason, although the analytic model postulates that the gas should transition to supersonic near the Bondi radius, the black contours in Figure \ref{fig:m0_density} show that the transition occurs at a smaller radius, again in agreement with previous numerical simulations. Finally, the accretion rate has already converged to its $0.78 \times \dot{M}^\mathrm{BHL}_\infty$ value even at the comparatively modest resolution of $N=32$, in simulation m0.0n32a.

The partially resolved case, m0.0n1.6a, where $R^{S}_\infty \sim \Delta x_\mathrm{min}$ shows intermediate behaviour, with a shallower central density feature and a smaller evacuated region. This simulation also shows more noticeable grid effects, both in the stream lines and in the central density peak, as spherical symmetry is poorly described by the small number of Cartesian resolution elements when $0.1 < N < 4$. The resulting steady state accretion rate is lower than the converged value, with $  \dot{M}_\bullet/\dot{M}_\infty^\mathrm{BHL}\sim 0.6 $ as the local density feature feeding the black hole is not replenished efficiently. 
  
\begin{figure}
	\centering
	\includegraphics[width=\columnwidth]{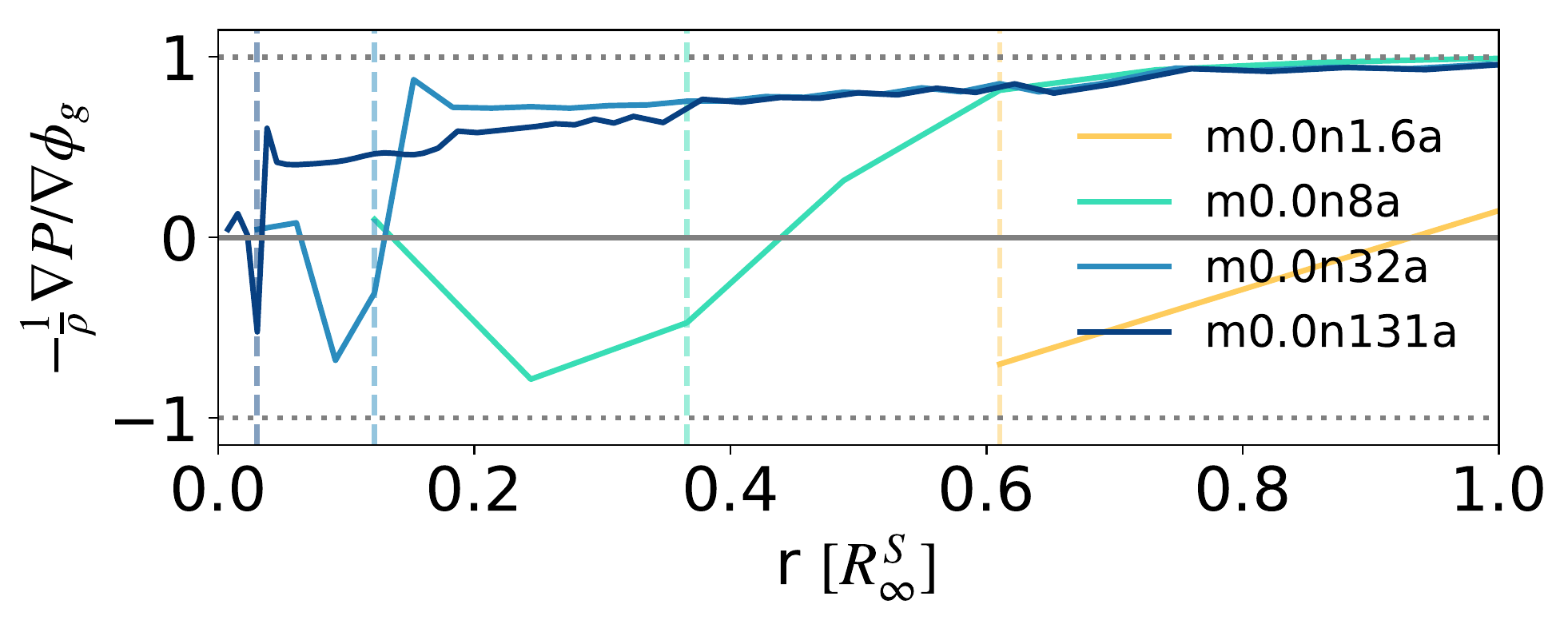}
	\caption{Acceleration due to the pressure gradient within the Bondi radius, compared to the gravitational acceleration at a range of radii $r$. The dashed vertical lines mark the point of highest pressure, i.e. the edge of the accretion region.}
	\label{fig:pressure_mach0}
\end{figure}

Arguably the most worrisome numerical aspect of transitioning to SLA is that the force due to the pressure gradient artificially created by the low density region developing in the immediate vicinity of the sink might dominate the gravitational force on the gas. Figure \ref{fig:pressure_mach0} shows that while such a pressure gradient does reinforce the gravitational pull on the gas at the edge of the accretion region, it is not the dominant force. Morevover, the contribution of this pressure force decreases for simulations with higher resolution as the gravitational acceleration $-\nabla \phi_{g}$ increases faster than the pressure gradient, $\nabla P$, within the Bondi radius. As previously mentioned, this effect is also insensitive to our choice of density floor, as long as $\rho_\mathrm{min} \ll \rho_\infty$, as for sufficiently small pressure inside the accretion region, $\nabla P \simeq \Delta P / \Delta x = (P_\mathrm{edge} - P_\mathrm{in}) / \Delta x \rightarrow P_\mathrm{edge} / \Delta x$, and therefore only depends on the cell size $\Delta x$ and the pressure at the edge of the accretion region, $P_\mathrm{edge}$, not the actual pressure inside the accretion region $P_\mathrm{in}$. 

From the pure Bondi problem presented here we conclude that the accretion algorithm is well behaved at all resolutions. Indeed, at low resolution, when $N < 0.05 $, local gas properties measured in the vicinity of the sink, using cloud particles, produce an accretion rate in excellent agreement with the analytic BHL formula. At high resolution, where $N > 8 $, accretion is driven by the local gas supply into the accretion region, and the algorithm transitions to SLA. The accretion rate onto the sink converges to the correct value of $\dot{M}_\bullet/\dot{M}^\mathrm{BHL}_\infty \sim 0.78$ in that case, fed by supersonically free-falling gas well within the Bondi radius. For intermediate resolutions of $ 0.05 < N < 8$, grid effects lead to poorer spherical symmetry and lower accretion rates, as gas neither reflects the values at infinity nor forms a central gas profile able to efficiently feed the sink, a conclusion also reached by \citet{Krumholz2004}. However, independently of resolution, the accretion rate onto the black hole estimated using the sub-grid algorithm remains within $20 \%$ of the correct $0.78 \times \dot{M}^\mathrm{BHL}_\infty$ value.
    
\subsection{The Hoyle-Lyttleton problem}
 \label{sec:hoyle_lyttleton} 
  
The other analytic solution was developed by \citep{Hoyle1939} for an accretor moving supersonically through a uniform medium, where the bulk velocity dominates over the local sound speed, such that $\dot{M}^\mathrm{BHL}_\infty $ approaches $\dot{M}^\mathrm{HL}_\infty = G M_\mathrm{sink} / v_\infty^2$. In this section we investigate accretion onto the sink in the highly supersonic case where $\mathcal{M}_\infty=10$.

\subsubsection{The adiabatic case}
\label{sec:quasi_adiabatic}

\begin{figure*}
	\begin{tabular}{cc}
	\includegraphics[height=0.8\columnwidth]{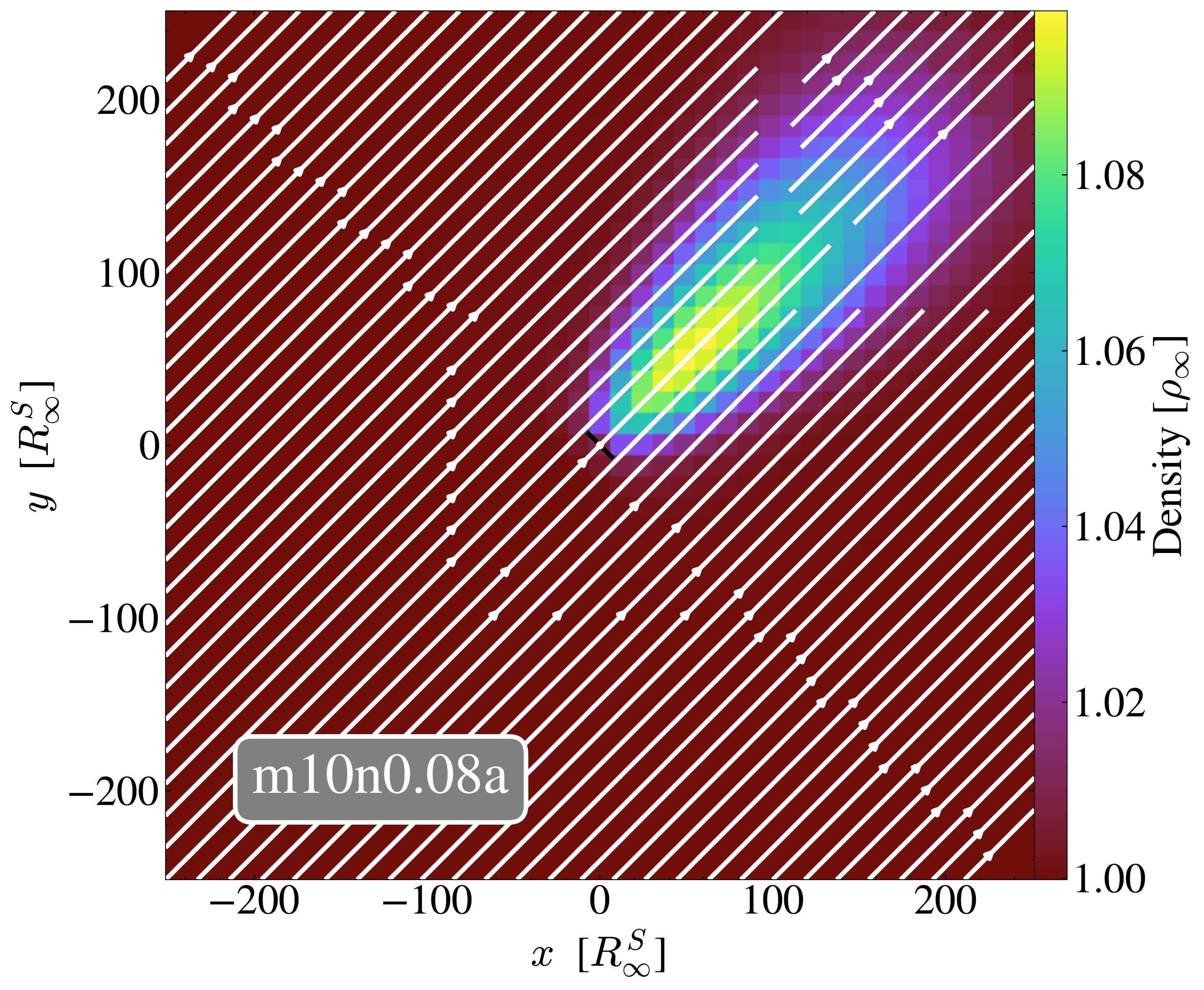} &
	\includegraphics[height=0.8\columnwidth]{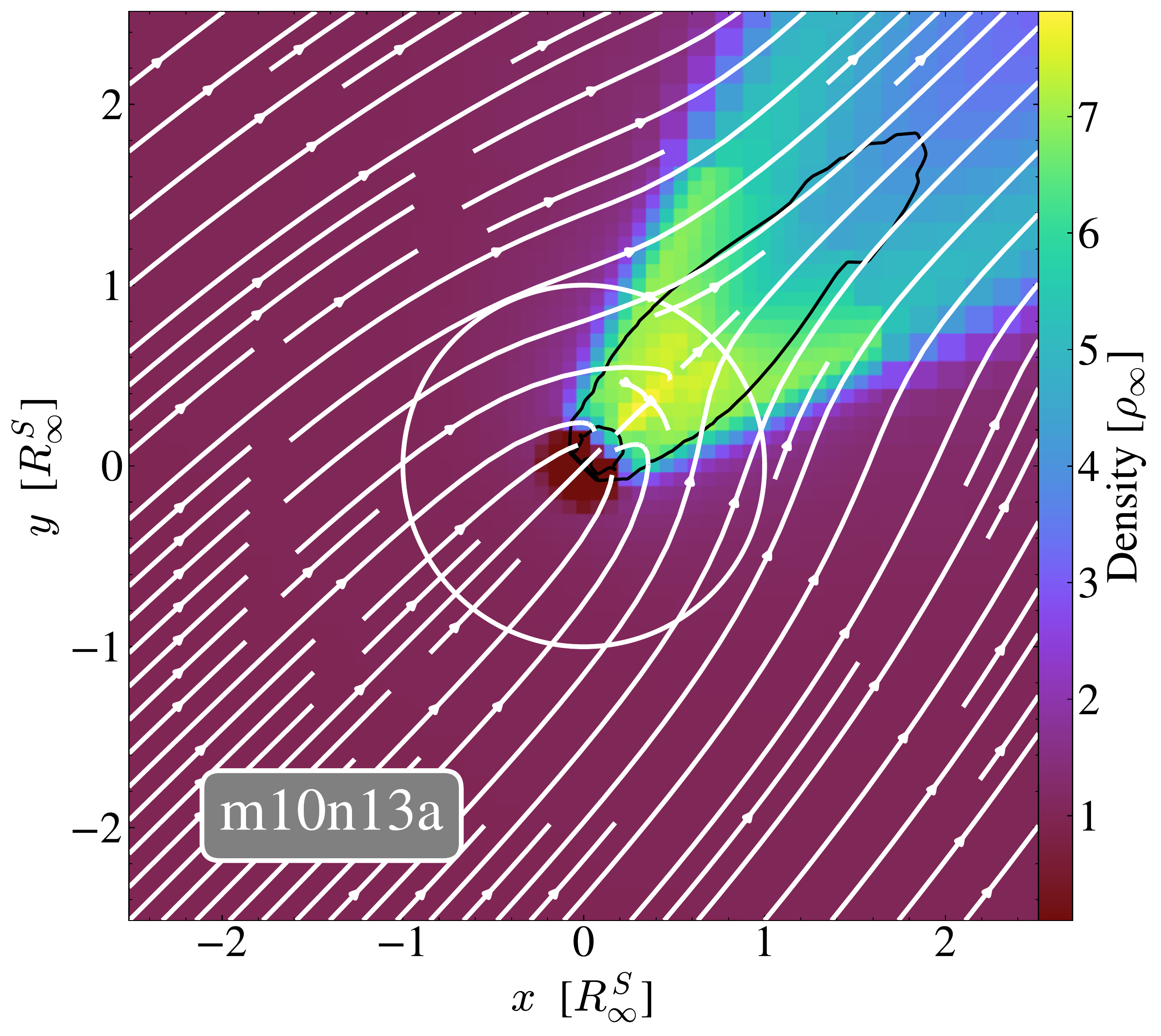}\\
	\includegraphics[height=0.8\columnwidth]{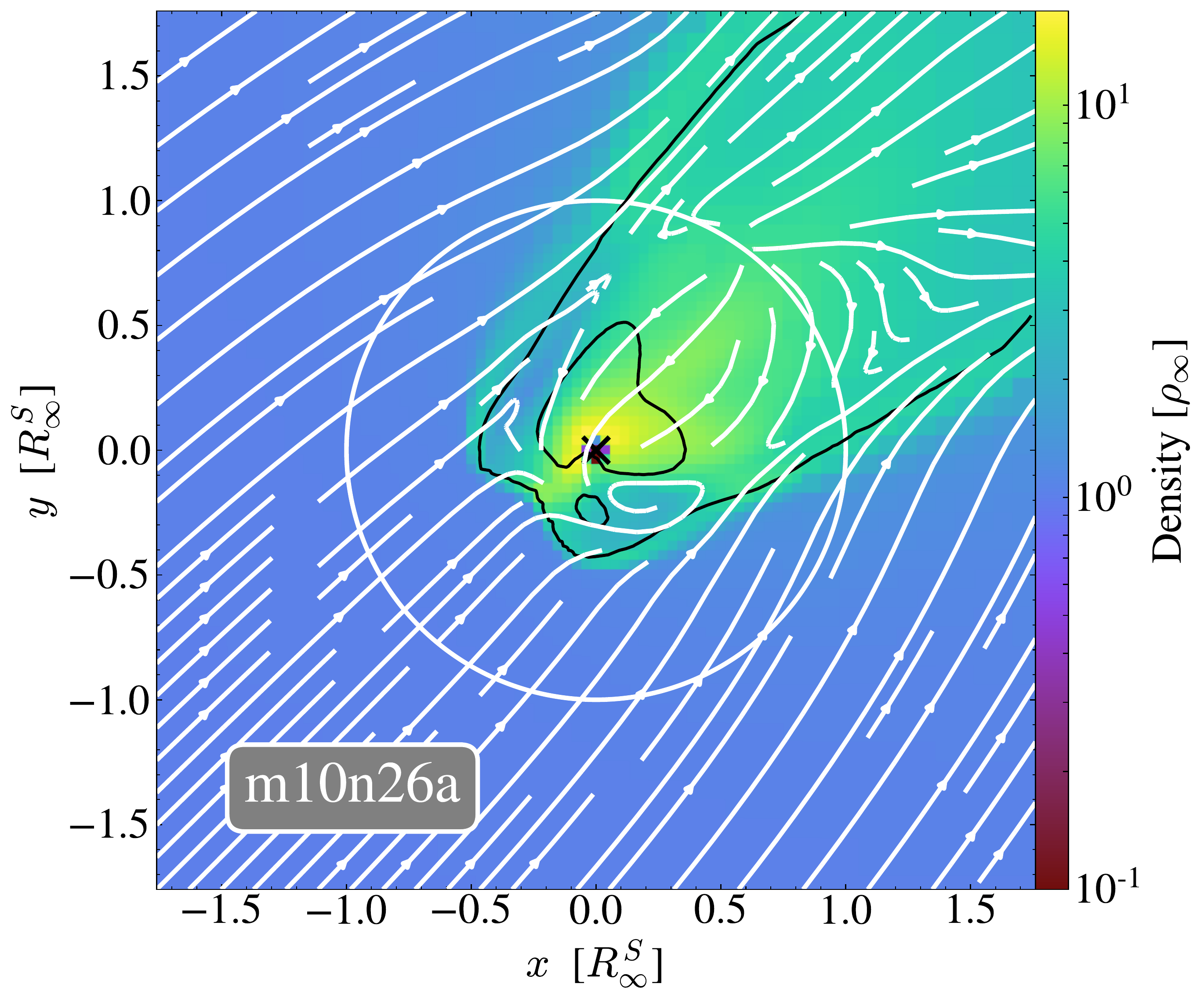}&
	\includegraphics[height=0.8\columnwidth]{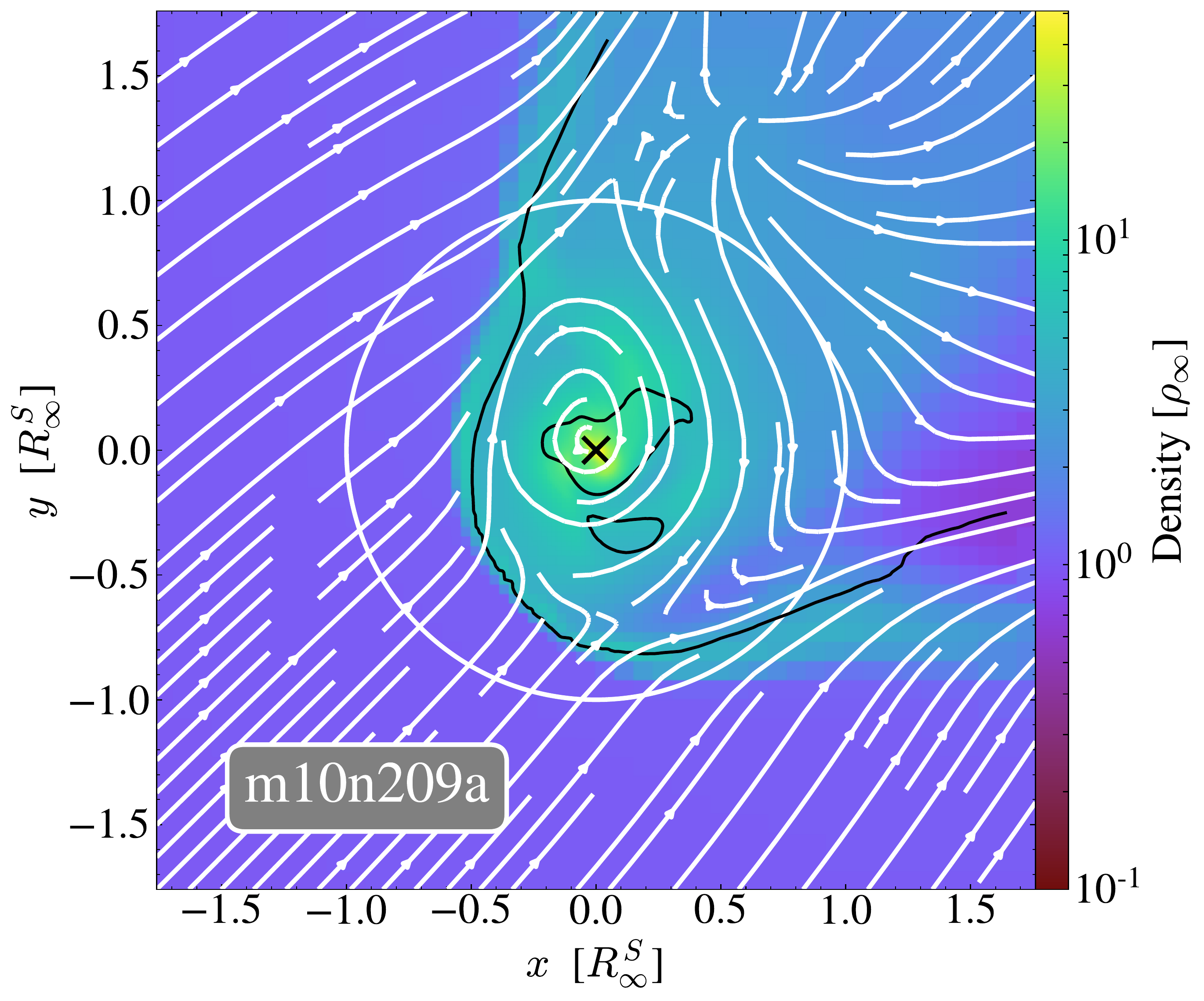}
	\end{tabular}
    	\caption{Slices of the central density feature for resolutions of $N=0.08,13,26$ and $209$ respectively. The  white circle denotes the size of the accretion radius $R^S_\infty$. The sonic surface for each slice is delineated by a black contour. Each simulation is shown at $t=25$.}
    \label{fig:m10_density}
\end{figure*}

  \begin{figure}
	\includegraphics[width=\columnwidth]{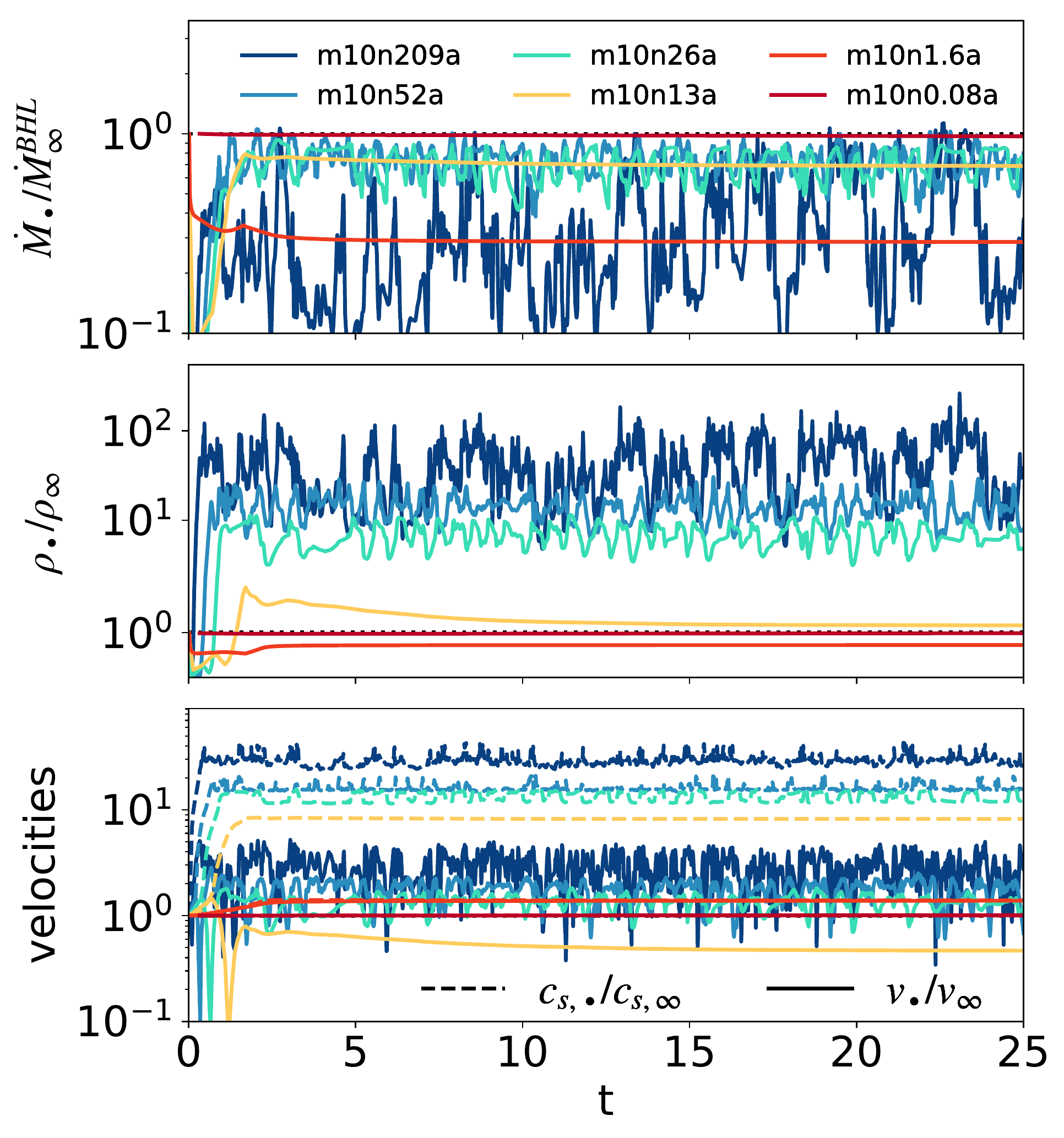}
    	\caption{Accretion rates (top panel) and mass weighted gas properties (density, velocity and sound speed)  at a variety of resolutions for  the supersonic Hoyle-Lyttleton case, where $\mathcal{M}_\infty=10.0$.  All gas properties are measured by the kernel-weighted cloud particles within the accretion region of the sink.}
    \label{fig:m10_profiles}
\end{figure}

For the adiabatic case, when $\gamma=1.3334$, in agreement with expectations from the analytic solutions \citep{Hoyle1939,Ostriker1998}, we find that a conical wake develops downstream of the sink, as is evident in Figure \ref{fig:m10_density}. In the unresolved case, m10n0.08a, the overdensity around the sink is small, and the streamlines and properties of the gas are only mildly perturbed by the presence of the sink. Therefore, $\dot{M}_\infty^\mathrm{BHL}\approx \dot{M}_\bullet^\mathrm{BHL}\approx\dot{M}_\bullet$, and the sink accretes according to the analytic solution (top panel in Figure \ref{fig:m10_profiles}). The gravitational wake is especially prominent because the simulations presented here are isolated. If the sink was embedded in a non-uniform medium, such as is typically found for black holes in galaxy simulations, we expect that local inhomogeneities would quickly wash out the gravitational focusing effect of the black hole.

With increasing resolution, such as m10n13a in Figure \ref{fig:m10_density}, the flow patterns resembles the analytic solution by \citet{Hoyle1939}, with bent streamlines, stagnation point, and accretion column clearly visible in the density slice. The bow-shock with the characteristic increase in density towards the edge of the shock, predicted by \citet{Ostriker1998}, also becomes apparent. At these intermediate resolutions, the shock is attached to the accretor, and the solution is stable (see Figure \ref{fig:m10_profiles}). As in the pure Bondi case, $\dot{M}_\bullet < \dot{M}^\mathrm{BHL}_\bullet$, and the accretion algorithm transitions to SLA, with the under-dense accretion region visible in the density slices (Figure \ref{fig:m10_density}).

\begin{figure*}
	\centering
	\begin{tabular}{ccc}
		\includegraphics[width=0.31\textwidth]{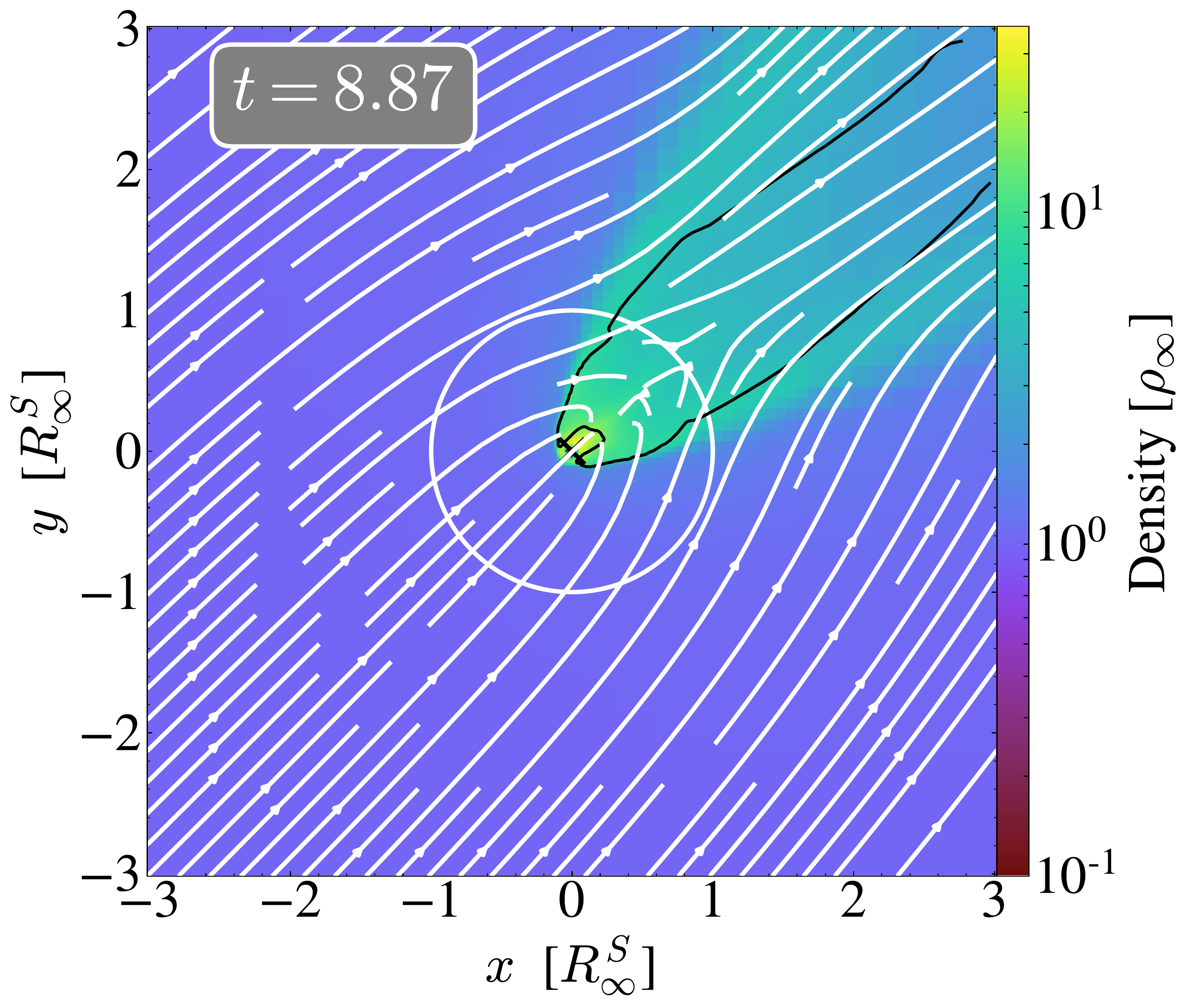} &
		\includegraphics[width=0.315\textwidth]{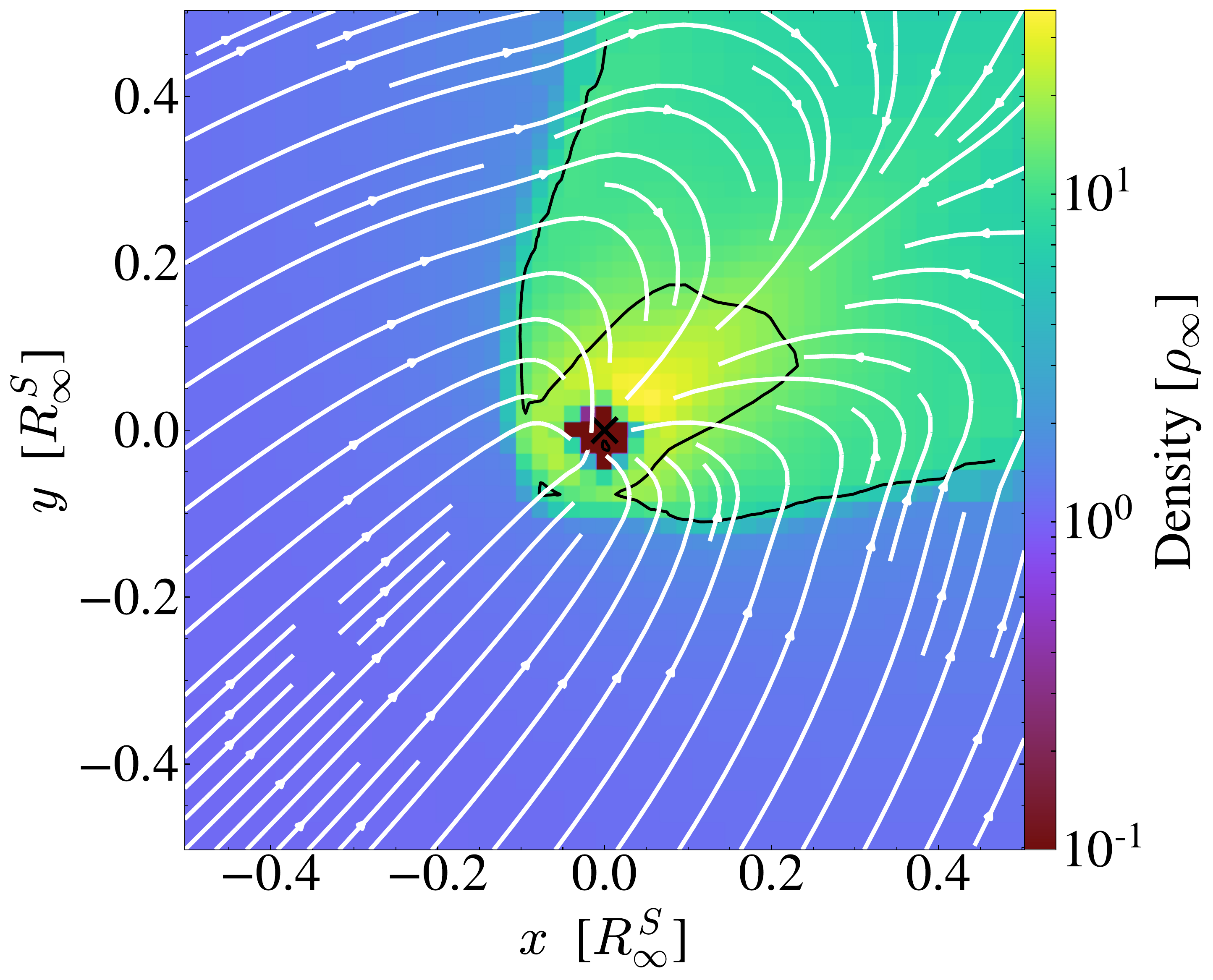} &
		\includegraphics[width=0.325\textwidth]{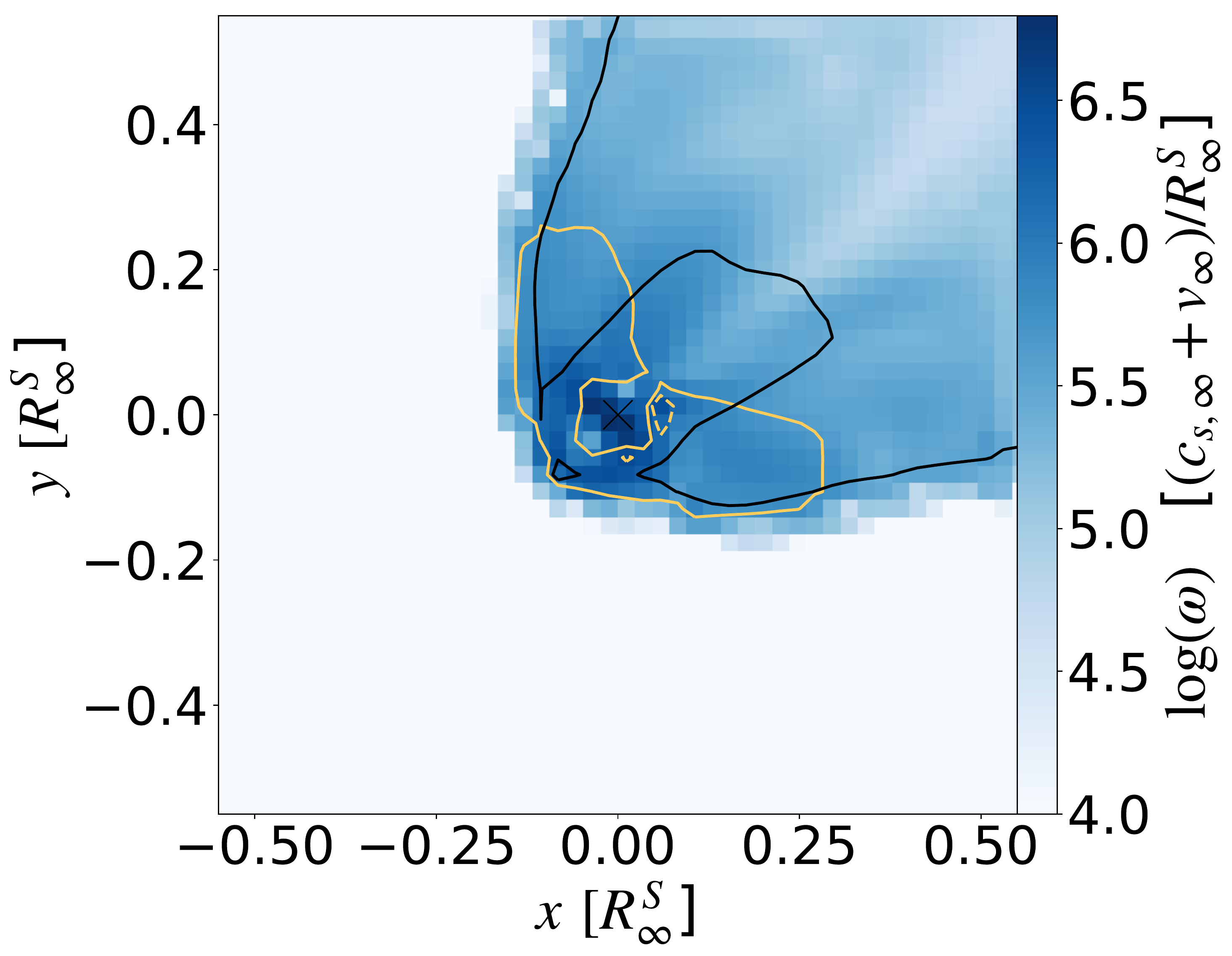} \\
		\includegraphics[width=0.31\textwidth]{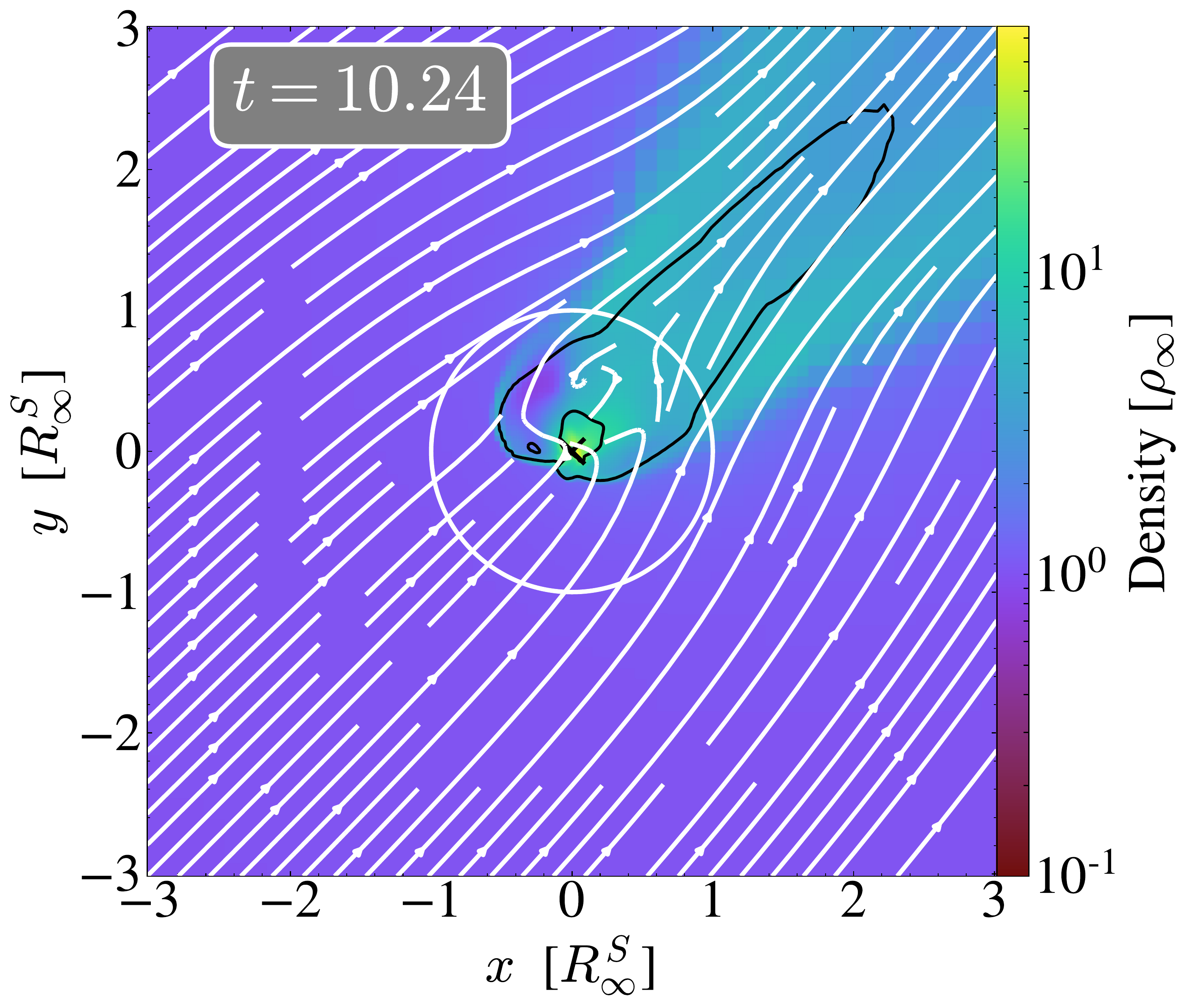} &
		\includegraphics[width=0.315\textwidth]{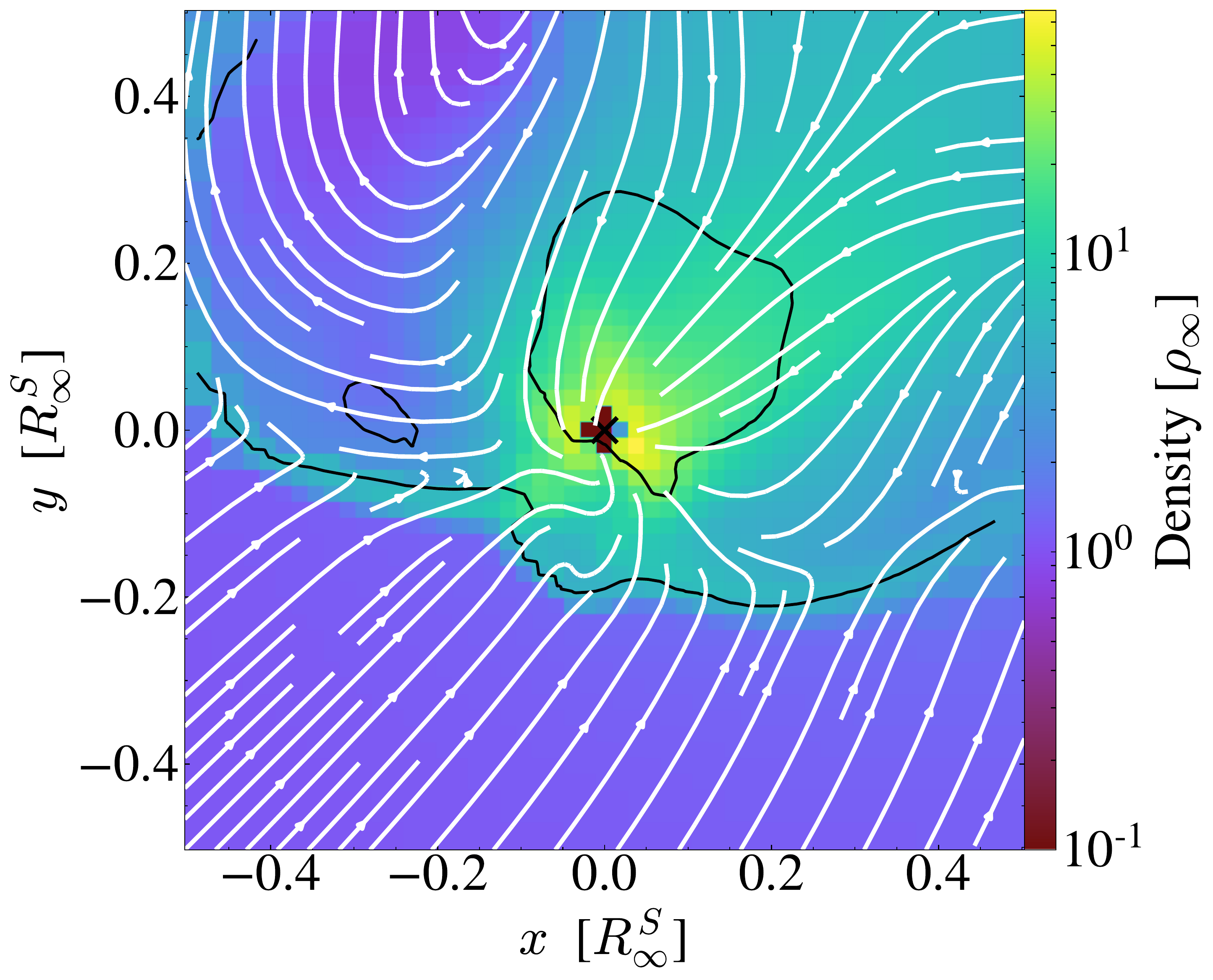} &
		\includegraphics[width=0.325\textwidth]{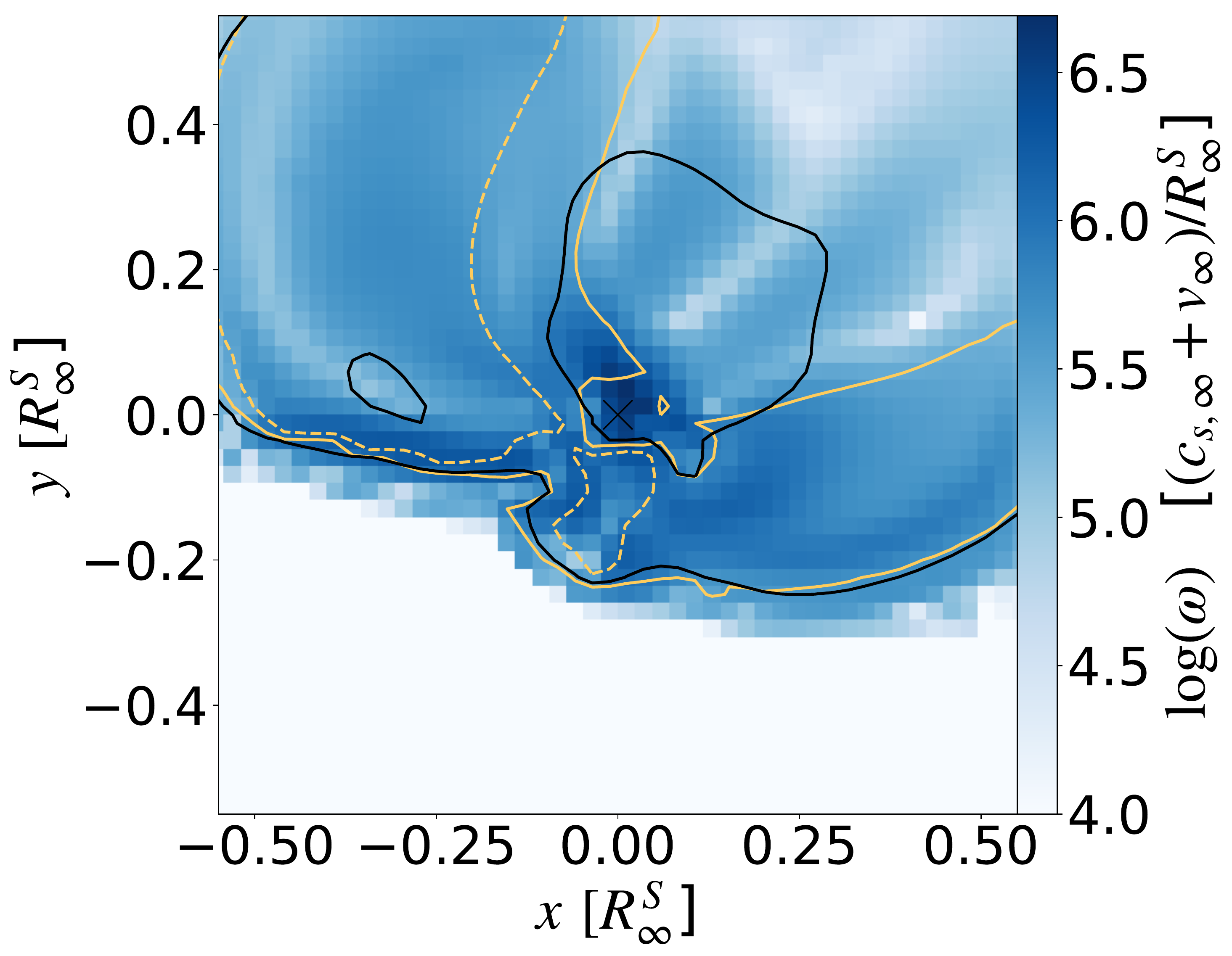} \\
	\end{tabular}
	\caption{Zoom in on flow patterns in and around the wake at two different times for m10n50a. {The left two columns show density}, with $R^S_\infty$ shown as a white circle when smaller than the box size. Streamlines are indicated in white, and the sonic surface as black contours. {The right hand column shows vorticity $\omega$, with entropy contours overplotted in yellow and the sonic surfaces again in black.} The global wake is stable, but the front of the bowshock alternates between periods of symmetric flow (top row, $t=8.87$) feeding the black hole efficiently, and instability dominated flow (bottom row, $t=10.24$).}
	\label{fig:m10_zoom}
\end{figure*}

With even more resolution, and an even smaller accretor, such as in m10n26a, the bowshock detaches from the accretor, and the solution becomes unstable, in agreement with results by \citet{Ruffert1996}. In this regime, the wake alternates between episodes when instabilities die down and the flow returns to a more symmetric configuration (top row, Figure \ref{fig:m10_zoom}), and episodes when instabilities dominate flow patterns, which are severely disruptive (bottom row, Figure \ref{fig:m10_zoom}). {While the accretion rate begins to fluctuate due to local instabilities, the time-averaged value remains similar to the case before the bowshock detached (compare m10n13a and m10n26a in Figure \ref{fig:m10_profiles}), as the accretion column feeding the black hole, characterised by a low vorticity region along the axis of symmetry of the wake continues to reform between unstable episodes (see top right hand panel in Figure \ref{fig:m10_zoom}).}
When not disrupted by instabilities, the gravitational attraction of the sink creates a peaked density profile around the accretion region, which, like in the Bondi case previously discussed, leads to a resolution dependence of the sound speed at the edge of the accretion region.

\begin{figure*}
	\centering
	\begin{tabular}{ccc}
		\includegraphics[width=0.315\textwidth]{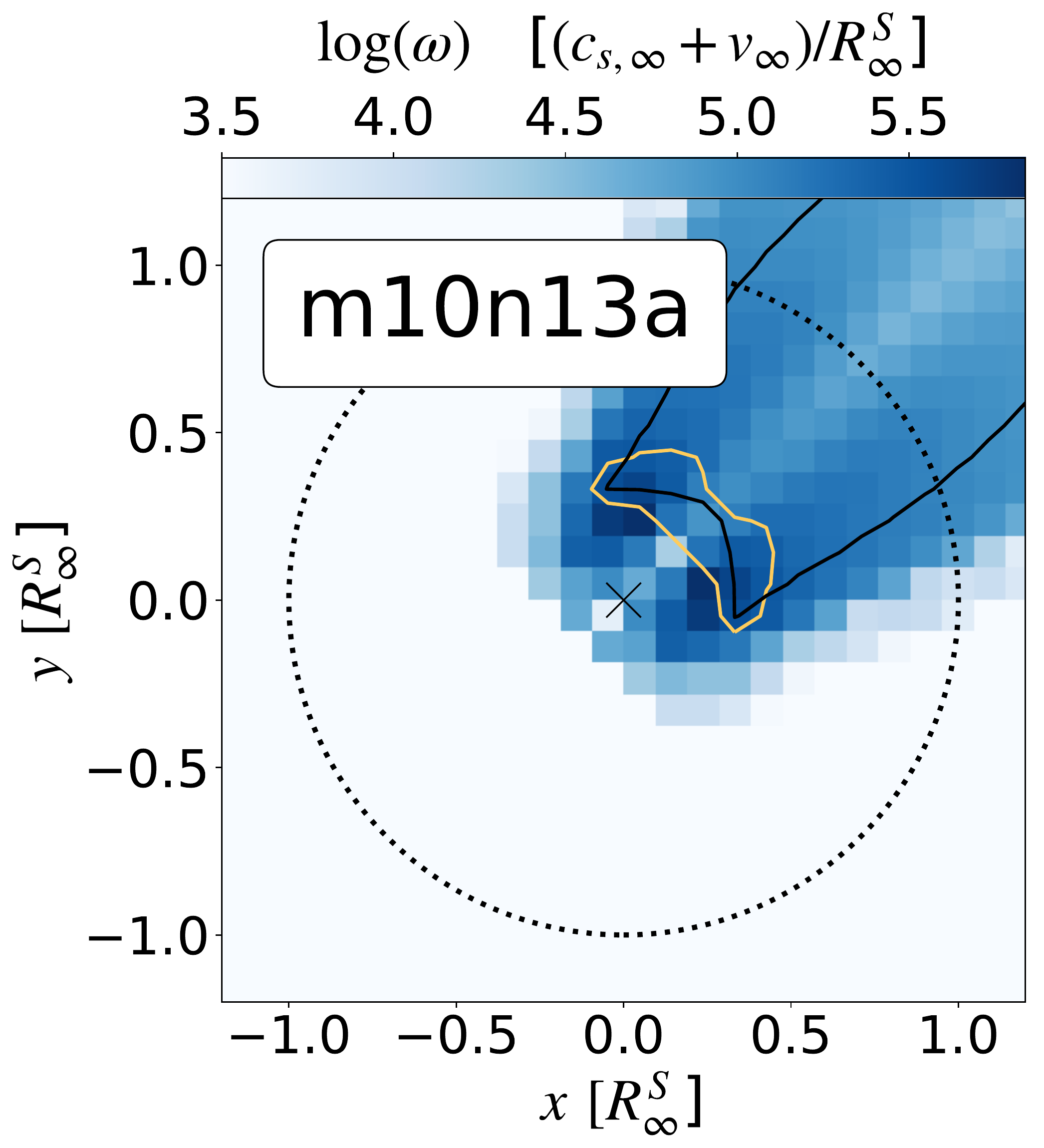}&
		\includegraphics[width=0.315\textwidth]{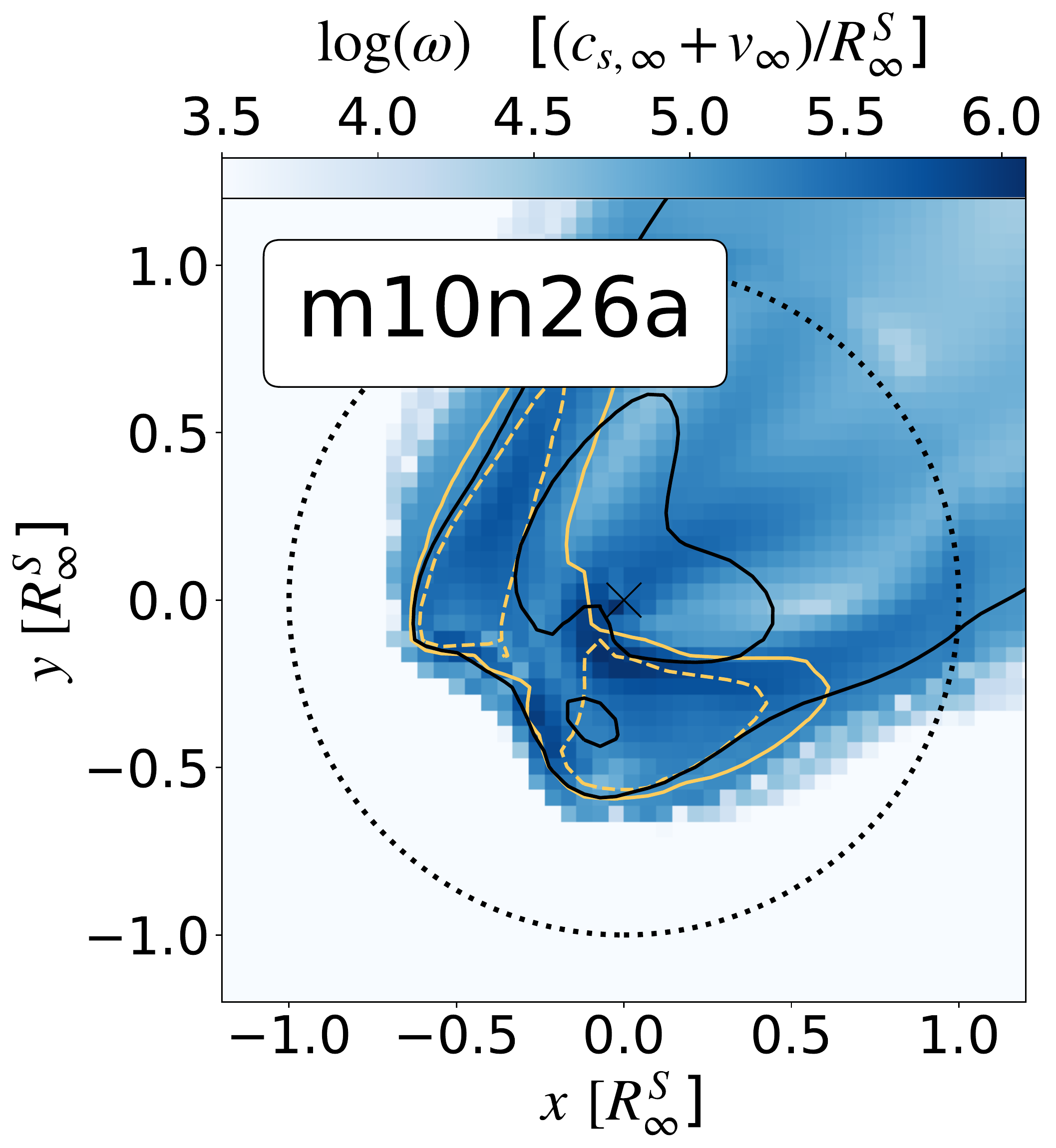}&
		\includegraphics[width=0.315\textwidth]{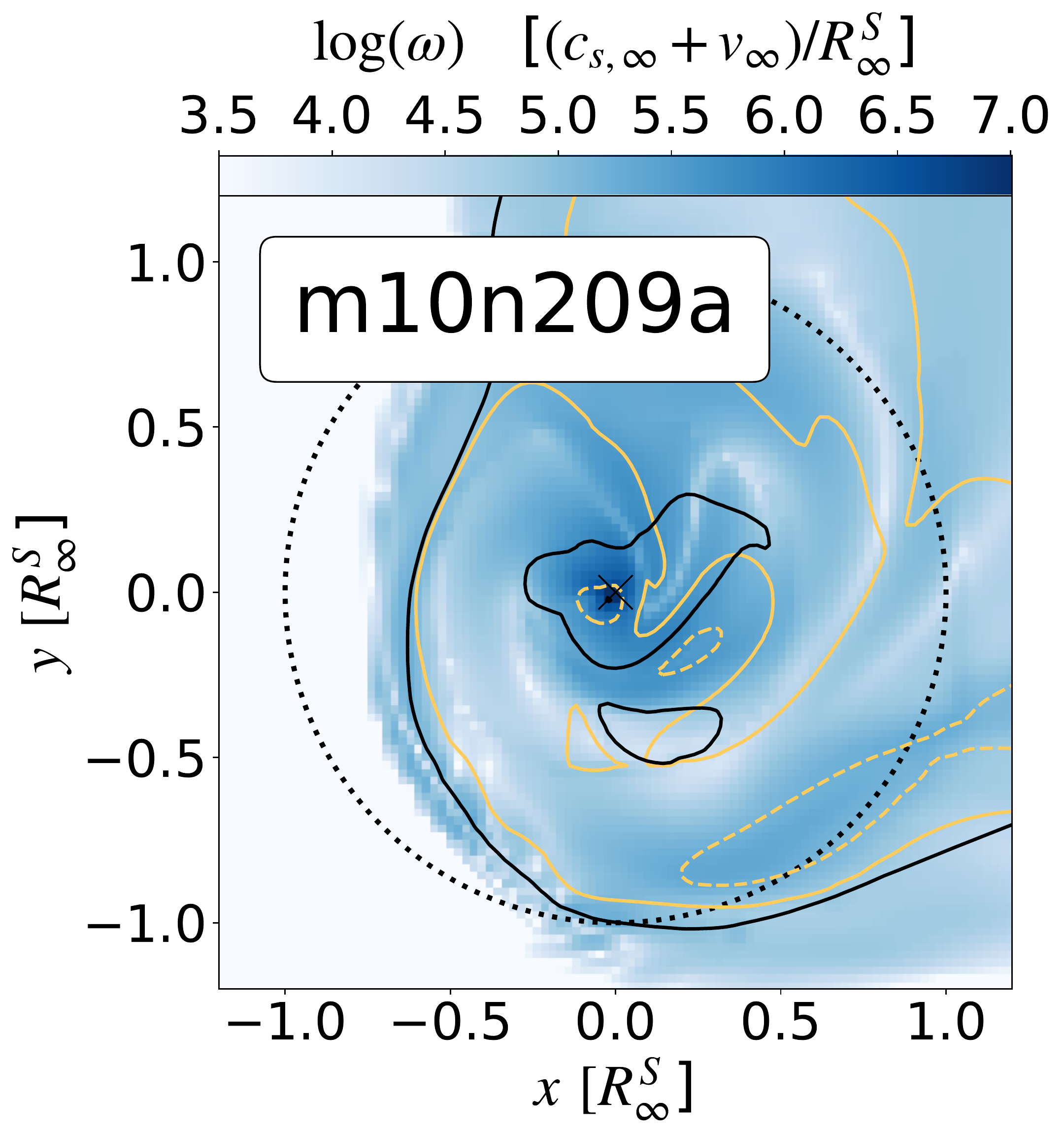}\\
	\end{tabular}
	\caption{{Vorticity $\omega$ profiles of the flow in the vicinity of the sink, for the snapshots shown in Figure \ref{fig:m10_density}. Sonic surfaces are annotated as solid black lines and entropy contours are shown in yellow. The dotted circle denotes the scale radius $R^S_\infty$. The lower limit of the colorbar was set to 3.5 in order to show the vorticity structure within the wake. All simulations are shown at $t=25$.}}
	\label{fig:vorticity_mach10}
\end{figure*}

While the global wake remains stable, eddies develop behind the shock, suggesting that the instability is caused by the physical acoustic-advection instability reported in \citet{Foglizzo2005} for $\gamma > 4/3 $, $\mathcal{M}_\infty>3$ and ``sufficiently small accretors''.  According to these authors, the instability is caused by entropy perturbations that advect from the shock to the sonic surface around the accretor, where they excite acoustic waves due to the inhomogeneity of the flow, which in turn propagate outwards back towards the shock, where they excite new entropy perturbations. We would therefore expect the instabilities to occur in the subsonic region, where the gas has been decelerated by the shock and not yet sufficiently reaccelerated by the gravitational potential of the sink. 

{In the simulations presented here, entropy perturbations (yellow contours) form in the subsonic region between the accretor and the bowshock (see m10n26a and m10n209a in Figure \ref{fig:vorticity_mach10}) for unstable simulations, as expected from the theory by \citet{Foglizzo2005}. For stable simulations, entropy perturbations are concentrated downstream of the accretor instead and are much smaller in magnitude (m10n13a in Figure \ref{fig:vorticity_mach10}). The origin of instabilities is explored further in Section \ref{subsec:discussion_instability}.}

The accretion algorithm creates a low density region around the sink, surrounded by a high density shell replenished by inflowing material. While most of the mass that enters the accretion region is removed, the sink particle algorithm does not implement a strict inflow criteria, with any surplus gas free to leave the accretion region during the next timestep. Accretion onto the sink varies on short timescales, as the turbulent flow feeds the black hole intermittently (see m10n26a in Figure \ref{fig:m10_profiles}), with an average value of $\dot{M}_{\bullet}/\dot{M}_\infty^\mathrm{BHL}\simeq 0.8$, in agreement with simulations discussed in \citet{Edgar2004}. This is also the value recovered for the steady state solution of extremely small accretors ($N=1000$) in the 2D axisymmetric simulations of \citet{Mellah2015}. 

\begin{figure}
	\centering
	\includegraphics[width=\columnwidth]{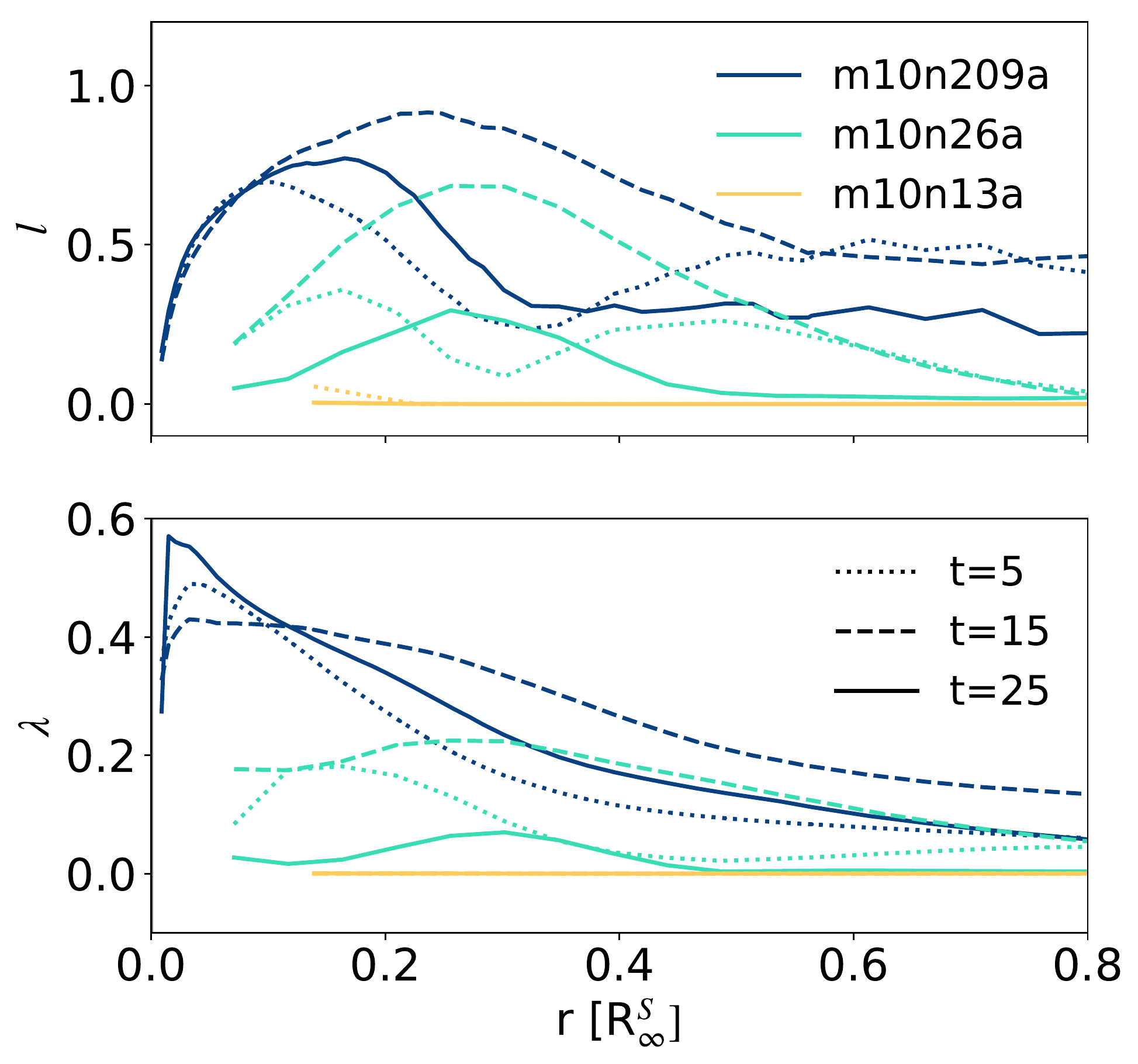}
	\caption{{Radial profiles of the specific angular momentum $l$ of  gas around the accretor (top), in units of $ ({c_{(s,\infty)}+v_\infty})/({\rho_\infty  {R^S_\infty}^4})$, and the gas spin parameter $\lambda$ according to \citet{Bullock2001} (bottom). Profiles are calculated using concentric spheres at radius $r$ centred on the sink particle. Each simulation is sampled at three points in time to highlight the range of profiles encountered in the unstable simulations. The stable m10n13a shows zero spin parameter and zero specific angular momentum around the accretor due to the symmetry of the flow.}}
	\label{fig:angular}
\end{figure}

The picture changes drastically for the highest resolutions, and smallest accretors, (m10n209a in Figure \ref{fig:m10_density}), as the instabilities become more severe. {The bow shock opening angle increases, supported by a strong rotational flow around the sink, with vorticity near the sink increasing by up to an order of magnitude (see Figure \ref{fig:vorticity_mach10}). The accretion column, which feeds the sink at lower resolutions, is permanently disrupted and the solution is entirely dominated by instabilities, as evident by the oscillations in Figure \ref{fig:m10_profiles}. Due to the increasing vorticity behind the bow shock, the gas within $R<0.15 R^S_\infty$ builds up coherent  angular momentum  (top panel, Figure \ref{fig:angular}). This angular momentum provides non-negligible rotational support against gravitational collapse for m10n209a (the spin parameter, $\lambda = \frac{L} {\sqrt{2} M(r) v_c(r) r} \gtrsim 0.5$, where $M(r)$ and $L(r)$ are the total mass and angular momentum contained within a sphere of radius $r$, and  $v_c(r)$ is the circular velocity at a given radius \citep{Bullock2001}, bottom panel, Figure \ref{fig:angular}), preventing gas from accreting efficiently. Periods of lower angular momentum around the black hole, seen for resolutions $14 \leq N \leq 50$, disappear. As a result, the time averaged accretion rate drops to $\dot{M}_\bullet / \dot{M}_\infty^\mathrm{BHL} \simeq 0.11$, with order of magnitude fluctuations around this value, so that extrapolating results from lower resolution runs provides a poor estimate of the accretion rate onto the sink.}

These results are in agreement with work by \citet{MacLeod2015}, who investigate Hoyle-Lyttleton type accretion in the presence of a density gradient at infinity, and find that the resulting circularisation of gas behind the (warped) shock reduces the accretion rate onto the sink by over an order of magnitude. In the simulations presented here, the circularisation is driven by the advective-acoustic instability, not a global density gradient. However, both cases shows that the presence of significant angular momentum in the gas behind the shock reduces the accretion onto the sink by more than order of magnitude. 

Overall, we conclude that, in contrast with the Bondi case, the accretion onto the sink for an adiabatic Hoyle-Lyttleton flow seems to converge to a value well below the analytic BHL solution, as it becomes entirely dominated by instabilities for very small accretors, i.e. $r^*/R^S < 0.01$ (equivalent to $N \geq 200$). 

\subsubsection{The quasi-isothermal case}
\label{subsec:isothermal}

\begin{figure}
	\includegraphics[width=\columnwidth]{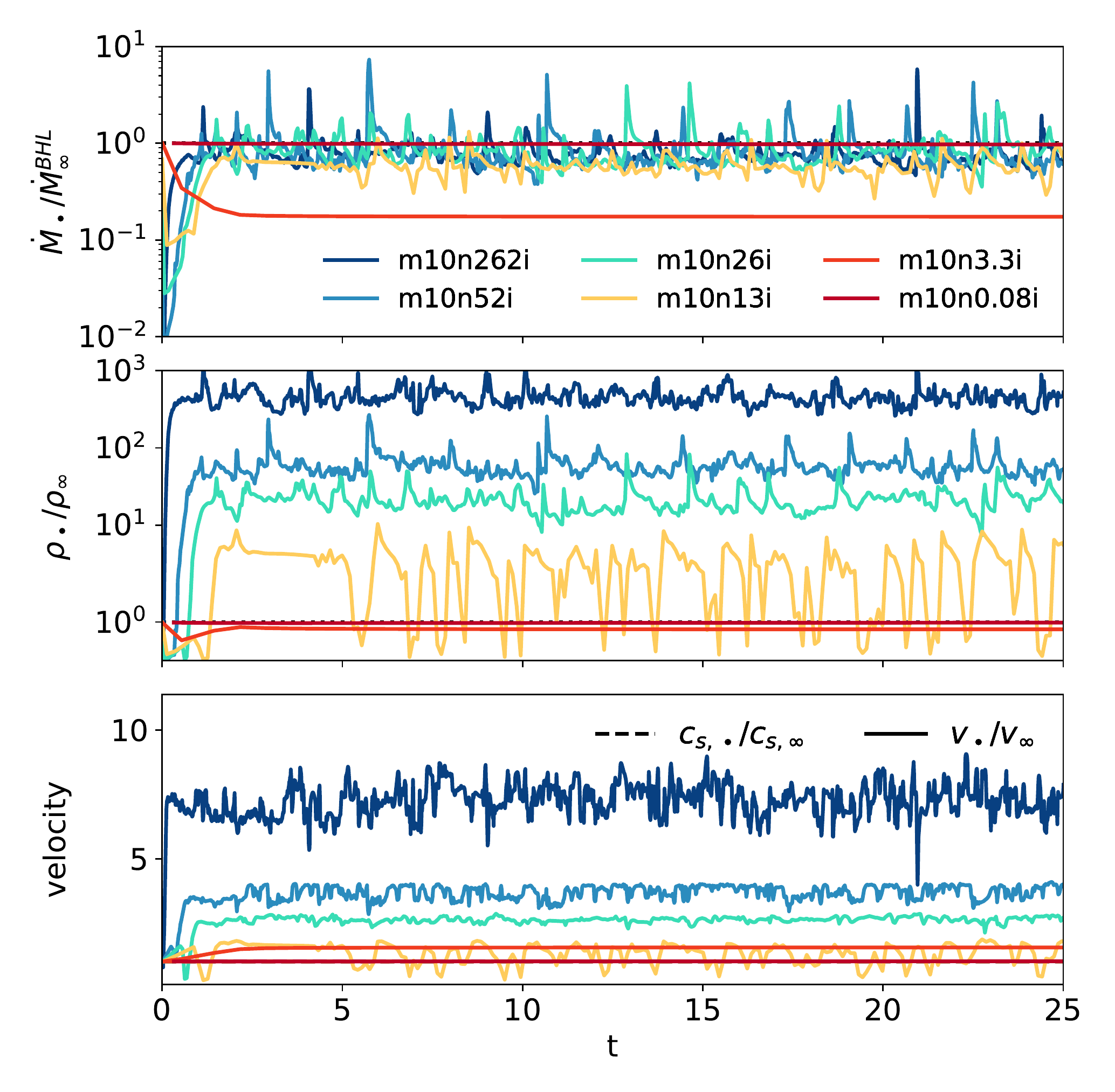}
    	\caption{Accretion rates and gas properties within the accretion region in the supersonic Hoyle-Lyttleton case, where $\mathcal{M}_\infty=10.0$. Gas is quasi-isothermal i.e. $\gamma=1.0001$, and each line represents a different resolution, as indicated in the top panel of the figure. The bottom panel shows both velocities (solid lines) and sound speed (dashed lines), but as the simulations are quasi-isothermal $c_{(s,\bullet}/c_{(s,\infty)} = 1$ for all simulations and the lines overlap.}
    \label{fig:gamma1_profiles}
\end{figure}

\begin{figure*}
	\begin{tabular}{cc}
	\includegraphics[height=0.8\columnwidth]{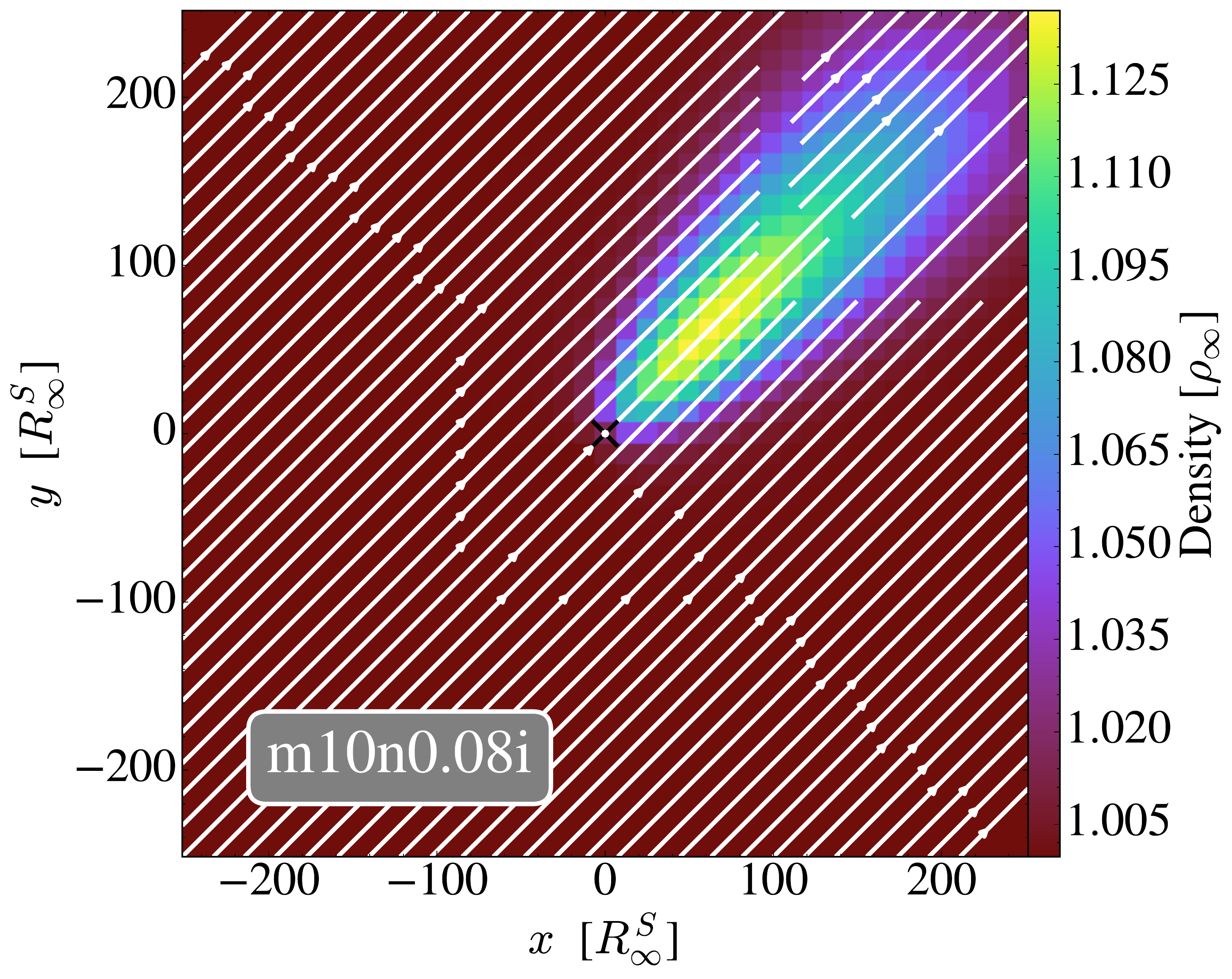} &
	\includegraphics[height=0.8\columnwidth]{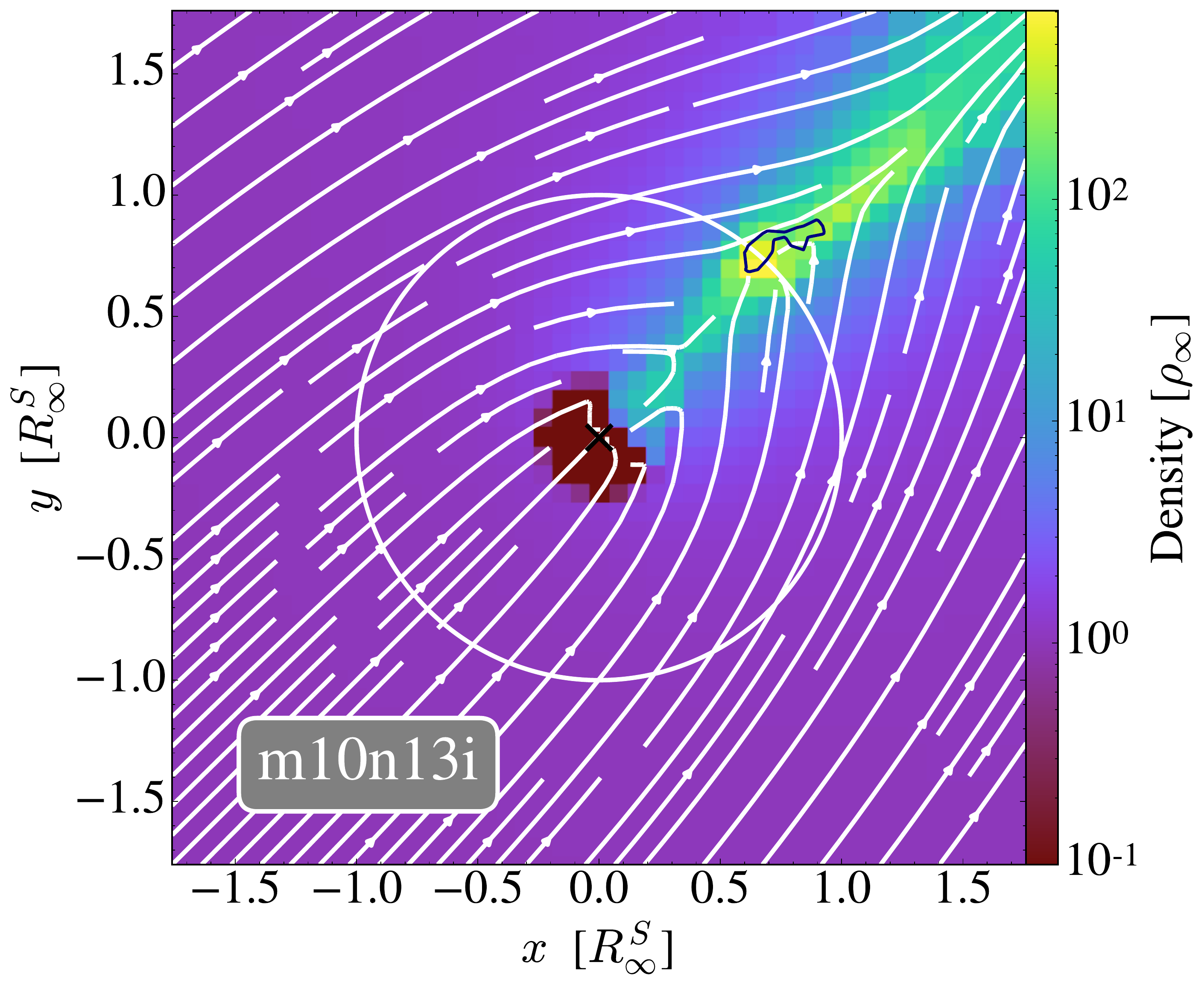}\\
	\includegraphics[height=0.8\columnwidth]{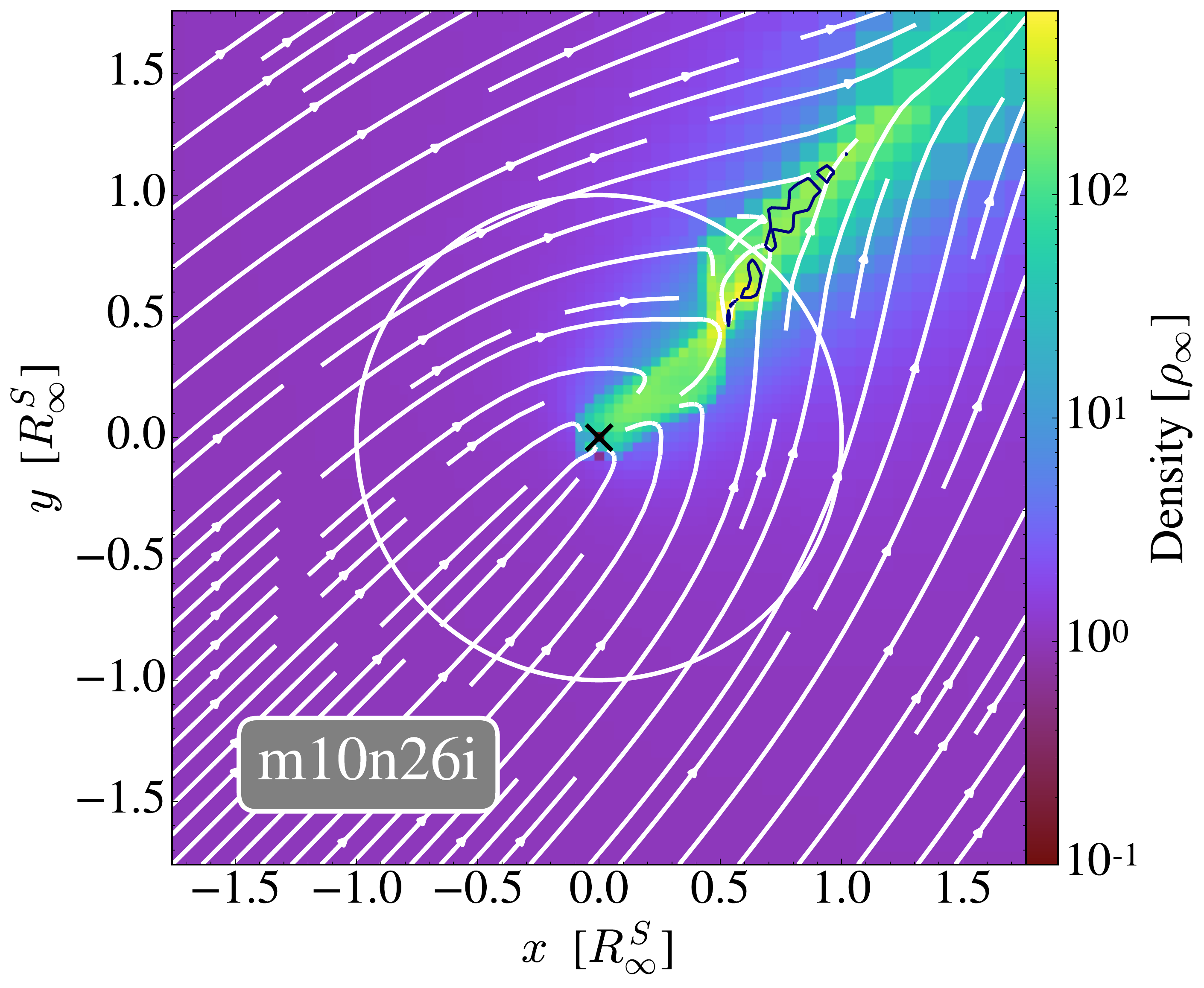}&
	\includegraphics[height=0.8\columnwidth]{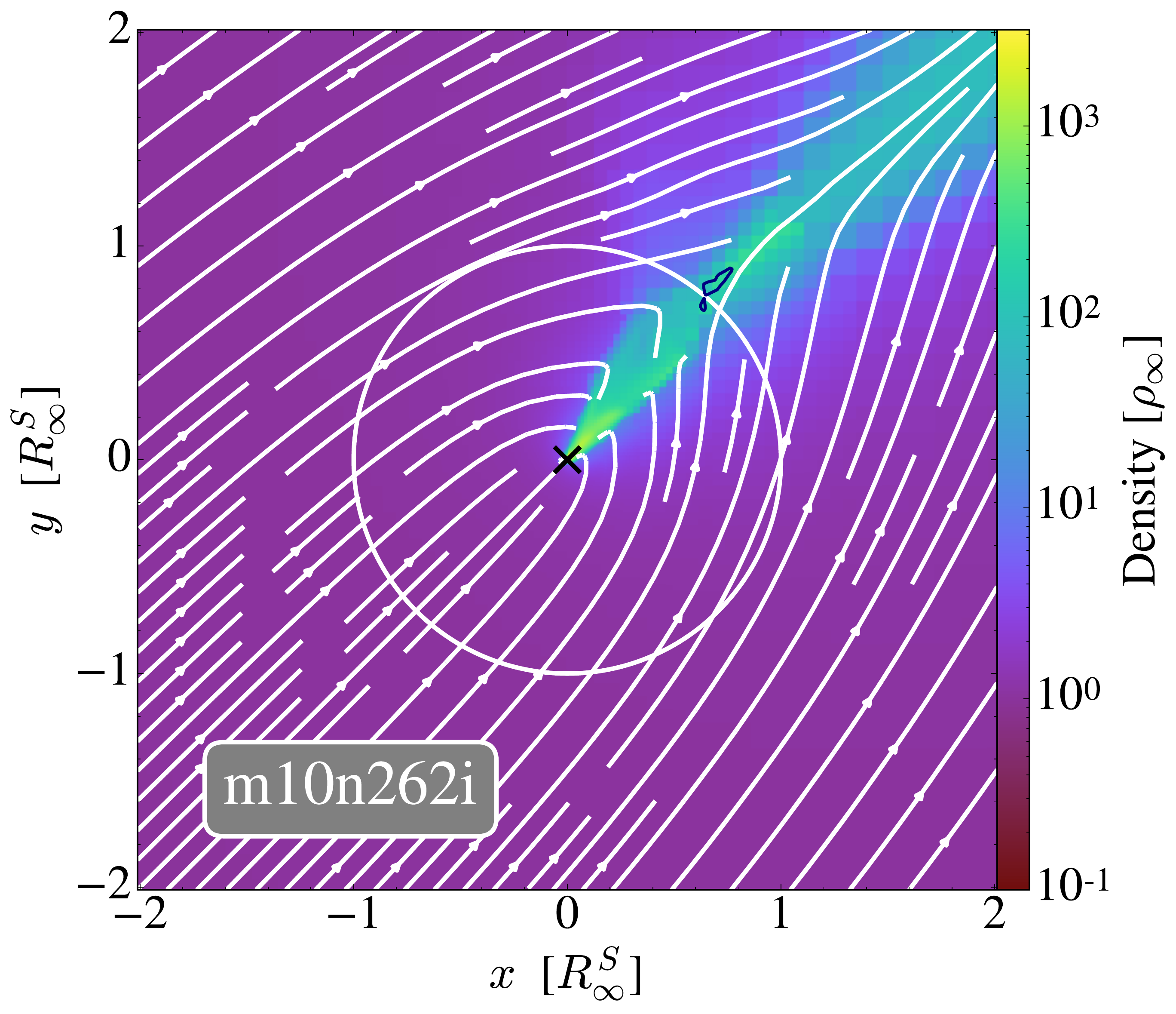}\\
	\end{tabular}
	\caption{Density slices for the quasi-isothermal runs, where $\gamma=1.0001$, for a variety of resolutions. The accretion radius $R^S_\infty$ is shown as a white circle. The sink is marked by a black cross. Sonic surfaces are represented as black contours and located near the stagnation points but are difficult to notice as they are small and narrow: the flow remains supersonic almost everywhere. Streamlines are shown in white. All simulations are shown at $t=25$.}
	\label{fig:gamma1_density}
\end{figure*}

\citet{Foglizzo2005} argue that the advective-acoustic instability should disappear for $\gamma < 4/3$, whereas \citet{Ruffert1995} report unstable flow for $\mathcal{M}_\infty=10$ at any $\gamma$. To further investigate the possible presence of instabilities, we ran a suite of quasi-isothermal simulations with $\gamma = 1.0001$. We find a gas flow pattern broadly similar to the adiabatic case presented above, albeit with very different consequences for the accretion onto the sink. As can be seen in Figure \ref{fig:gamma1_density}, at very low resolution, for m10n0.08i, only a small overdensity forms downstream of the sink, which accretes according to the analytic BHL formula (see the top panel in Figure \ref{fig:gamma1_profiles}).  For higher resolutions, $N \geq 3$, the local flow pattern again shows curved streamlines, stagnation point, shock and SLA. However, the shock opening angle is significantly smaller. Figure \ref{fig:gamma1_profiles} also shows that the accretion onto the sink is unstable for $N > 5$, a significantly lower resolution threshold than for the $\gamma > 4/3$ runs. {Like in the adiabatic case, simulations with an intermediate resolution (e.g. m10n3.3i) have a reduced accretion rate as the resolution does not yet allow the characteristic flow pattern to fully emerge. As is discussed further in Section \ref{subsec:discussion_instability}, at a resolution of $N \sim 3.3$, the stagnation point lies within the accretion region of the accretor, slowing down gas flows onto the sink. }By definition, the sound speed (bottom panel of Figure \ref{fig:gamma1_profiles}) remains constant in the quasi-isothermal case, so as resolution increases, the relative velocity of the gas with respect to the sink increases much more than in the adiabatic case (bottom panel of Figures \ref{fig:m10_profiles} and  \ref{fig:gamma1_profiles}). 

As a result, and contrary to the adiabatic case, the shock remains attached to the accretion region regardless of resolution and the flow stays supersonic everywhere except in a small narrow region around the stagnation point (see Figure \ref{fig:gamma1_density}). Despite this, we find that strong instabilities develop in the wake, particularly for small accretors, which also lead to accretion rate variations on short timescales. Instead of originating at the bowshock, these instabilities occur downstream of the sink and affect the wake globally, disrupting the accretion column onto the sink. No eddies are visible in the stream lines (see Figure \ref{fig:gamma1_density}). Rather, the narrow wake clumps and is distorted in the direction perpendicular to the axis of symmetry of the wake, in agreement with the seminal work of \citet{Ruffert1996}. Taken at face value, the persistent instabilities for $\gamma \simeq 1$ seem to contradict predictions by \citet{Foglizzo2005}, who argue that the acoustic-advective instability should disappear for $\gamma < 4/3$. However these authors do point out that a different type of instability could occur at low $\gamma$, based on vorticity (rather than entropy) perturbations between shock front and accretor. We note that the instabilities seen here are also reminiscent of the ones discussed in \citet{Cowie1977} and \citet{Soker1990}, who study the fate of small overdensities when the wake is modelled as an accretion line, { but postpone a further discussion of the origin of these instabilities to Section \ref{subsec:discussion_instability}.} Beyond the exact nature of the instabilities, the main difference between the adiabatic and the quasi-isothermal simulations is that the strong drop in average accretion rate for the smallest accretors, which have  $N>200$, is absent for quasi-isothermal simulations. At the highest resolution probed here, the $\gamma = 1.0001$ accretion rate converges to $\dot{M}_\bullet / \dot{M}_\infty^\mathrm{BHL} \simeq 0.74$: while instabilities exist, they do not efficiently prevent accretion onto the sink. 

In summary, we conclude that for black hole moving supersonically, with $\mathcal{M}_\infty=10$, accretion proceeds via the BHL algorithm at low resolution, $N<8$, and transition to SLA at higher resolution. The wake is unstable for both values of $\gamma$, with the instabilities in the adiabatic case originating in the subsonic region between the sink and the detached bowshock, and near the stagnation point in the quasi-isothermal case.  With increasing resolution, and decreasing size of the accretor, the flow becomes unstable on progressively shorter timescales. In the adiabatic case, instabilities dominate for $N>100$, which leads to an order of magnitude reduction in the time-averaged accretion rate onto the sink, whilst in the quasi-isothermal case this averaged accretion rate converges to a value close to that of the BHL case.
 
 
\subsection{Exploring the full parameter space in $\mathcal{M}$}
\label{sec:machs}

\begin{figure*}
	\centering
	\begin{tabular}{cc}
		\includegraphics[width=0.38\textwidth]{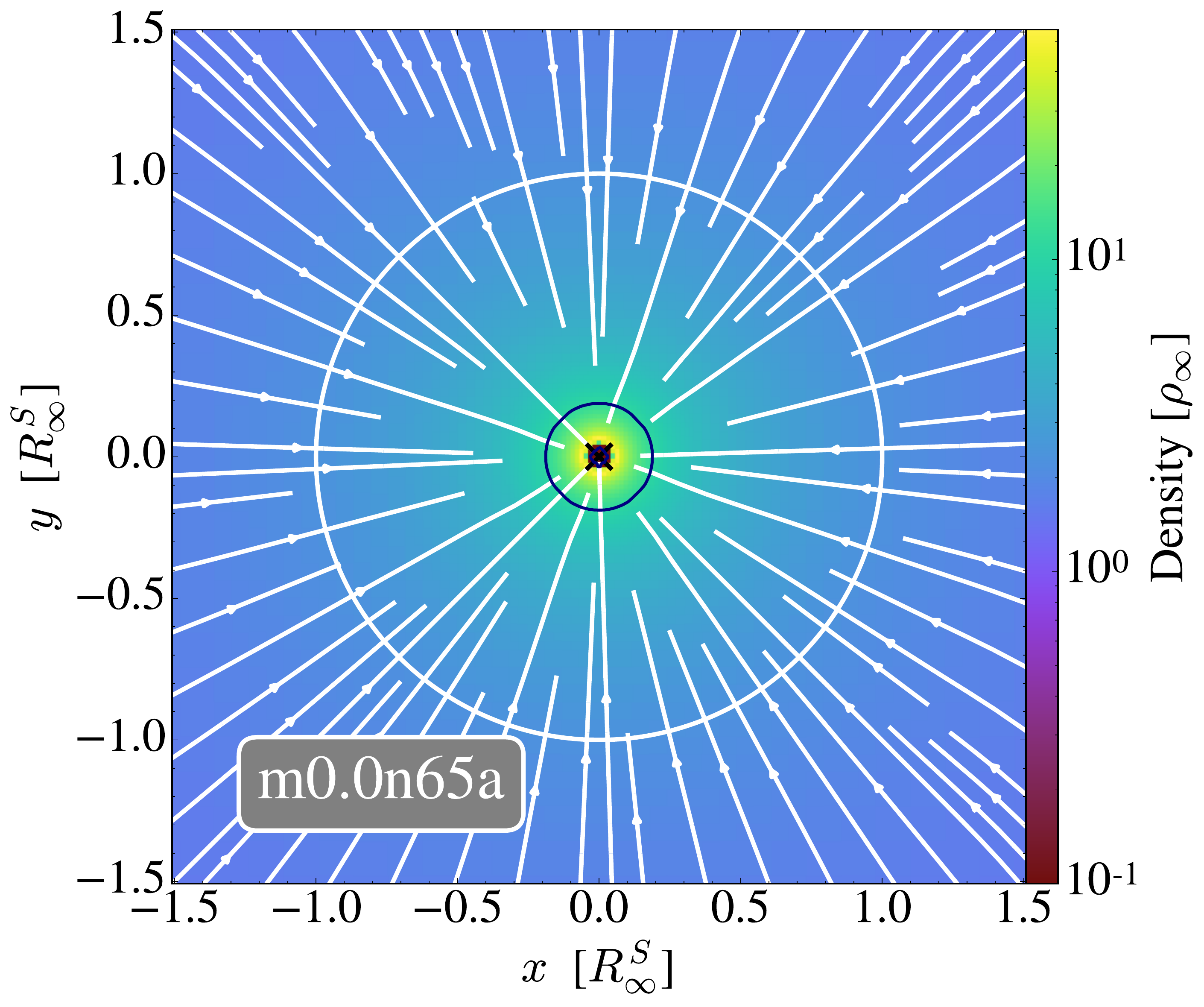} &		        		
		\includegraphics[width=0.38\textwidth]{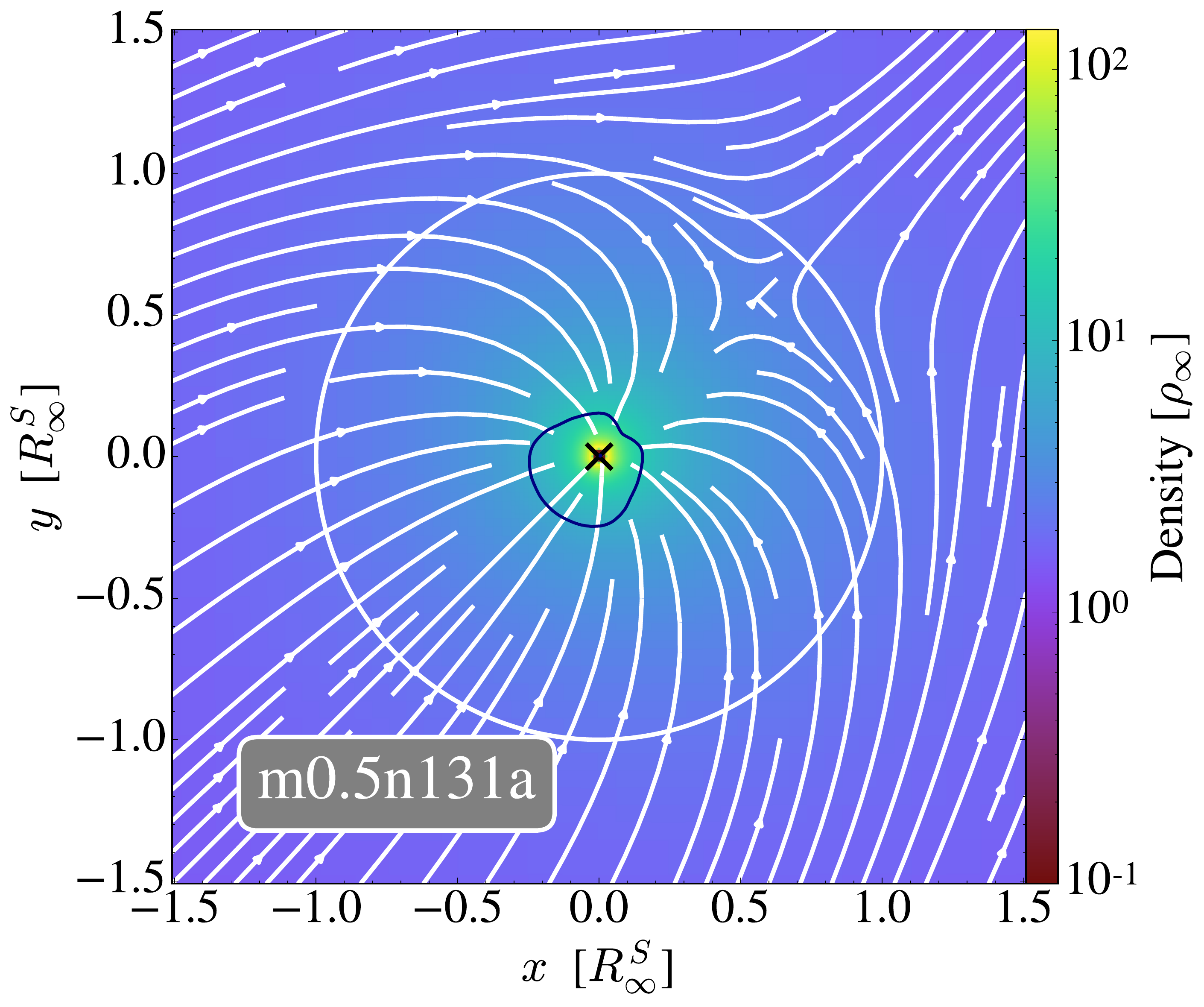}\\
		\includegraphics[width=0.38\textwidth]{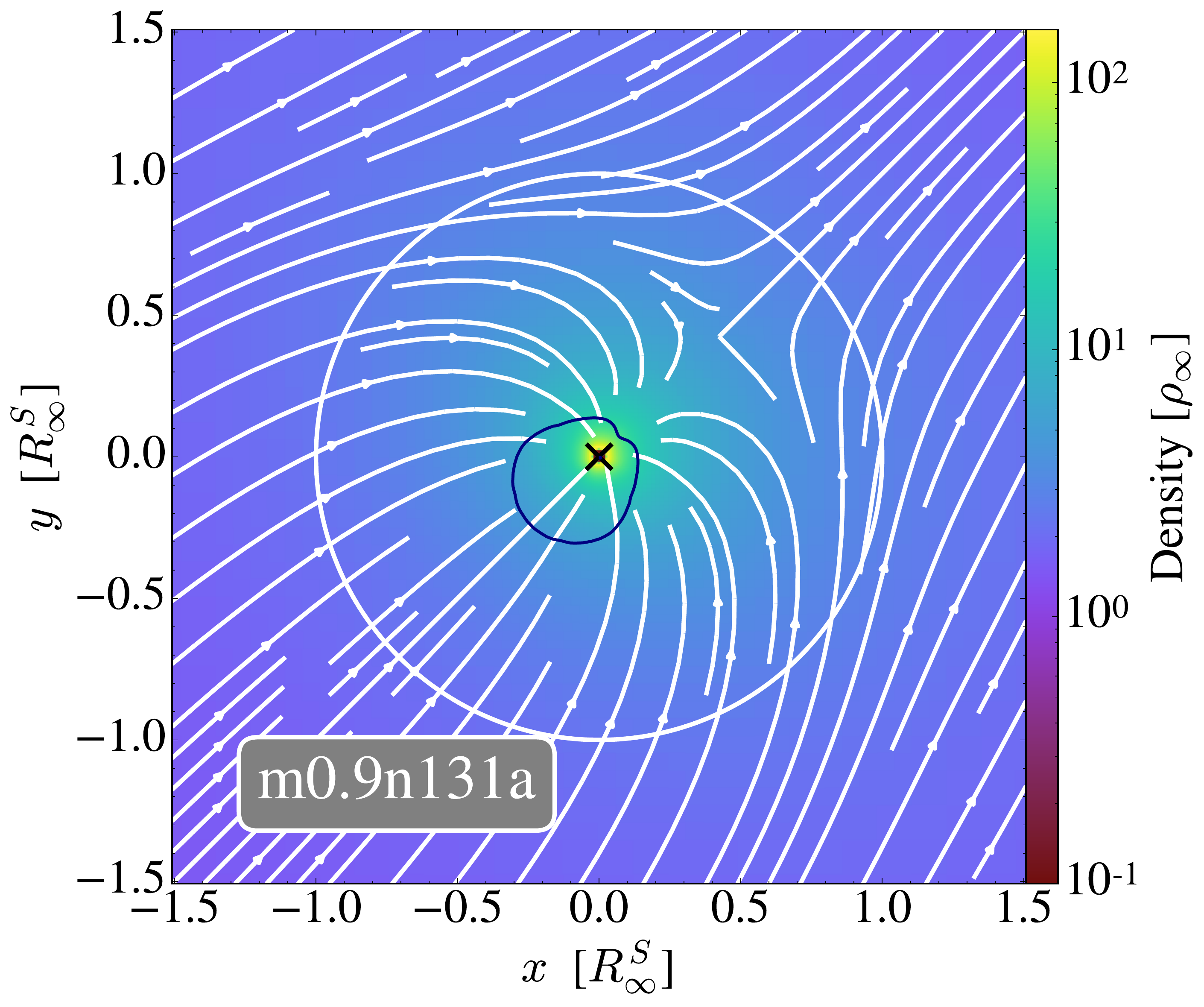}&
		\includegraphics[width=0.38\textwidth]{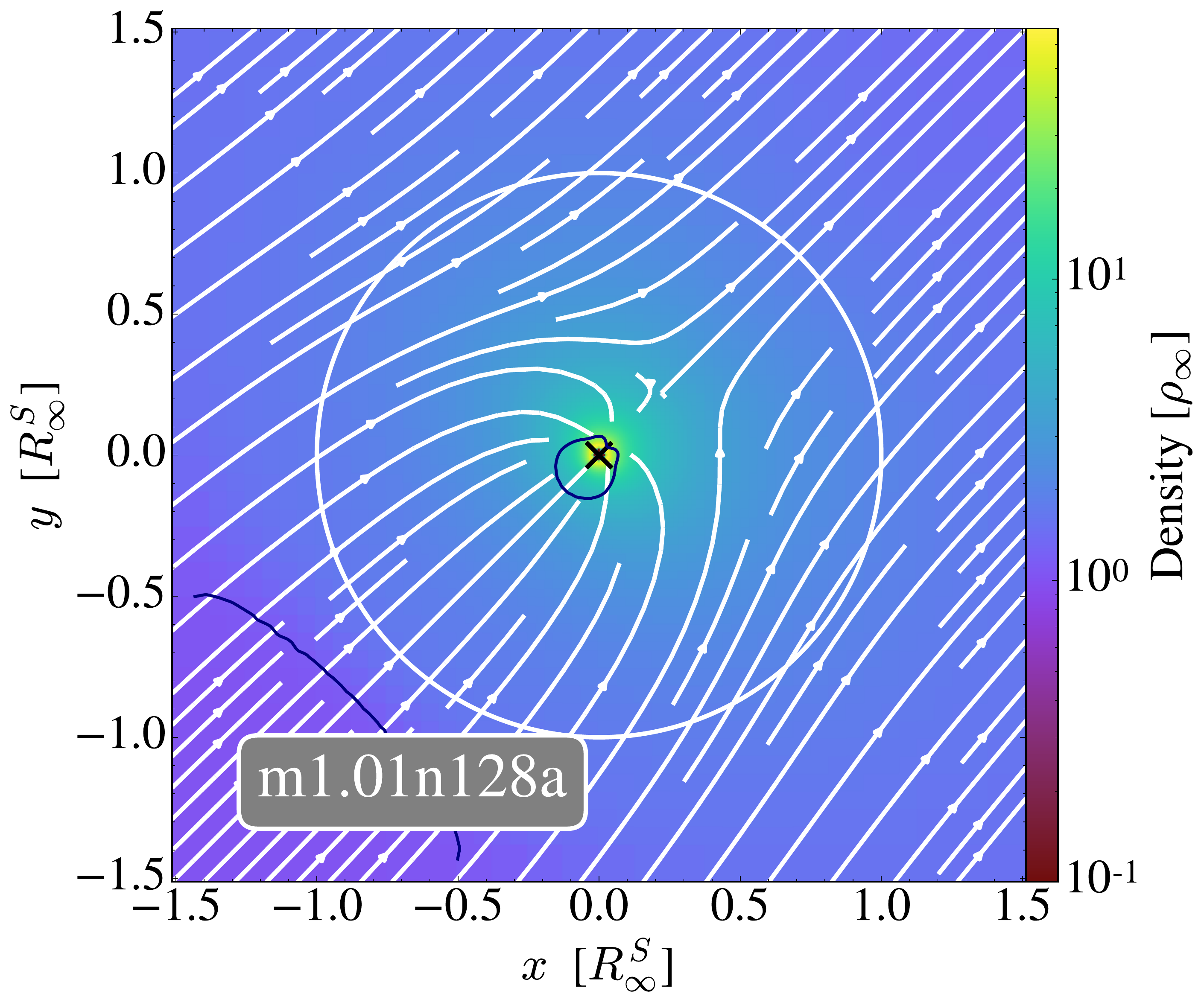}\\
		\includegraphics[width=0.38\textwidth]{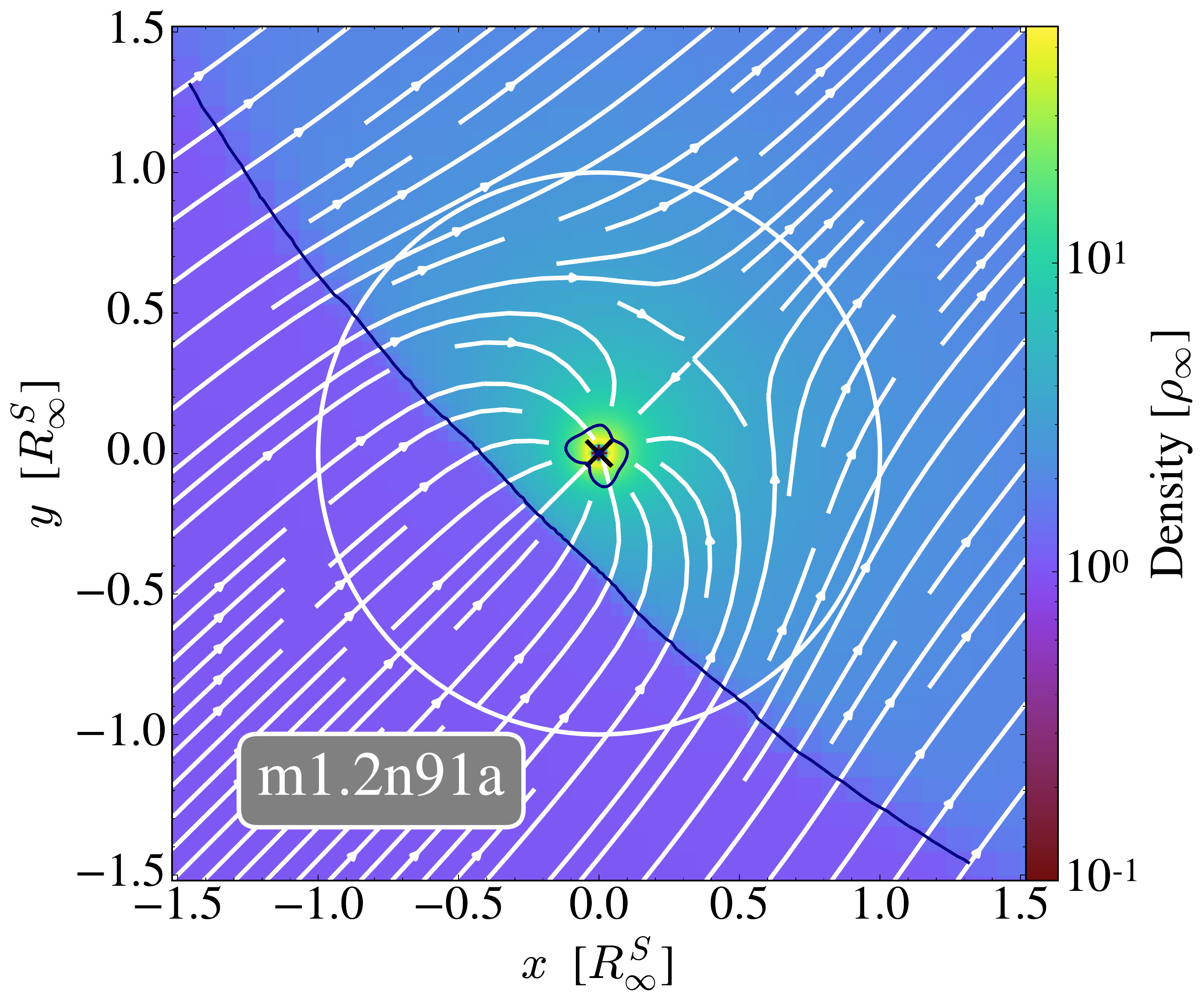}&
		\includegraphics[width=0.38\textwidth]{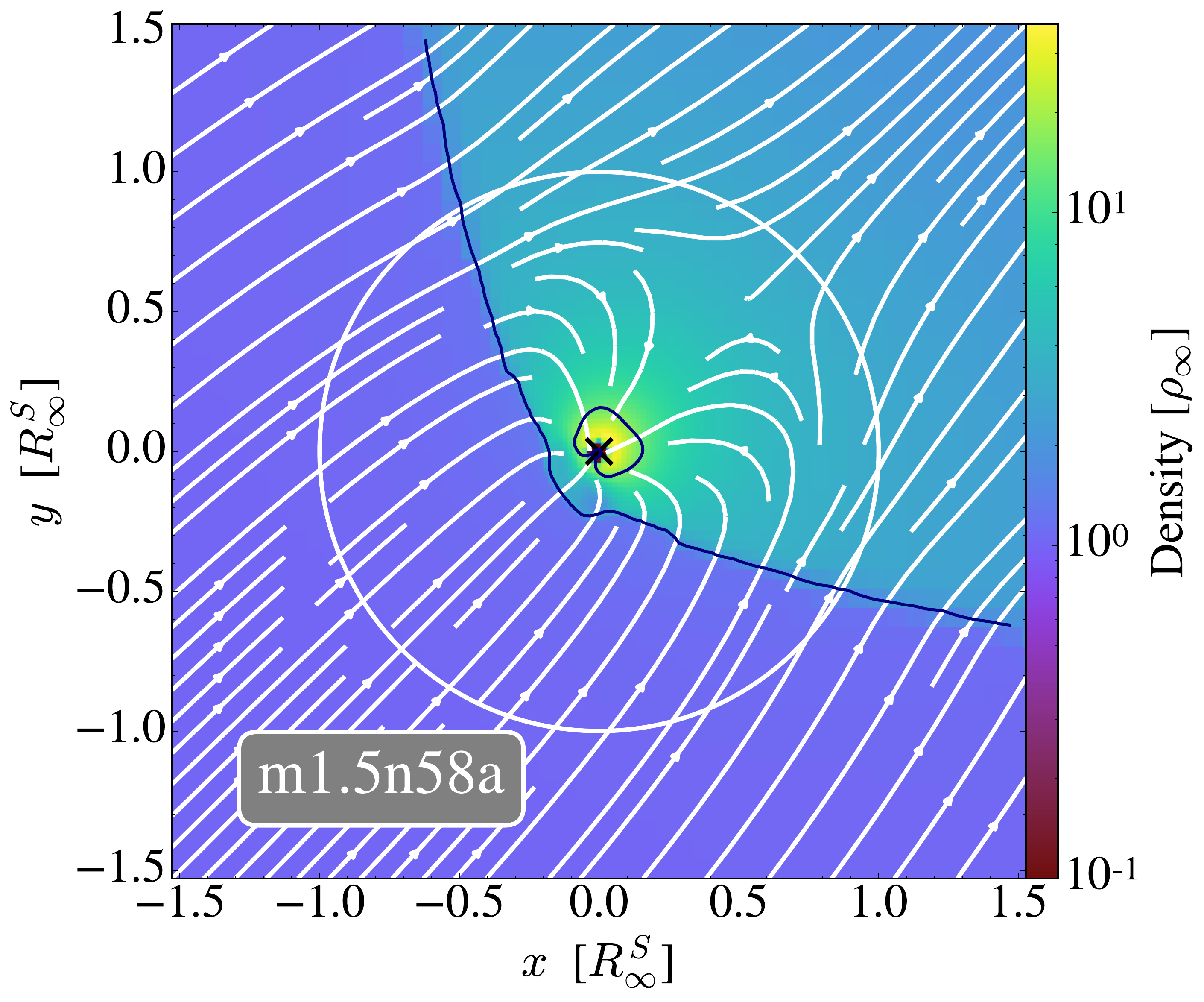}\\
		\includegraphics[width=0.38\textwidth]{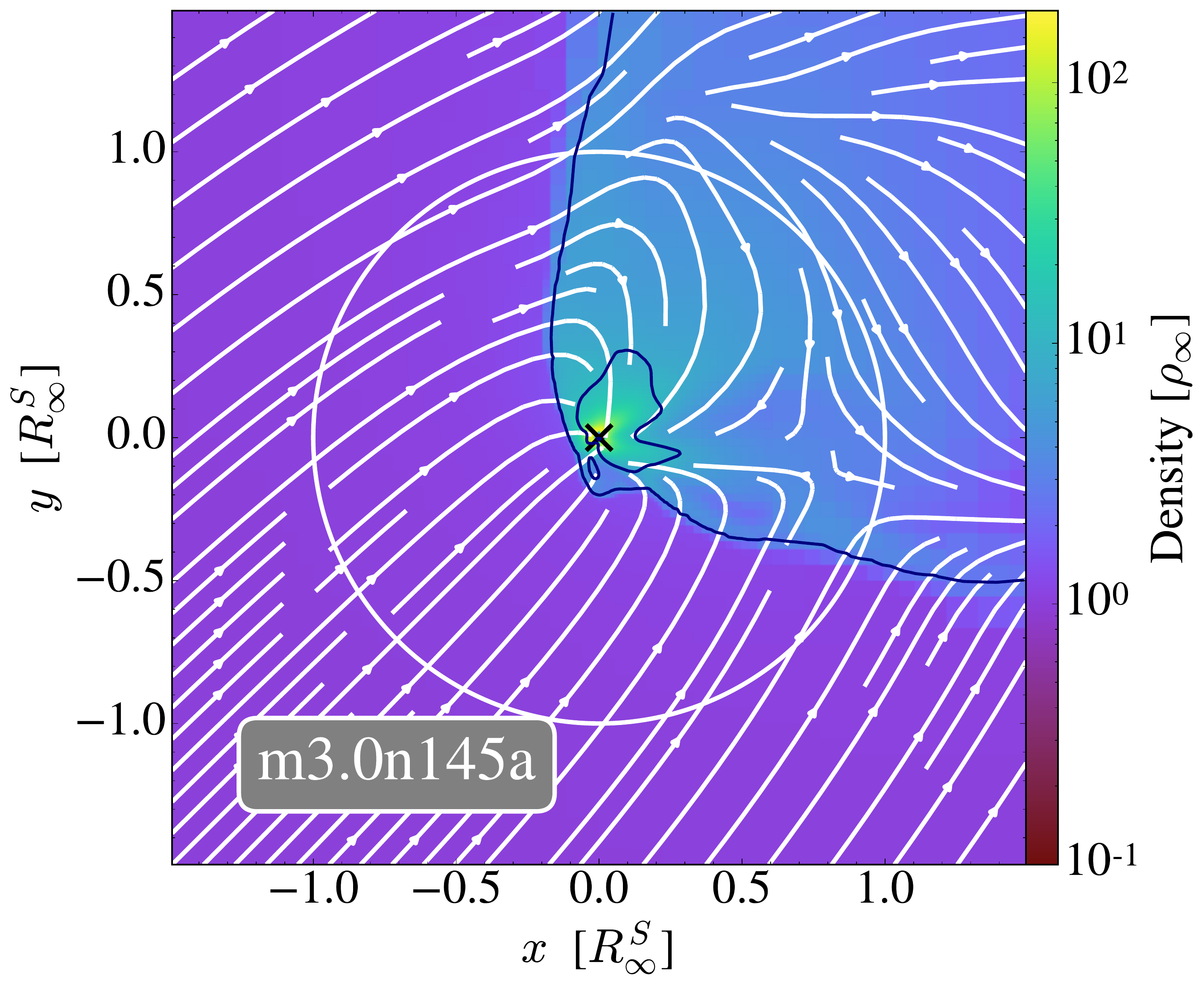}&
		\includegraphics[width=0.38\textwidth]{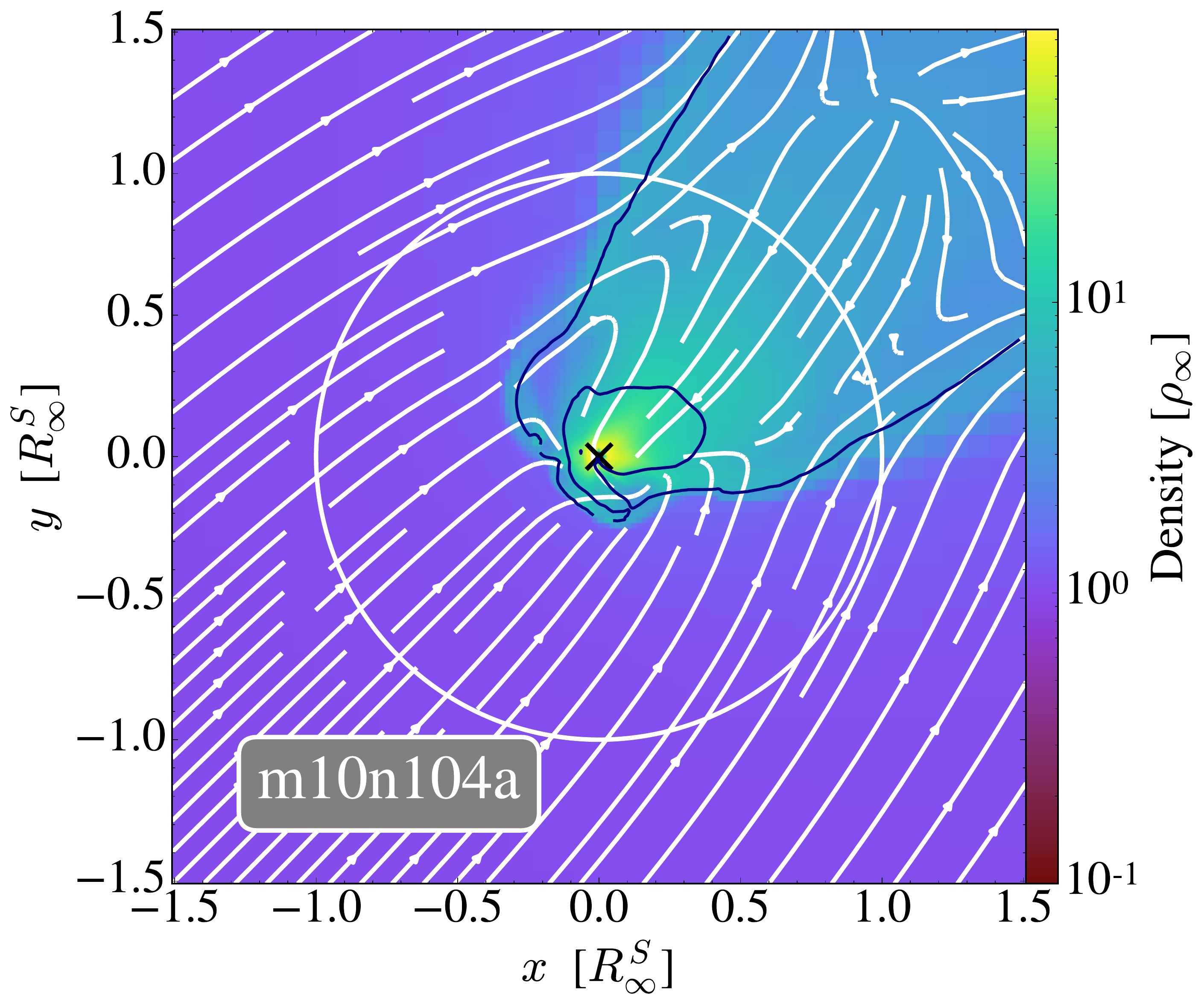}\\
	\end{tabular}
	\caption{Density slices for a range of Mach numbers, ordered from the lowest $\mathcal{M}_\infty$ in the top left to the highest in the bottom right. The sink is denoted by the black cross at the centre of each panel. Flow lines are represented by white lines, $R^S_\infty$ is indicated by a white circle and black contours mark sonic lines. All simulations are shown at $t=25$.}
	\label{fig:density_slices}
\end{figure*}

We now explore the evolution of flow patterns and accretion rates in the adiabatic case for a wider range of Mach numbers and different resolutions. Note that in the case of intermediate Mach numbers ($ 0.3 \lessapprox \mathcal{M}_\infty \lessapprox 3.0 $), we do not expect the analytic BHL formula (Equation \ref{eq:bondi_hoyle}) to yield as accurate an estimate of the accretion onto the sink as in the low and high Mach number cases previously studied as it is merely an educated interpolation between these two extremes cases.  

\begin{figure}
	\includegraphics[width=\columnwidth]{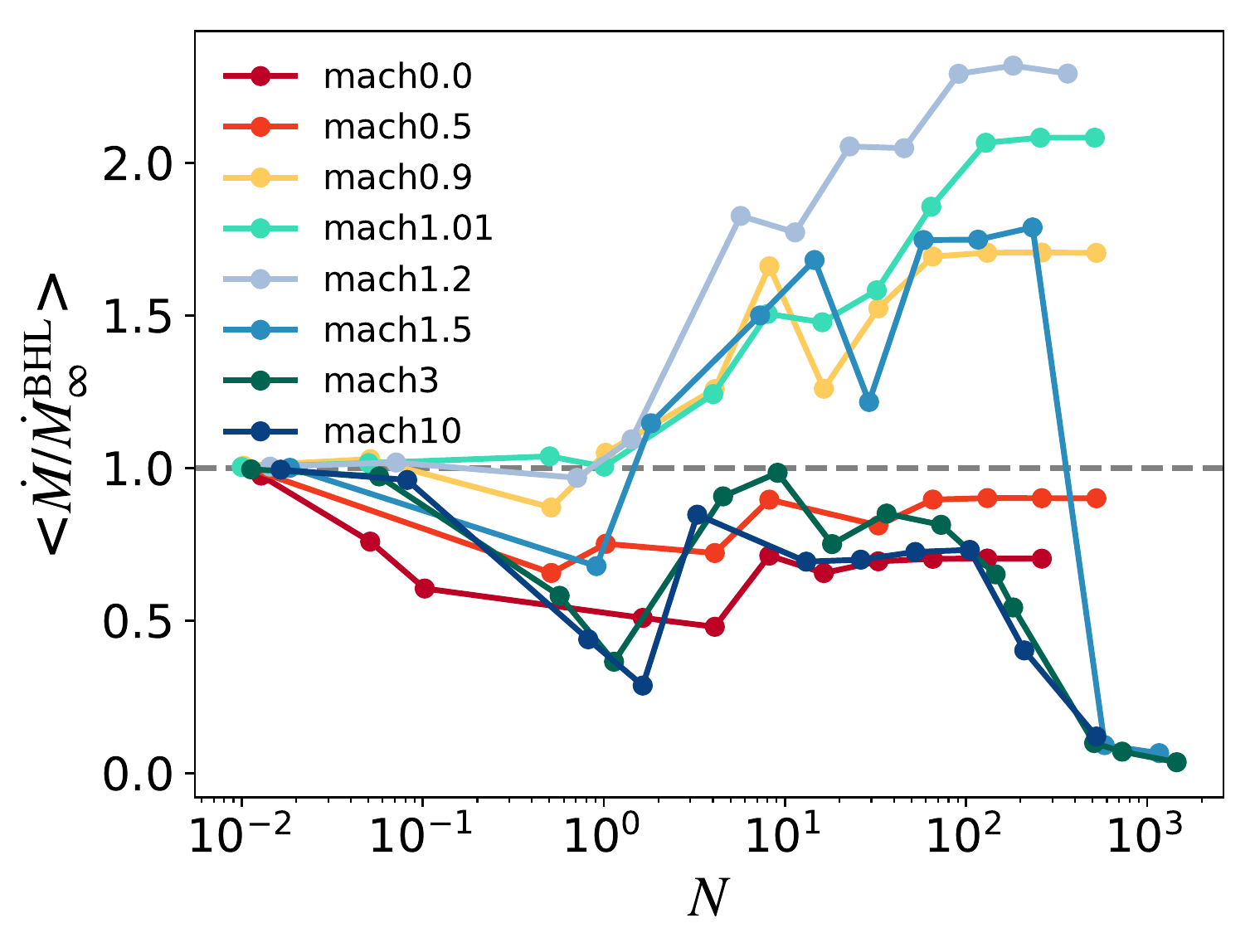}
    	\caption{Dependence of the accretion rate on resolution for a variety of Mach numbers in adiabatic simulations with $\gamma = 1.3334$. 
Each data point represents the time averaged value for $t > 10$ of a specific run, as indicated on the panel. }
    	\label{fig:dotM_dx}
\end{figure}

Having said that, Figure \ref{fig:dotM_dx} presents the average accretion rate, in units of the analytic BHL accretion rate $\dot{M}^\mathrm{BHL}_\infty$, for a variety of Mach numbers, as a function of resolution. As could already be seen for the Bondi and the Hoyle-Lyttleton problem (Section \ref{sec:bondi} and \ref{sec:hoyle_lyttleton} respectively), the unresolved case with $N< 0.01$ closely follows the analytic formula. As resolution increases, the behaviour becomes more complicated. In the intermediate regime, where $0.01 < N \leq 50 $, the simulations diverge from the analytic formula in a way that depends non-monotonically on the Mach number. Sub- and supersonic simulations ($\mathcal{M}_\infty = 0.5$ and $\mathcal{M}_\infty = 3$ respectively) show average accretion rates systematically lower than the BHL formula, by up to a factor $5$. Trans-sonic simulations, on the other hand, with $0.9 \leq \mathcal{M} \leq 1.5$ feature accretion rates which are larger by up to a factor of $2.3$. The BHL formula, used here to normalise results, is most uncertain for the transsonic regime, where both the bulk velocity $v_\infty$ and the sound speed $c_{s,\infty}$ have a significant influence on the flow. Our high resolution results ($ N > 50$) support this conclusion as the accretion rates indeed converges to higher values in the trans-sonic regime. We caution that at intermediate resolutions, the pressure force into the low resolution region around the sink can dominate over the local gravitational force, possibly funnelling extra gas into the accretion region and thus leading to an overestimate of the accretion rate. However, this effect is very localised as it only occurs at the edge of the accretion region, and is alleviated by the kernel function which creates a gradual transition of density within the region covered by the cloud particles. However at high resolution ($N>100$), the gravitational force comfortably dominates over the pressure force for all cases studied here (see Figure \ref{fig:pressure_mach0} for a measure in the Bondi case). Moreover, all resolved simulations with Mach numbers $\mathcal{M}_\infty \leq 1.2$ show steady state solutions (Figure \ref{fig:density_slices}). 

From $\mathcal{M}=1.5$ on, eddies begin to form behind the shock, and when the accretor becomes small enough, instabilities becomes stronger and begin to influence accretion onto the sink more significantly. The density slices for both $\mathcal{M}_\infty=3$ and $\mathcal{M}_\infty=10$ show strong instabilities that disrupt the flow patterns and decrease the time-averaged accretion rate onto the sink by up to an order of magnitude below the analytic value. We caution that while a lot of care has been taken to minimise the impact of initial conditions (see Appendix \ref{sec:initial_conditions}), the seeding of the instabilities could still be due to the way the simulations are initialised. This is likely to affect the exact resolution and/or Mach number at which the instability dominated regime appears, but unlikely to make it vanish altogether.

In summary, we conclude that the sink particle algorithm, using a locally evaluated BHL accretion rate as described in Section \ref{sec:accretion}, is a versatile sub-grid model that smoothly adapts to a variety of resolutions.  For highly resolved simulations ($N >100$), the kernel function used to remove mass ensures that the maximum accreted mass per timestep, dominated by the dense cells at the edge of the accretion region, always exceeds the local gas supply and the sub-grid model automatically transitions to SLA. Intermediate and low resolutions ($N \leq 50$) lead to mixed results and appear to be a difficult regime when approximating the accretion rate onto the black hole from local gas properties, but still manage to capture the accretion rate within a factor $\approx 2$, at least in the moderate Mach number regime ($\dot{M}^\mathrm{BHL}_\infty < 1.5$). It is only at higher Mach numbers that they deviate from resolved time averages by more than an order of magnitude, an effect that can significantly impact the final mass of the sink. This is potentially important when simulating the cosmological growth of supermassive black holes, where an early accretion boost is crucial to enable the black holes to reach observed masses within the limited timeframe available \citep[See][for a review]{Volonteri2010}.


\section{Drag force}
\label{sec:fdrag}

The gravitational force of the wake formed downstream of sink particles in the presence of a significant bulk flow exerts dynamical friction on the sink particle, opposite to the direction of motion, causing it to reduce its relative velocity with respect to the gas over time. However, the total drag force of the wake depends sensitively on the mass contained in the wake, particularly close to the sink. 

\begin{figure}
	\includegraphics[width=\columnwidth]{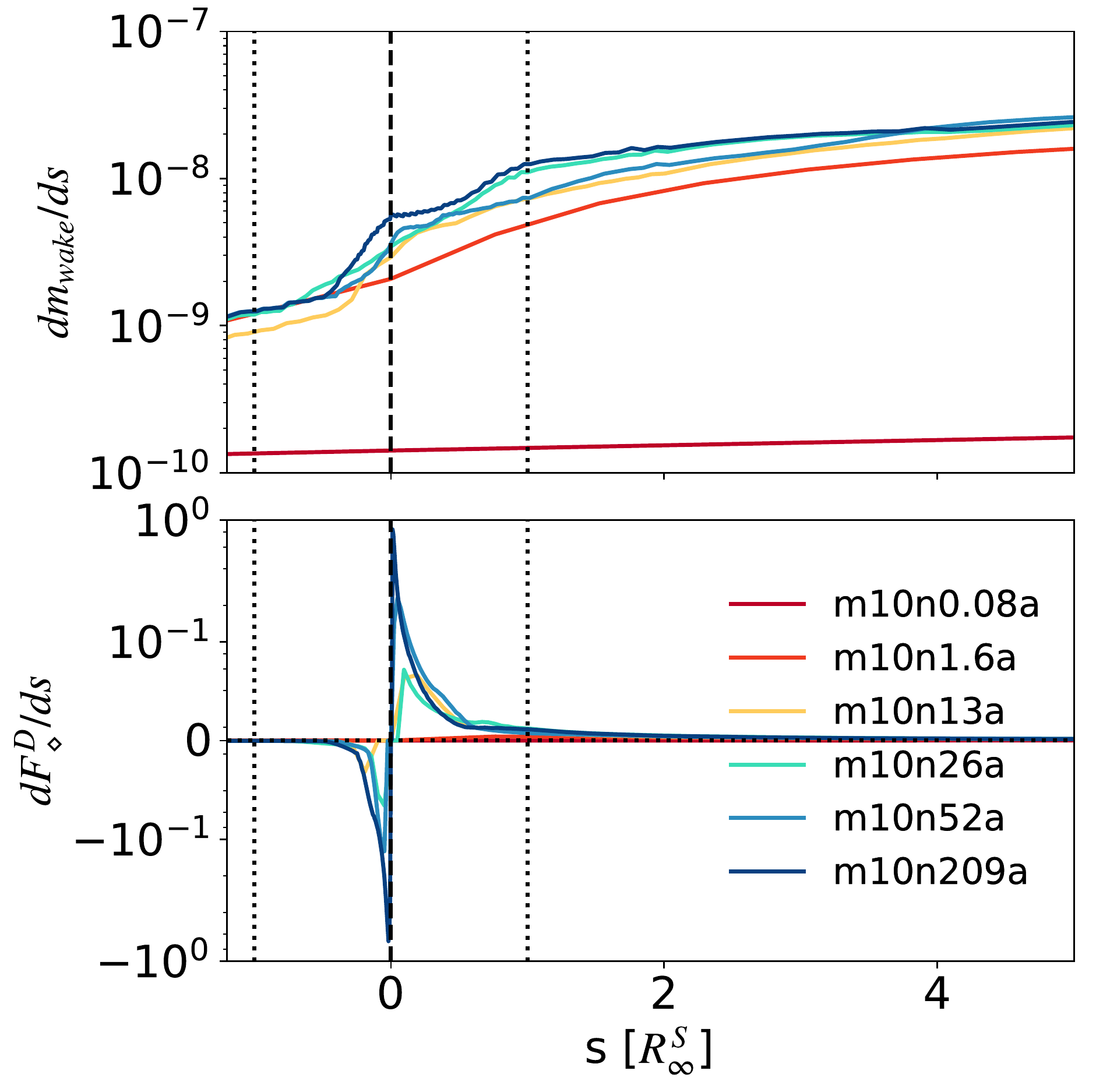}
    	\caption{Differential mass (top panel, in units of $[M_{sink}/R^S_\infty]$) and drag force (bottom panel, in units of [$4 \pi \rho_\infty (GM_{sink})^2 /(R^S_\infty c_{s,\infty}^2)$]) profiles at time $t=25$ for a highly supersonic adiabatic flow ($\mathcal{M}_\infty=10$) and a variety of resolutions, as labelled on the bottom panel. $s$ is the distance of a given mass slice to the sink particle measured along the axis of symmetry of the wake. Negative values indicate density slices upstream of the sink, and the sink location is denoted by the vertical dashed line. Dotted vertical lines stand for $s=R^S_\infty$.}
    \label{fig:m10_wake}
\end{figure}

Returning to the $\mathcal{M}_\infty = 10$ adiabatic simulations discussed in detail in Section \ref{sec:quasi_adiabatic}, one can see in the density slices in Figure \ref{fig:m10_density} that the wake develops even for low resolution simulations. Figure \ref{fig:m10_wake} shows the mass distribution and drag force profile of each wake at $t=25$, plotted against the distance $s$ to the sink measured along the axis of symmetry of the wake. The mass distribution close to the sink, where $\lvert s \rvert < 0.5 R^S_\infty$, shows some variation with increasing resolution, due to the locally unstable flow. However, the global structure of the wake at larger radii converges quickly, for $N\geq13 $, and remains stable. The highest contributions to the drag force are found in the immediate vicinity of the accretor, but as symmetric contributions upstream and downstream of the sink cancel out, the larger scale structure of the wake (within $s \approx R^S_\infty$ in this case) contributes the bulk of the net drag force onto the sink. As a result, the overall drag is adequately captured even with moderate resolution: integrating $\mathrm{d}F^D_\diamond/\mathrm{d}s$ over $s$ for simulations m1013a, m10n26a, m10n52a, m10n209a yields values for $F^D_\diamond$ of 0.057, 0.063, 0.071, 0.059 respectively in the dimensionless units used in this work, so that even in the most violently unstable case (m10n209a), the drag force fluctuates by less than 20\%. 

\begin{figure}
	\centering
	\includegraphics[width=\columnwidth]{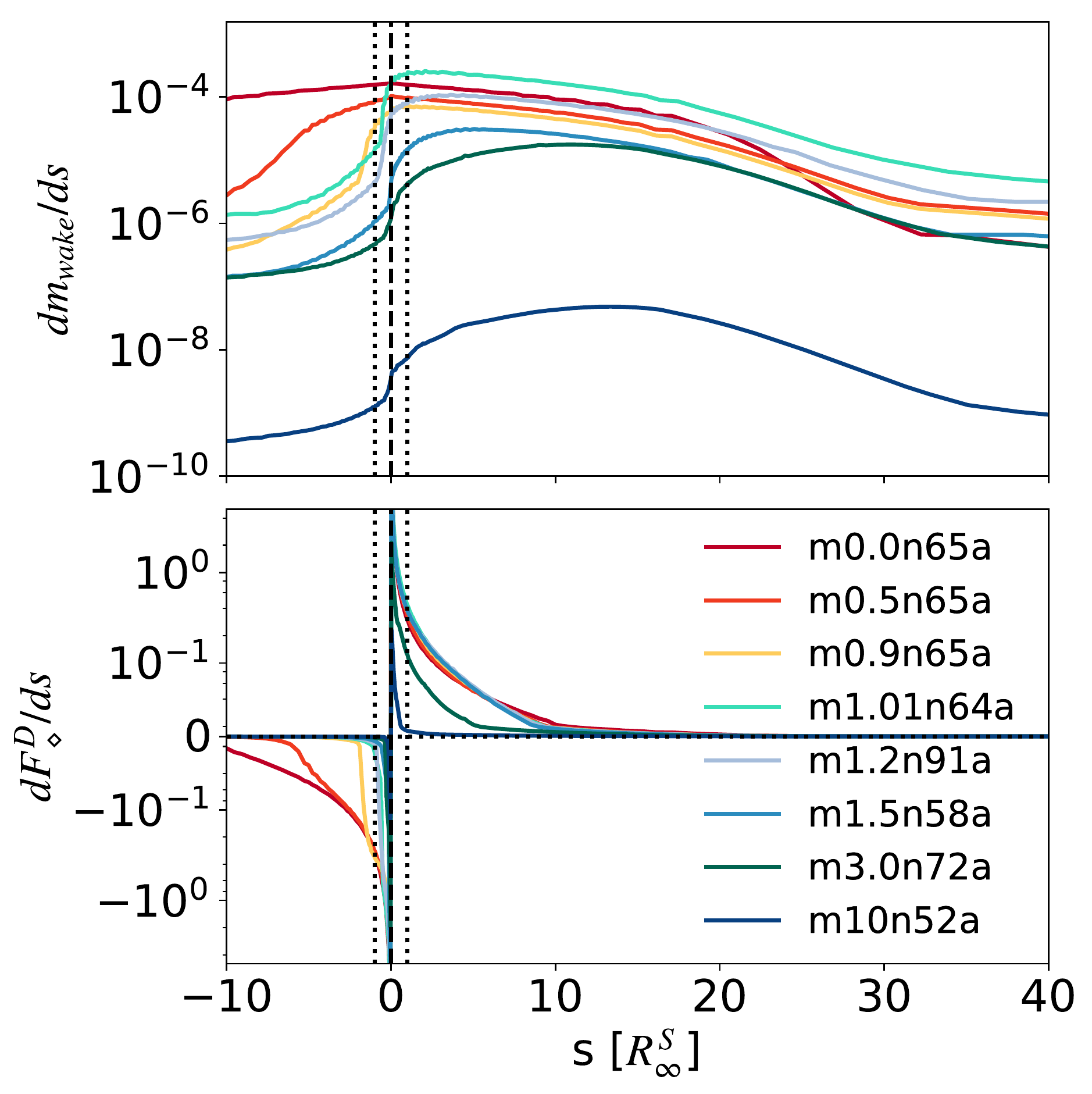} 
	\caption{Same differential mass (top panel, in units of $[M_{sink}/R^S_\infty]$) and drag force (bottom panel, in units of [$4 \pi \rho_\infty (GM_{sink})^2 /(R^S_\infty c_{s,\infty}^2)$]) profiles as Fig \ref{fig:m10_wake} measured at time $t=25$, but for different Mach numbers, as labelled on the bottom panel. Note the increased upstream contribution to the drag force as the Mach number decreases.} 
	\label{fig:fdrag_profiles}
\end{figure}

Figure \ref{fig:fdrag_profiles} shows the spatial contribution of density slices along the wake for adiabatic simulations at $t=25$ for a range of Mach numbers, the density slices for which can be found in Figure \ref{fig:density_slices}. As expected, the wakes contain a significant amount of mass on relatively large scales (up to $s \approx 25 \times R^S_\infty$), especially for transsonic ($\mathcal{M}_\infty \simeq 1$) configurations. However, the inverse square dependence on the distance to the sink means that for all Mach numbers investigated here, most of the gravitational drag force comes from a region within $r_\mathrm{max} \simeq 10 \times R^S_\infty$. Note that the intensity of the force also depends on the opening angle of the wake, with a similar mass profile exerting a stronger pull in the direction of motion if confined to a narrower cone. Finally, we also measure a non-negligible contribution of material in front of the sink particle, pooling behind the detached shock, that exerts a gravitational force in the opposite direction and reduces the overall drag, especially in the sub- and transsonic regimes. While this feature was also observed in \citet{Chapon2013}, it appears more prominently in the simulations presented here, and is completely absent in the analytic solutions for supersonic black holes by \citet{Ostriker1998}, who state that the sink particle only generates a density wake within the rear Mach cone. 

\begin{figure}
	\centering
		\includegraphics[width=0.5\textwidth]{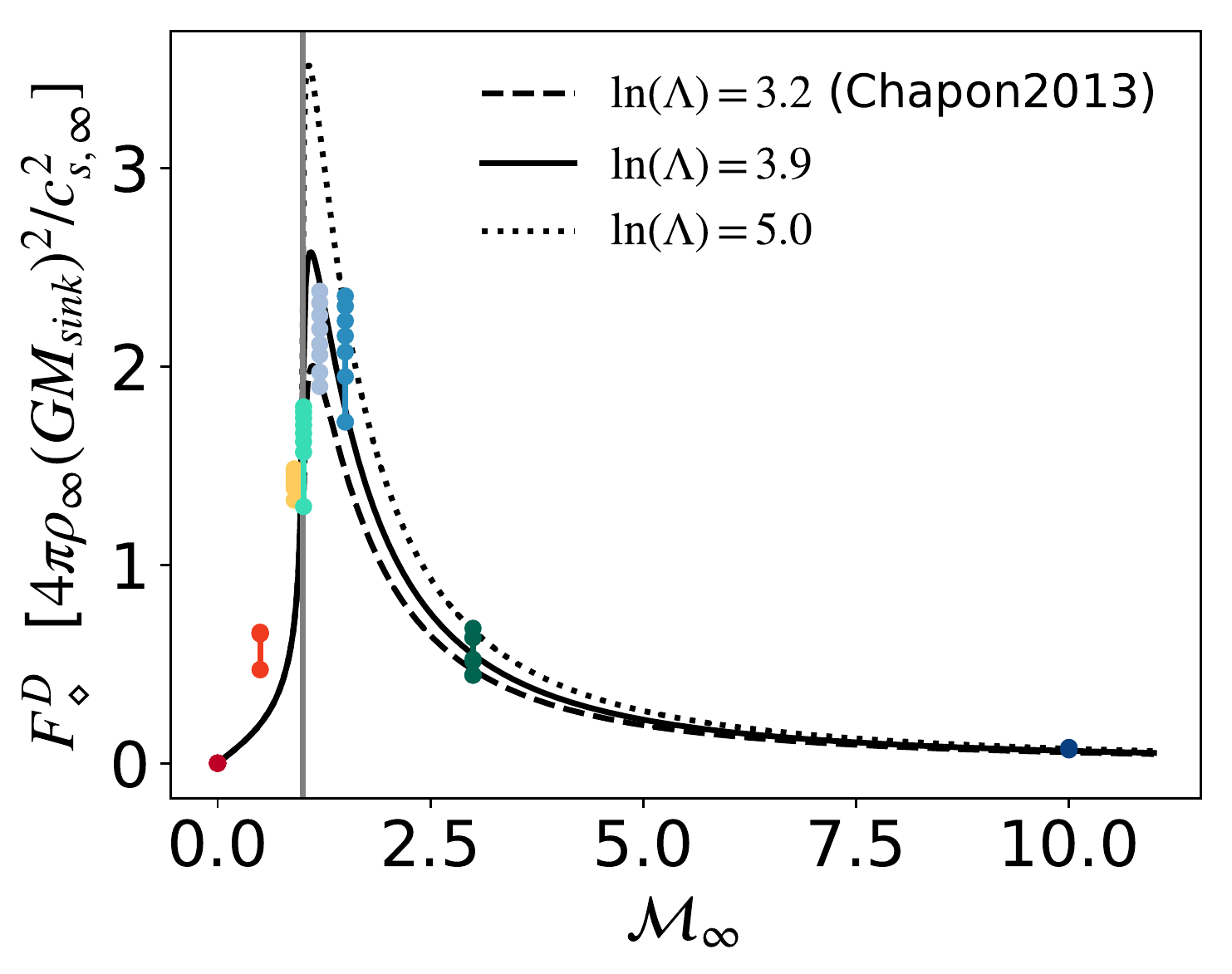} 
	\caption{Drag force due to the gravitational wake as a function of the Mach number of the flow, for simulations with a resolution $N\approx 65$ (see bottom panel of Figure \ref{fig:fdrag_profiles} for the exact resolution). The different curves represent the analytic formula of \citet{Ostriker1998} (Equation \ref{eq:fdrag}) for different values of the Coulomb logarithm, as indicated on the panel. Each simulation was sampled at times $t=[10,12,14,16,18,20,22,24]$ to give an idea of the dispersion in the drag force measurements, hence the multiple data points for any given Mach number.}
	\label{fig:fdrag_mach}
\end{figure}

Comparing the total net gravitational drag for the set of resolved simulations in Figure \ref{fig:fdrag_profiles} to analytic estimates in Figure \ref{fig:fdrag_mach}, 
and considering that the size of the accretor sets the smallest scale $r_\mathrm{min} \simeq r^* \simeq 2 \Delta x_\mathrm{min} \simeq \frac{1}{32} R^{S}_\infty$, the Coulomb logarithm evaluates to $ \ln( \Lambda ) = \ln (r_\mathrm{max}/r_\mathrm{min}) \simeq 5.8 $, larger than the value of $3.2$ reported by \citet{Chapon2013} in the transsonic 
regime. However it is clear from the bottom panel of Figure \ref{fig:fdrag_profiles} that $r_\mathrm{max}$, the characteristic size of the medium which the accreting object traverses, 
is quite a sensitive function of Mach number, as it drops from $\approx 10 \times R^S_\infty$ in the transsonic regime to a value of 
about $R^S_\infty$ at Mach $\mathcal{M}_\infty = 10$, corresponding to $\ln (\Lambda) \simeq 3.5$. Fitting over the whole range of Mach numbers covered by our simulations, 
we find a time averaged best fit Coulomb logarithm of $\ln(\Lambda) = 3.9$, with $\ln(\Lambda) = 3.2$ (5.0) providing an adequate lower (upper) bound to individual time 
measurements. We suspect that this discrepancy with the \citet{Chapon2013} results is partly due to the different value of $\gamma$ used in these authors' simulations, but more likely caused by the absence of accretion onto their black holes. 
This is somewhat corroborated by the fact that we also find an excess in gravitational drag for subsonic sinks, which show a more prominently asymmetric density profile than 
in the non-accreting analytic solutions.

\begin{figure}
	\includegraphics[width=\columnwidth]{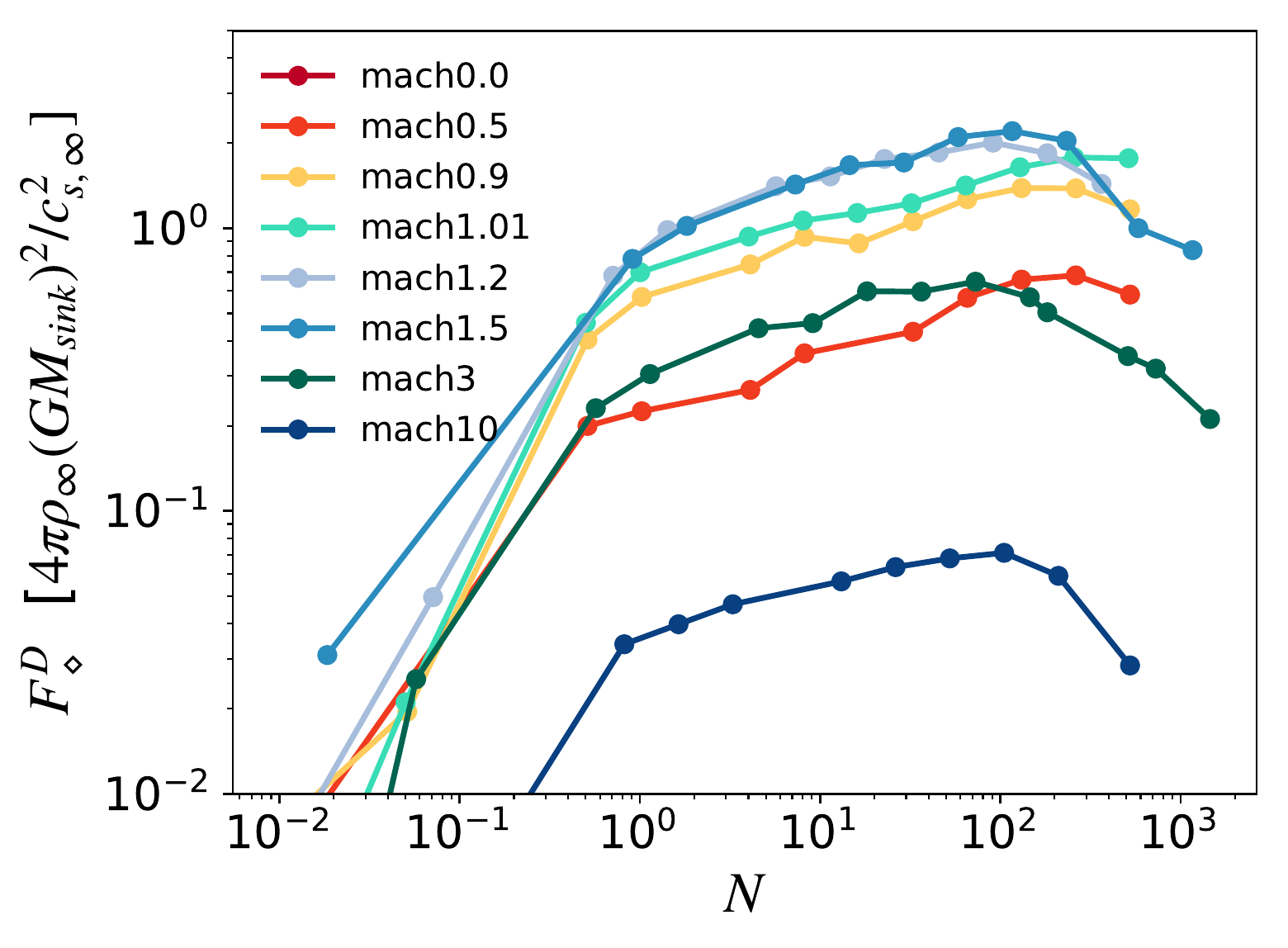}
    	\caption{Time averaged gravitational drag force as a function of resolution for a variety of Mach numbers, as indicated on the panel. 
          The drag force for $\mathcal{M}_\infty=0$ does not appear on the plot as it is negligible.}
    \label{fig:fdrag_dx}
\end{figure}

Contrary to the accretion rates in Figure \ref{fig:dotM_dx}, the drag force due to the wake forming behind the sink shows rapid  convergence at a surprisingly low resolution of $N \geq 1$, as shown in Figure \ref{fig:fdrag_dx}. It is only vastly underestimated at resolutions as low as $N <1$, i.e. when the scale length corresponding to the gravitational influence of the black hole, $R^S_\infty$, is small in comparison to the minimal cell size, so that gravitational focusing is inefficient. The drag force due to the wake is only moderately influenced (up to $50 \%$) by the instabilities developing behind the bow shock, as it is dominated by larger scale contributions. 

On top of the direct computation of the gravitational drag exerted on the sink by the overdensity in its wake, RAMSES includes a sub-grid algorithm that calculates the drag force based on the same local mass weighted average quantities used to estimate the accretion rate, $F^D_\bullet ( M_\mathrm{sink},\rho_\bullet,v_\bullet,c_{s,\bullet}) \hat{\bold{v}}_\bullet$, with $\ln(\Lambda)=3.2$. It is designed to compensate for the lack of drag force in low resolution simulations, clearly visible in Figure \ref{fig:fdrag_dx} for $N<1$. As the drag force on the sink is a vector quantity, it inherits its direction from the local relative velocity, $\bold{v}_\bullet$. 

\begin{figure*}
	\includegraphics[width=\textwidth]{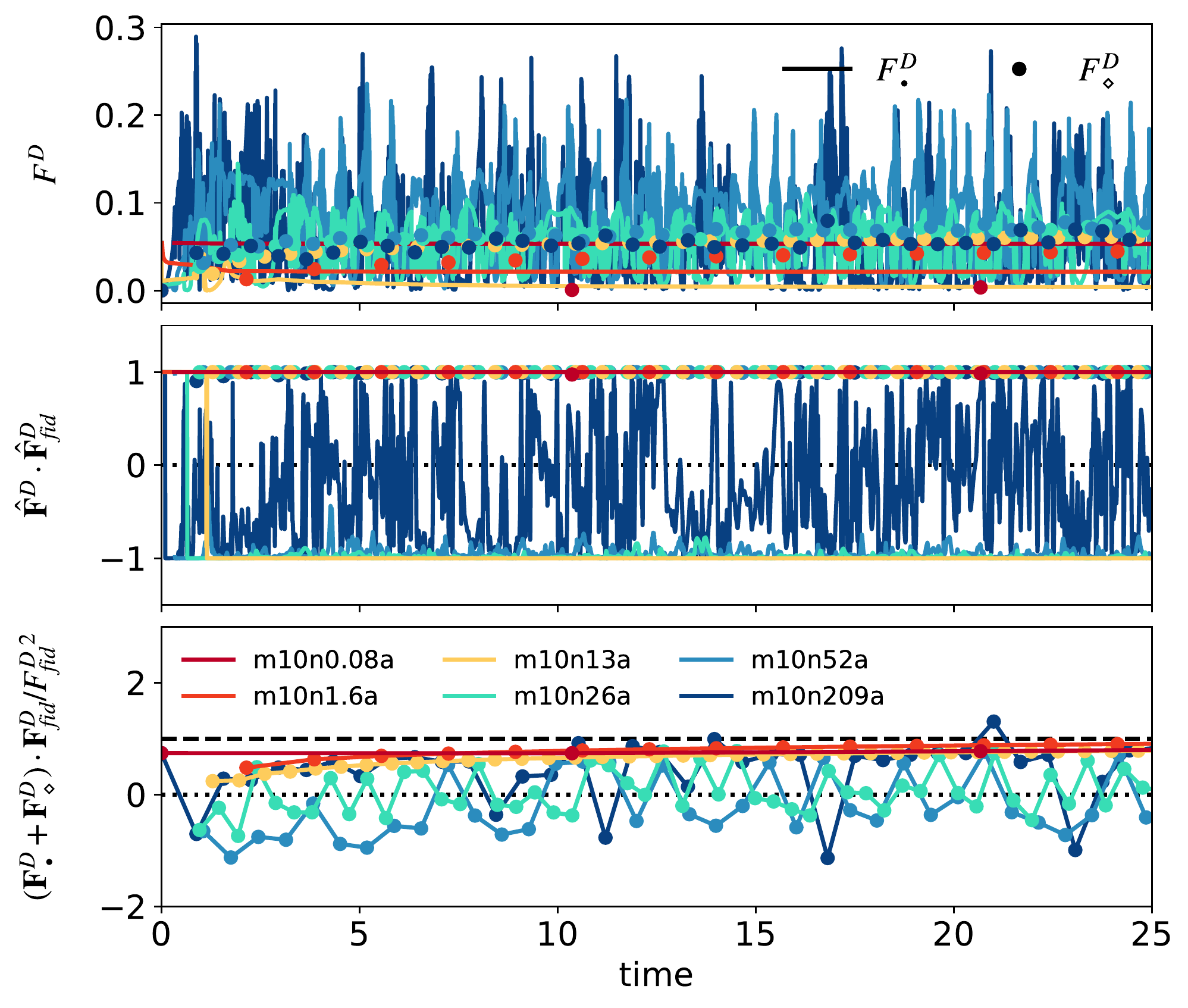}
	\caption{Magnitude of the drag force calculated in two different ways: i) directly from the contribution of each gas cell of the simulation, $F^D_\diamond$, shown as filled markers and ii) using the sub-grid algorithm based on local mass weighted quantities and $\ln(\Lambda) =3.2$ (Equation \ref{eq:Idrag}), $F^D_\bullet$, shown as solid lines. $F^D_\diamond$ is sampled at coarse timesteps of the simulation, whereas $F^D_\bullet$ is sampled at fine timesteps, hence the different number of datapoints. The horizontal dashed line denotes the point when the instantaneous force is equal to the fiducial force, $F^D_\mathrm{fid}$, taken to be the force measured in m10n209a simulation at $t=25$ (see text for details). The horizontal dotted line shows where the instantaneous force is either zero or perpendicular to the fiducial value.} 
	\label{fig:m10_fdrag}
\end{figure*}

For the supersonic case, $\mathcal{M}_\infty=10$, the top panel of Figure \ref{fig:m10_fdrag} shows that the magnitude of the cell based force, $F^D_\diamond$ converges for $t>10$ and $N>10$. The instabilities in the flow for $N>20$ cause small variations of about $10 \%$ in magnitude (filled symbols). To investigate the impact of resolution both on the magnitude and direction of the drag force, we define a fiducial drag \mbox{$\bold{F}^D_\mathrm{fid} = - F^D_\diamond (N=209,t=25) \hat{\bold{v}}_\infty$}, based on the flow velocity at infinity and the converged drag force intensity $F^D_\diamond(N=209,t=25)$. Figure \ref{fig:m10_fdrag} shows that the drag force due to the wake is always parallel to the fiducial force (filled markers, middle panel). As expected from the direction of the wake, the force acts in the opposite direction to the global flow velocity, and slows the sink down. For low resolution, $N<2$, the drag force due to the wake can be underestimated by a large factor, but remains steady. For high resolutions, $N > 10$, the magnitude of the force rapidly converges to within $10 \%$ of the fiducial value. At the highest resolutions, $N>20$, short term variations in the magnitude of the force are visible (filled markers, top panel) but no significant deviation from the axis of symmetry of the wake (middle panel). 

The sub-grid based drag force, $\bold{F}^D_\bullet$ (see top two panels of Figure \ref{fig:m10_fdrag}), by contrast, shows a very erratic behaviour for $N > 20$. Its magnitude, $F^D_\bullet$ (solid lines), varies considerably on the shortest timescale probed here, the finest timestep of the simulations, and can be both significantly larger or significantly smaller than the $F^D_\diamond$ value. Moreover, higher resolutions show larger fluctuations. 
Not only does the magnitude of the force fluctuate rapidly, but it is also directed in the opposite direction to the fiducial force most of the time (actually at all times for $10 < N < 200$). This can be easily understood, because the bulk of the mass enters the accretor through the accretion column, which has a flow direction directly {\em opposite} to the global flow (see Figure \ref{fig:m10_density} for some examples). As a result of mass weighting, the velocity of the accretion column thus dominates the local flow velocity, $\mathbf{v}_\bullet$, and as $\mathbf{F}^D_\bullet \propto \mathbf{v}_\bullet$, the sub-grid drag force flips direction as soon as the accretion column forms.  The full extent of the problem becomes apparent when calculating the total drag force in the presence of the sub-grid algorithm, $\mathbf{F}^D_\mathrm{tot} = \mathbf{F}^D_\diamond + \mathbf{F}^D_\bullet$ (bottom panel of Figure \ref{fig:m10_fdrag}). For the resolved cases, $N>20$, $\mathbf{F}^D_\bullet$ frequently is the dominant term, causing a net force that \textit{accelerates} the sink relative to the global gas flow. This is clearly unphysical, and entirely caused by the fact that the local mass weighted relative velocity, $\mathbf{v}_\bullet$ does not reflect either the direction and/or magnitude of the value at infinity, as soon as the accretion column forms and the bow shock detaches from the accretor. As expected, the total drag force for the unresolved cases $N<2$ is unaffected, as the local velocity reflects the value at infinity because the accretion column has not (fully) developed. The sub-grid based drag force therefore significantly and accurately contributes to the overall drag force in that case. This contribution naturally drops as the accretion column builds up, leading to a decrease in relative velocity and an increase in sound speed as the flow in the vicinity of the accretor becomes sub-sonic (see bottom panel of Figure \ref{fig:m10_profiles}). From Figure \ref{fig:m10_fdrag} (top and middle panel) the flip in drag force direction occurs around at $N \simeq 10$, which does not create a significant problem at this resolution as the resolved drag force term dominates. As a result, we find $\mathbf{F}^D_\mathrm{fid} > \mathbf{F}^D_\mathrm{tot} (N \leq 10) \geq 0.8 \times \mathbf{F}^D_\mathrm{fid}$ (bottom panel of Figure \ref{fig:m10_fdrag}) for the standard drag force sub-grid model used in RAMSES with $\ln (\Lambda) = 3.2$, and near perfect agreement at $N < 2$ with our best fit value of $\ln (\Lambda) = 3.9$ (not shown).

Based on these results, and taking a conservative approach, sub-grid algorithms for the drag force onto the sink should be avoided as soon as the characteristic scale radius, $R^S_\infty$, becomes larger than the size of the accretor, $2 \Delta x_\mathrm{min}$. Note that $R^S_\infty$ itself depends on the relative velocity of the black hole and the ISM and can therefore be difficult to determine in more complex simulations than the idealised Bondi-Hoyle-Lyttleton flows investigated here, where the values at infinity are known. The criteria at which the sub-grid drag force on black holes becomes unphysical might therefore have to be revised for galaxy evolution simulations, the analysis of which we leave for future work. 

\section{Are the instabilities at high Mach number physical?}
\label{subsec:discussion_instability}

\begin{figure*}
	\centering
	\begin{tabular}{ccc}
		\includegraphics[width=0.325\textwidth]{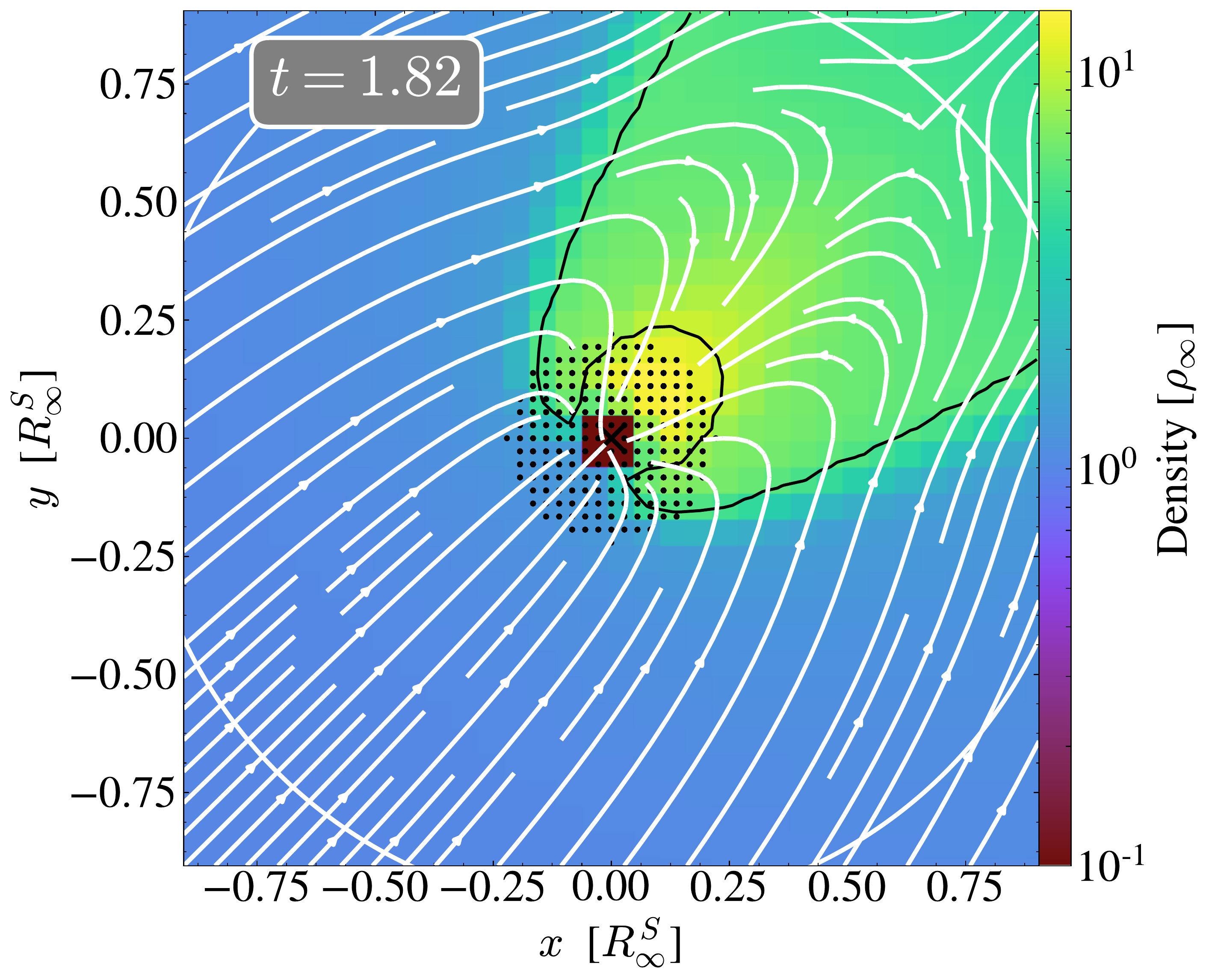}&
		\includegraphics[width=0.325\textwidth]{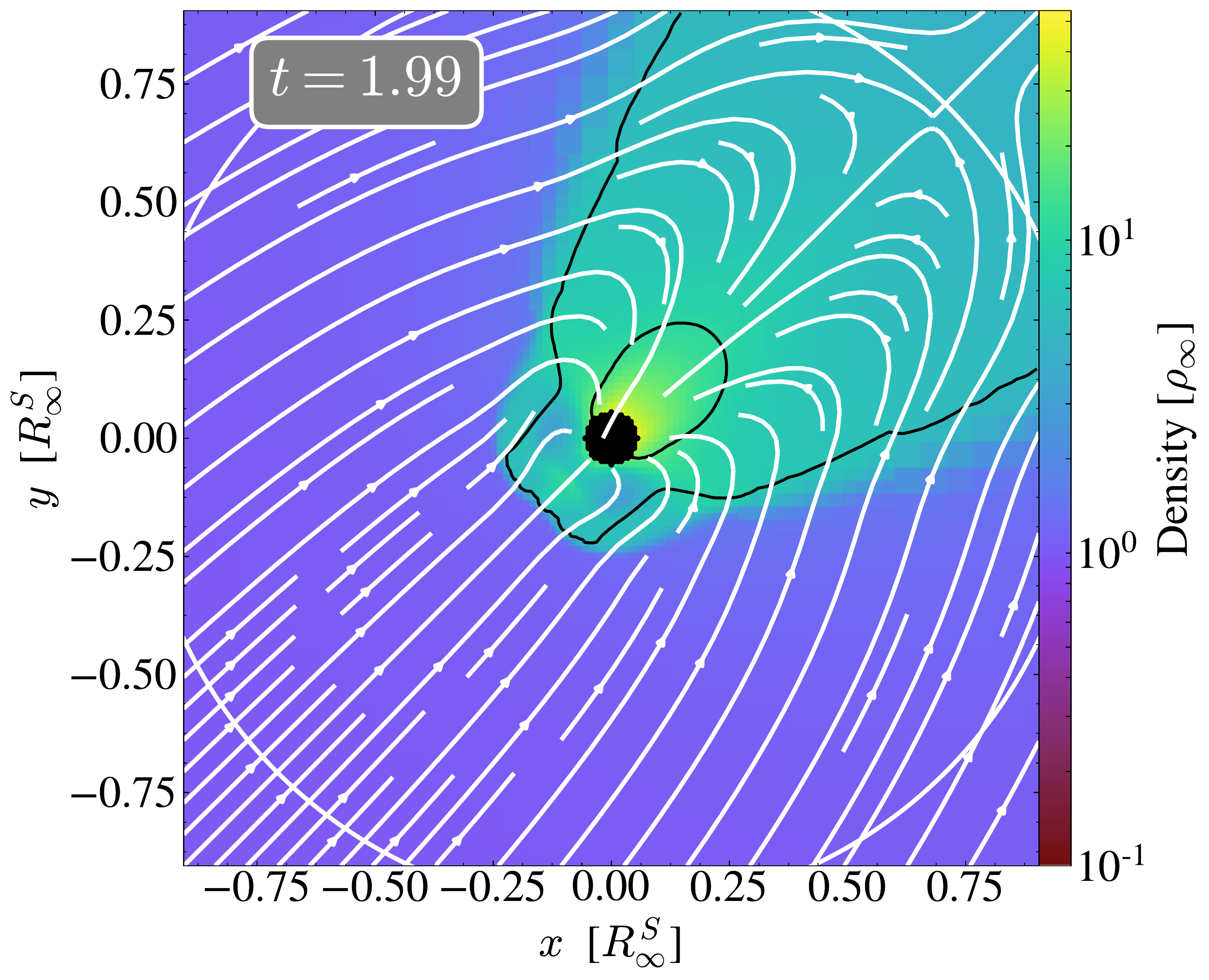}&
		\includegraphics[width=0.325\textwidth]{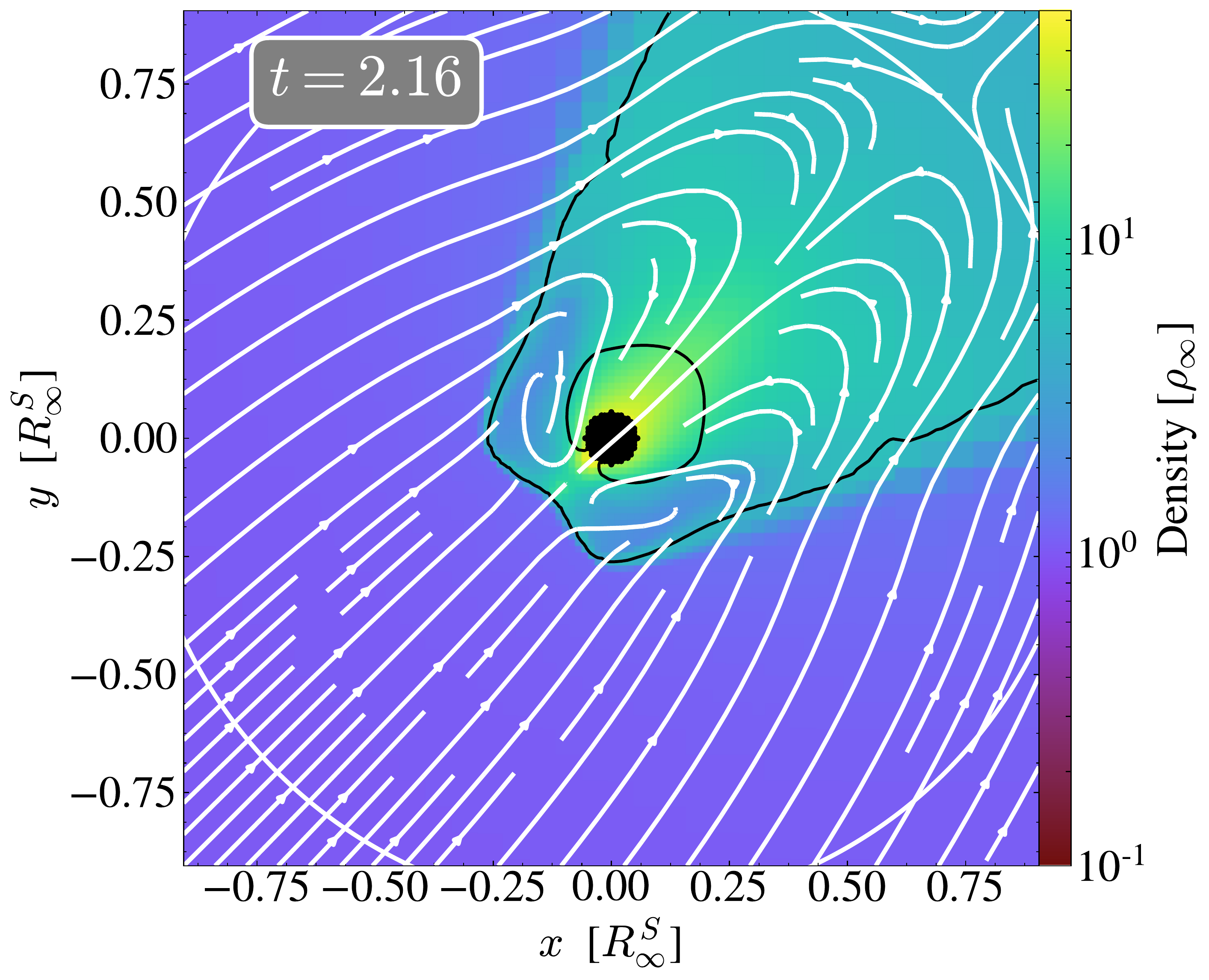}\\
	\end{tabular}
	\caption{Density slices across the characteristic radius for a shrinking accretor, here from $N=36$ to $N=72$, equivalent approximately to $r^* / R^S_\infty = 0.06 \rightarrow 0.028$ embedded in an adiabatic supersonic ($\mathcal{M}_\infty = 3$) flow. While the larger accretor shows a stable flow pattern, short lived eddies appear behind the bow shock for the smaller accretor.  The size of the accretor is annotated by the cloud particles (black dots), and $R^S_\infty$ by the solid white circle. Sonic surfaces are denoted as black contours. }
	\label{fig:eddies}
\end{figure*}

To the best of our knowledge, no complete analytic analysis yet exists to explain the instabilities regularly found for accretion wakes in the Hoyle-Lyttleton problem, so the question remains whether the observed instabilities are physical or numerical in nature. Our work on the danger of using small accretors with uniform initial conditions (see Appendix \ref{sec:initial_conditions}) shows that a numerical origin is extremely difficult to rule out. However, a vast amount of efforts to stabilise simulations have failed, as instabilities have been reported using a wide variety of codes, coordinate systems and models for the accretor \citep[See][for a review]{Foglizzo2005}. In agreement with the results presented here, a majority of authors report a clear link of the appearance of instabilities with the size of the accretor. 

Claims of the existence of steady state solutions for small accretors and high Mach numbers, such as the work by \citet{Mellah2015} ($N=1000$ and $\mathcal{M}=16$) and \citet{Pogorelov2000} ($N=20$ and $\mathcal{M}=20$) have also been made. However, all such results are based on 2D axisymmetric simulations, and it is well known that for Hoyle-Lyttleton simulations, certain instabilities only appear for particular configurations. For example the flip-flop instability, frequently observed in 2D, is entirely absent in 3D simulations \citep{Blondin2009}. 

To investigate how a shrinking accretor impacts the stability of the flow, we take an established, stable simulation with an intermediate accretor size (m3n32a), and shrink the accretor by adding refinement levels to reproduce the conditions in m3n72a, where instabilities unsettle the flow. Both simulations are supersonic, with $\mathcal{M}_\infty=3$, and m3n32a has reached a steady state before the accretor is shrunk. 

\begin{figure}
	\centering
	\includegraphics[width=\columnwidth]{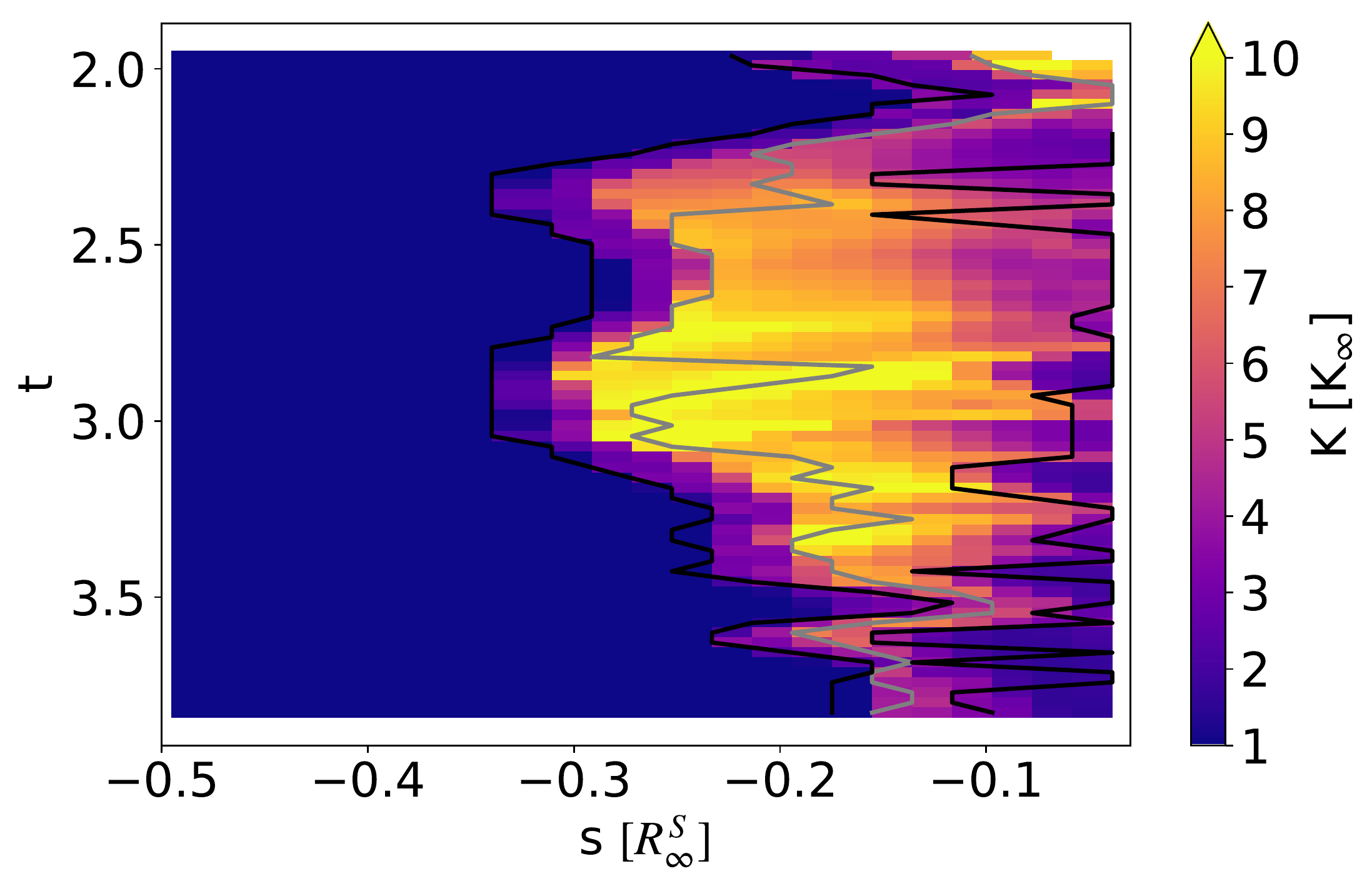}
	\caption{{Time evolution of entropy,  $K$, profiles measured in the direction of the initial flow velocity $\hat{\bf{v}}$ for the simulation shown in Figure \ref{fig:eddies}. $s$ denotes the distance to the accretor parallel to the axis of symmetry of the wake, with negative values measured upstream of the sink. Black lines denote the sonic points along the profile, while the grey line highlights the position of the entropy peak.}}
	\label{fig:entropy_timeseries}
\end{figure}

\begin{figure*}
	\centering
	\begin{tabular}{ccc}
		\includegraphics[width=0.32\textwidth]{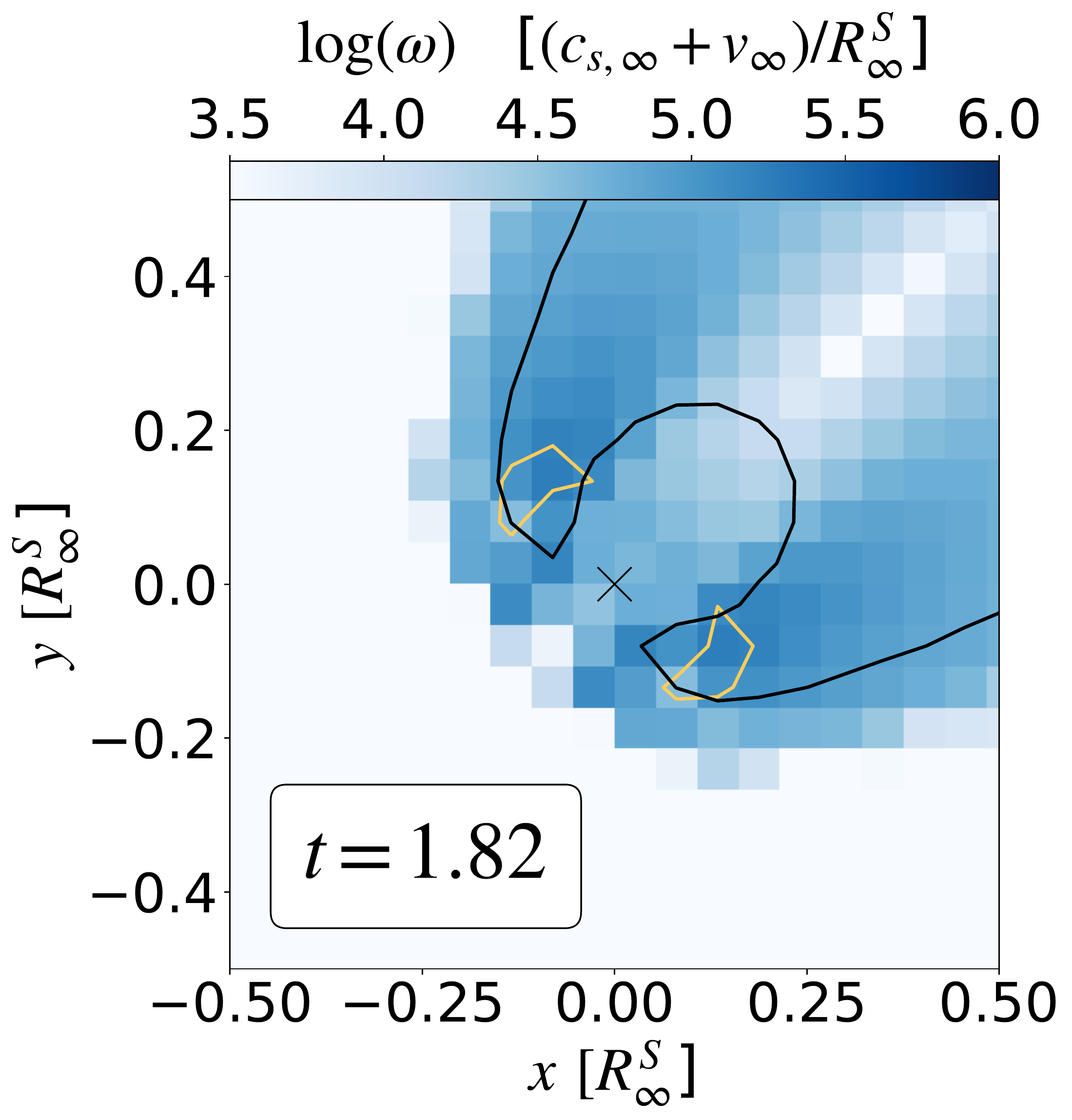}&	
		\includegraphics[width=0.32\textwidth]{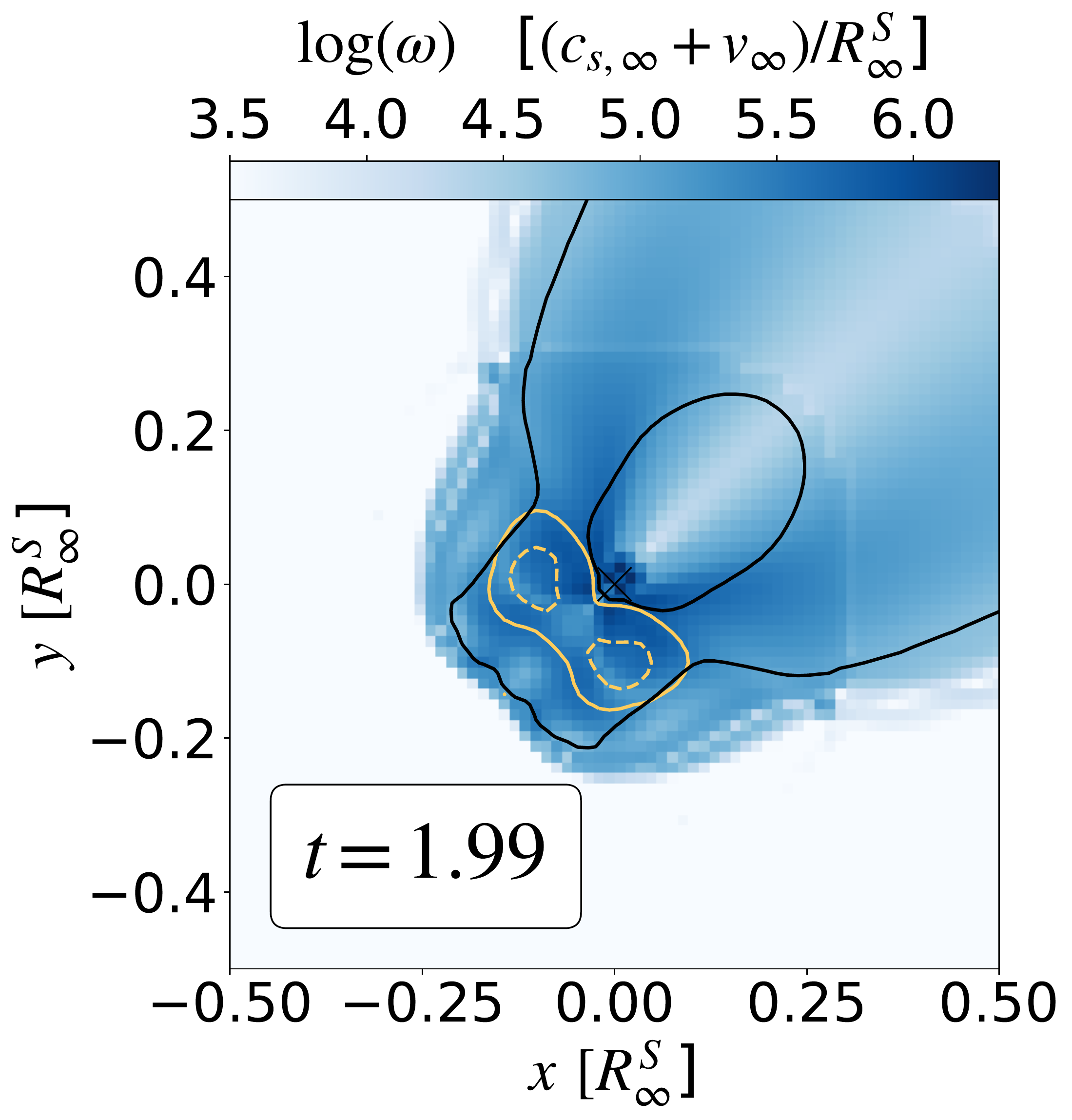}&	
		\includegraphics[width=0.32\textwidth]{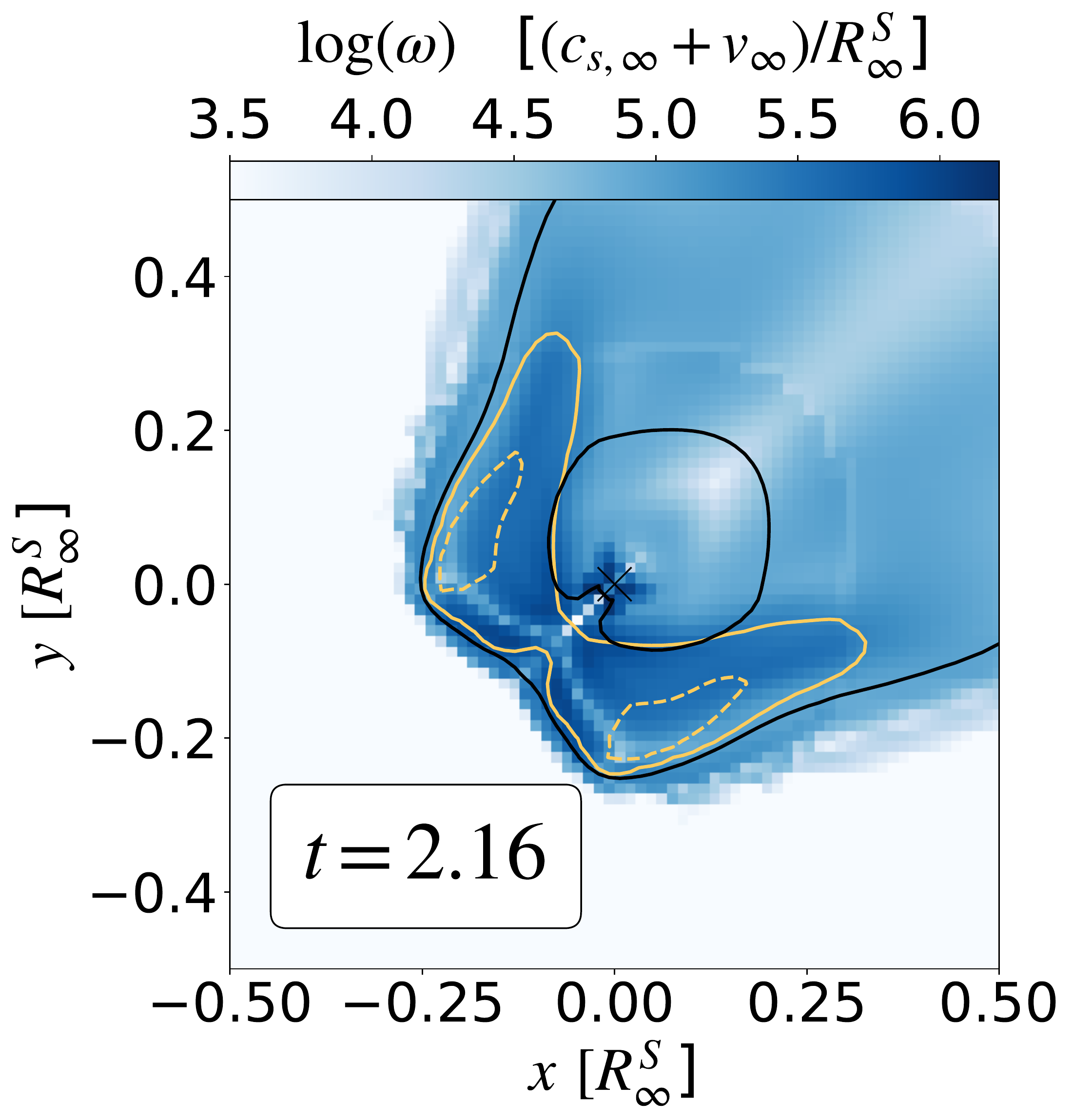}\\
	\end{tabular}
	\caption{{Vorticity slices of the same simulation as shown in Figure \ref{fig:eddies}. Black contours show the sonic surfaces, whereas yellow contours denote entropy $K/ K_\infty = 5$ (solid line) and  $K/ K_\infty = 10$ (dashed line). The location of the accretor is denoted by a black cross. A minimum value of $\log(\omega) = 3.5$ was enforced on the plot to bring out the vorticity structure within the bowshock and wake. The circular features in the vorticity at radii of $\sim 0.3 \ R^S_\infty$ in the right hand two plots are caused by discrete jumps in the spatial resolution of the nested grid.}}
	\label{fig:vorticity}
\end{figure*}

The density slices in Figure \ref{fig:eddies} confirm observations from the $\mathcal{M}_\infty=10$ case presented in Section \ref{sec:hoyle_lyttleton}. Shrinking the accretor creates a subsonic region, forming upstream between the supersonic region surrounding the accretor and the shock front, which is absent for the steady state solution at lower resolution, { as can be seen in Figure \ref{fig:eddies}. As soon as the subsonic region develops, entropy perturbations form close to the accretor upstream of the sink (Figure \ref{fig:entropy_timeseries}). They oscillate back and forth between the two sonic surfaces, disturbing the upstream bowshock in the process. As can be seen both in the profiles in Figure \ref{fig:entropy_timeseries} and in the contours in Figure \ref{fig:vorticity}, the entropy perturbations remain confined between the two sonic surfaces, and therefore are only able to form if the accretor is sufficiently small to allow this sub-sonic region to open. }  During the next 10 dynamical times, the simulation does not resettle into a steady state. 

As previously mentioned, \citet{Foglizzo2009} suggest that the advective-acoustic instability should unsettle the flow in exactly this manner. It is characterised by small entropy perturbations forming behind the shock, which are advected towards the accretor, where the local rise in density causes them to reflect back to the shock front, exciting further perturbations. {Contrary to \citet{Foglizzo2009}, who predict that entropy perturbations form at the bowshock and propagate towards the accretor, the simulations presented here show that the entropy perturbations first appear close to the accretor and then propagate forwards in the direction of the bow shock instead.}
As was shown in section \ref{sec:machs}, all of our simulations which have $\gamma > 4/3$, a sufficiently strong shock (which occurs for $\mathcal{M_\infty} \geq 1.5 $), and enough resolution for the shock to detach, show these types of instabilities, providing strong support for the physical origin advocated in the analysis of these authors. 

\begin{figure}
	\centering
	\includegraphics[width=\columnwidth]{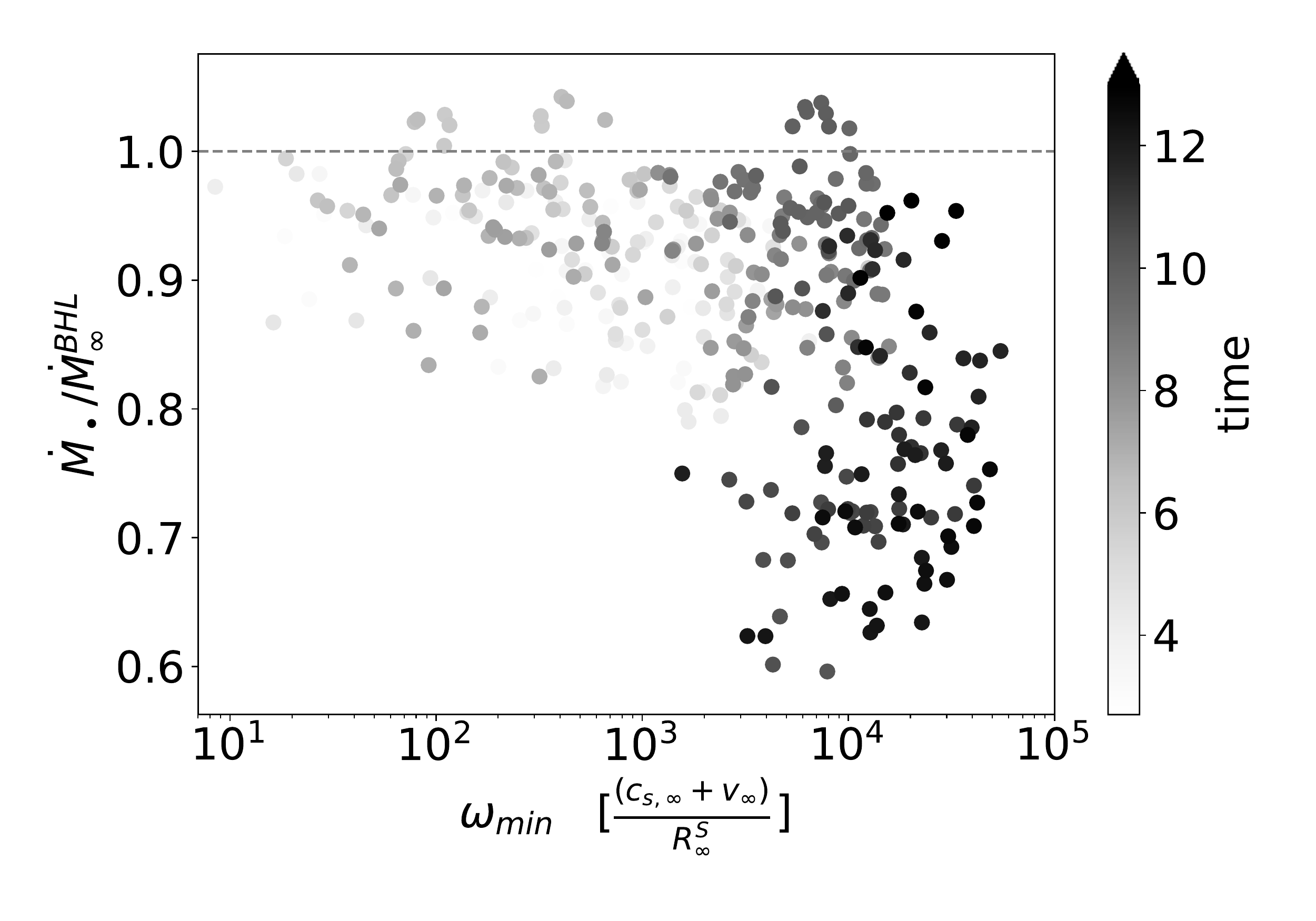}
	\caption{{Evolution of the instantaneous accretion rate onto the black hole versus the minimum vorticity, $\omega_{min}$ within the accretion column for the simulation shown in Figure \ref{fig:eddies}. The vorticity is measured within a hemisphere with a radius of $5 \Delta x$, centred on the accretor and oriented downstream of the accretor.}}
	\label{fig:accretion_column}
\end{figure}

{While the entropy perturbations upstream of the sink appear as soon as the sub-sonic region opens up, the flow downstream of the sinks takes longer to become unsettled. This can be seen in Figure \ref{fig:accretion_column}, which shows the minimum vorticity within a hemisphere of radius $5 \Delta x$, centred on the sink and oriented downstream, plotted against the instantaneous accretion rate of the sink particle. On timescales of order 5 flow crossing times of $R^S_\infty$, the flow at the bottom end of the accretion column evolves from a low-vorticity state, with $\omega_{min} < 10^3 $, in which the black hole is accreting efficiently at $\dot{M_\bullet}/\dot{M}^{BHL}_\infty>0.8$ to a high vorticity state with $\omega_{min} > 10^4 $, in which accretion rates vary strongly and can drop as low as $\dot{M_\bullet}/\dot{M}^{BHL}_\infty=0.6.$ (Figure \ref{fig:accretion_column}). While the vorticity perturbations behind the bowshock are difficult to quantify due to the non-uniform, resolution dependent vorticity pattern of the characteristic Hoyle-Lyttleton flow, the vorticity slices in Figure \ref{fig:vorticity} show that both vorticity troughs and peaks develop in the regions affected by entropy perturbations (yellows contours) as the flow forms drawn out eddies (see flow lines in Figure \ref{fig:eddies}). Over time, these eddies stretch backwards along the wake, eventually upsetting the symmetry of the flow downstream of the accretor, including the accretion column feeding the sink.}

For the isothermal case, the shock remains attached and the flow is supersonic everywhere, but the wake again becomes globally unstable for sufficiently small accretors. This is particularly important for simulations of SMBHs, as their galactic environment is strongly affected by radiative cooling. While the instabilities for the quasi-isothermal case appear to have a different --- and arguably less well analytically understood --- origin than the adiabatic ones, they lead us to conclude that Hoyle--Lyttleton accretion onto galactic black holes is very unlikely to reach a steady state, whether radiative cooling plays an important role or not. 

\begin{figure*}
	\centering
	\begin{tabular}{ccc}
		\includegraphics[width=0.32\textwidth]{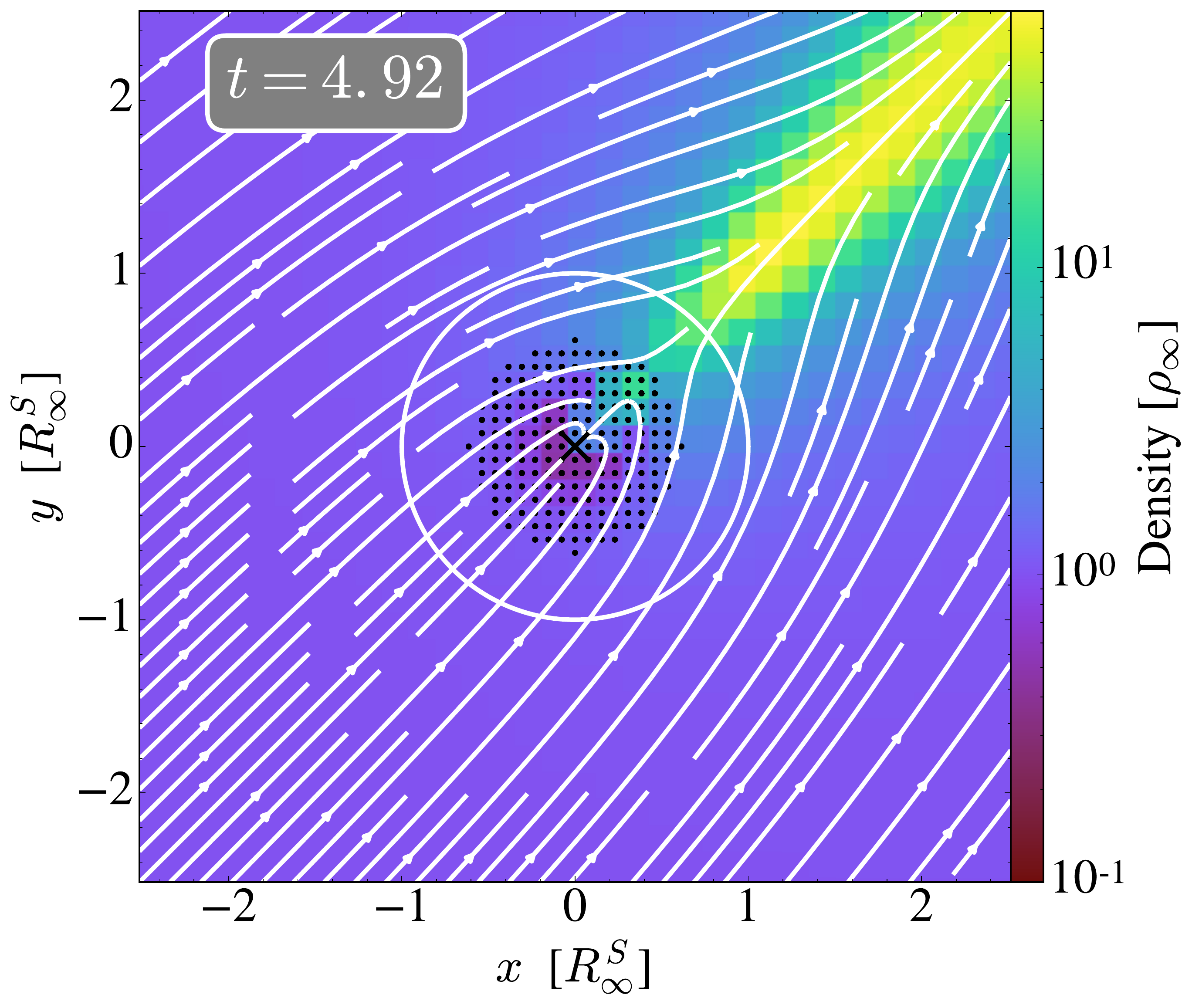} &
		\includegraphics[width=0.32\textwidth]{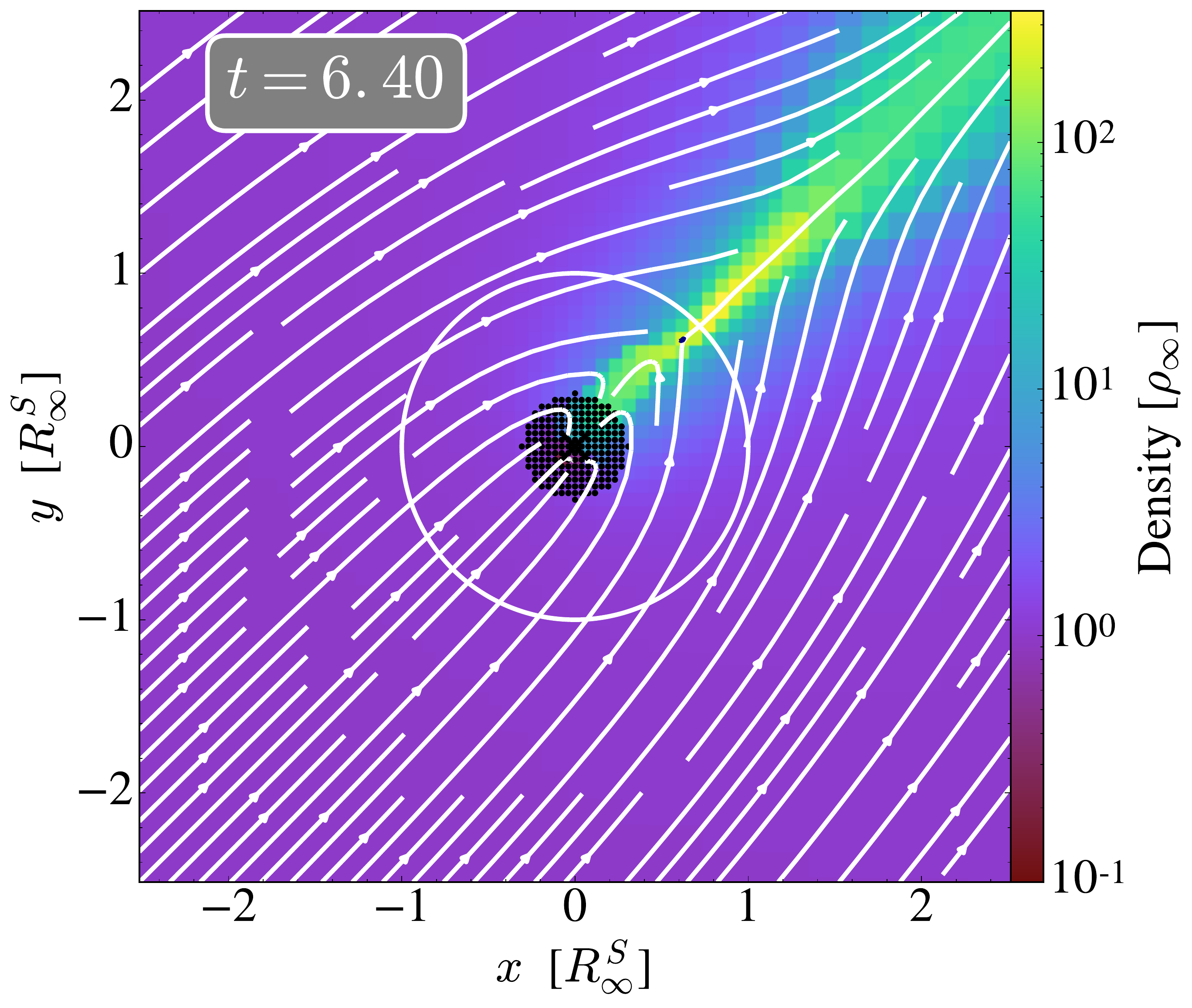}&
		\includegraphics[width=0.32\textwidth]{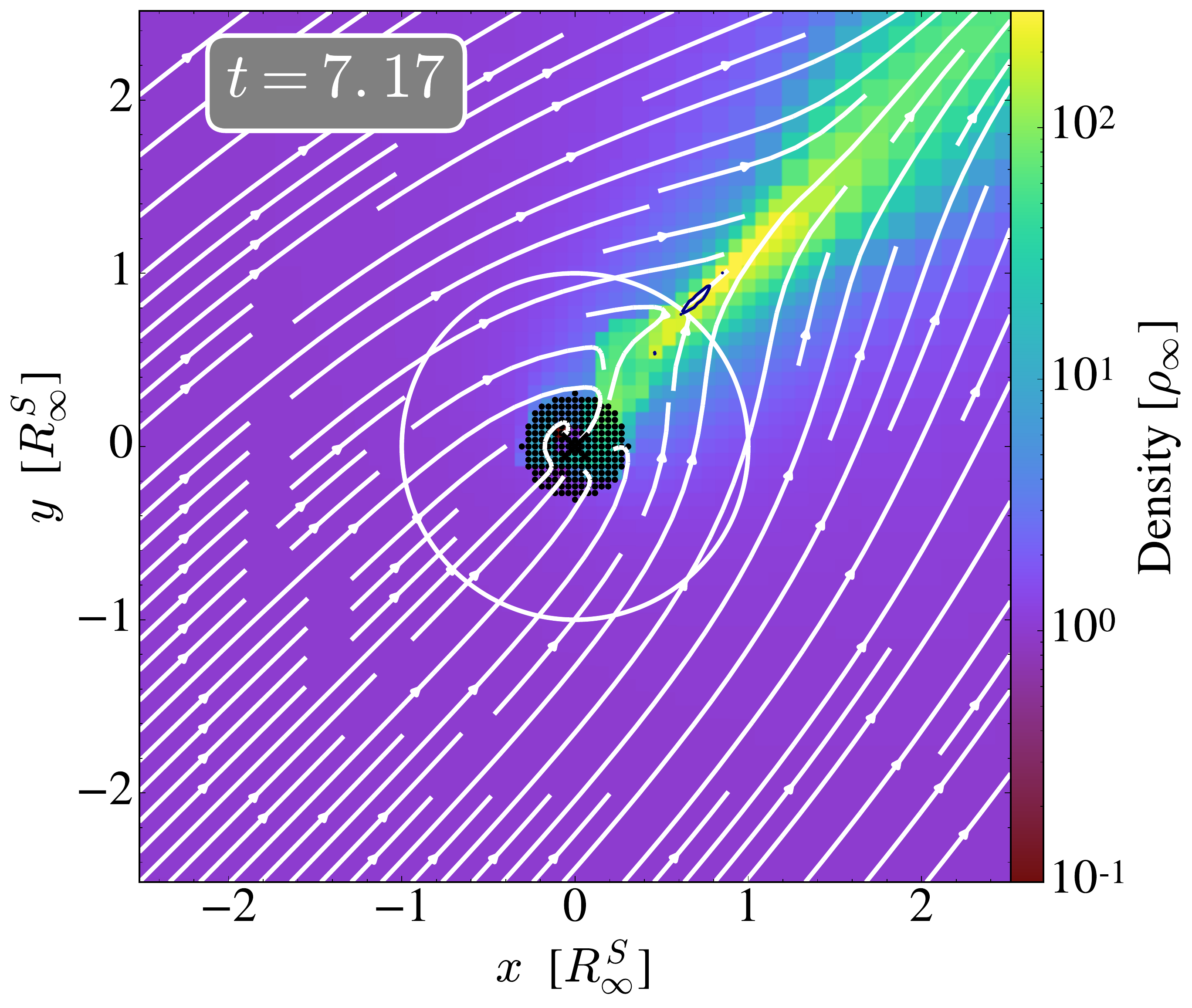}\\
	\end{tabular}
	\caption{Density slices through the characteristic scale radius for a shrinking accretor, here from $N=3.5$ to $N=7$, embedded in a quasi isothermal supersonic flow with $\mathcal{M}_\infty = 3$.  The size of the accretor is annotated by the cloud particles (black dots), and $R^S_\infty$ by the solid white circle. Sonic surfaces are denoted as black contours: the flow remains supersonic everywhere except in a very narrow region around the stagnation point. Flowlines are shown in white.}
	\label{fig:eddies_gamma1.0}
\end{figure*}

\begin{figure*}
	\centering
	\begin{tabular}{ccc}
		\includegraphics[width=0.32\textwidth]{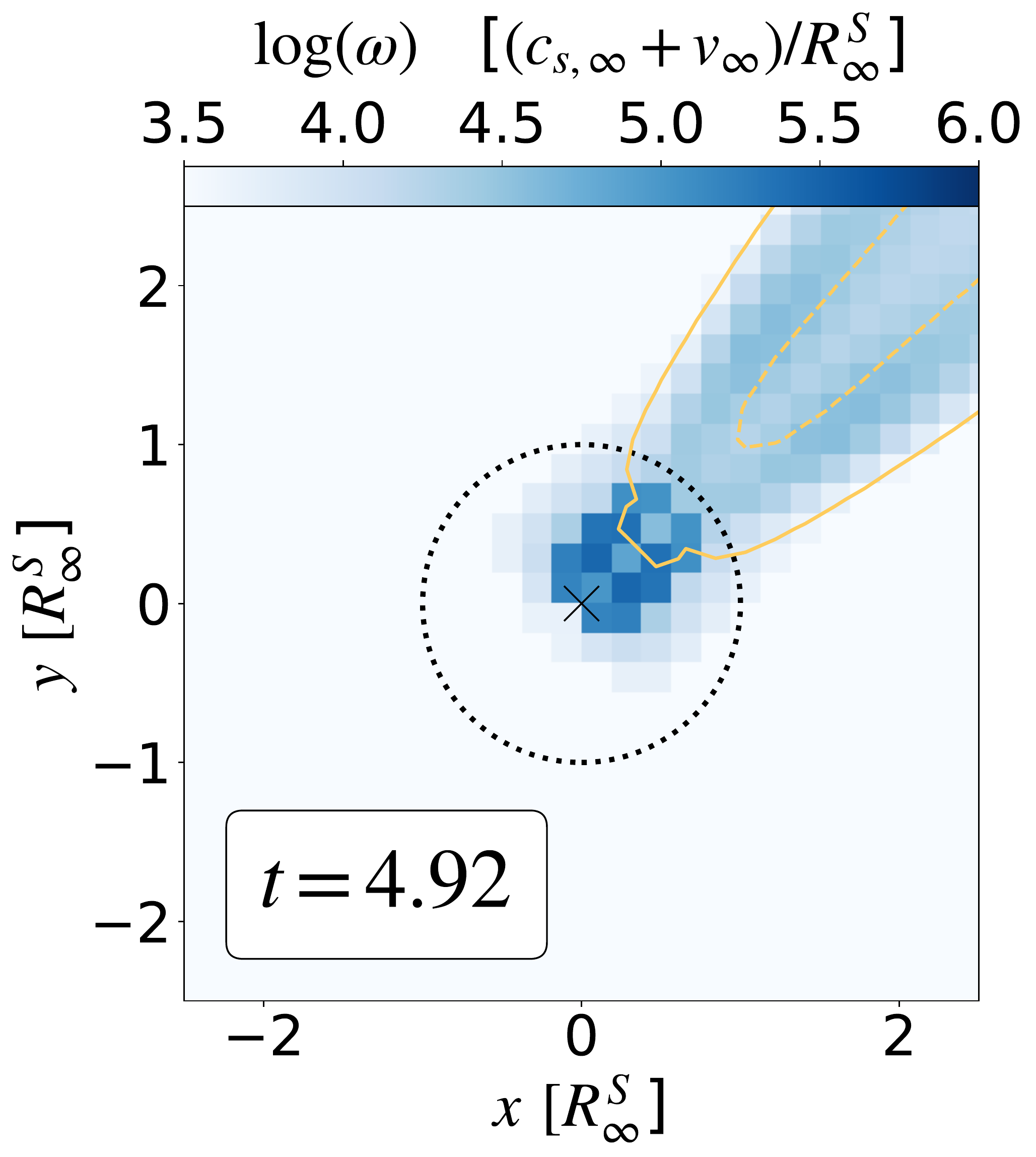}&	
		\includegraphics[width=0.32\textwidth]{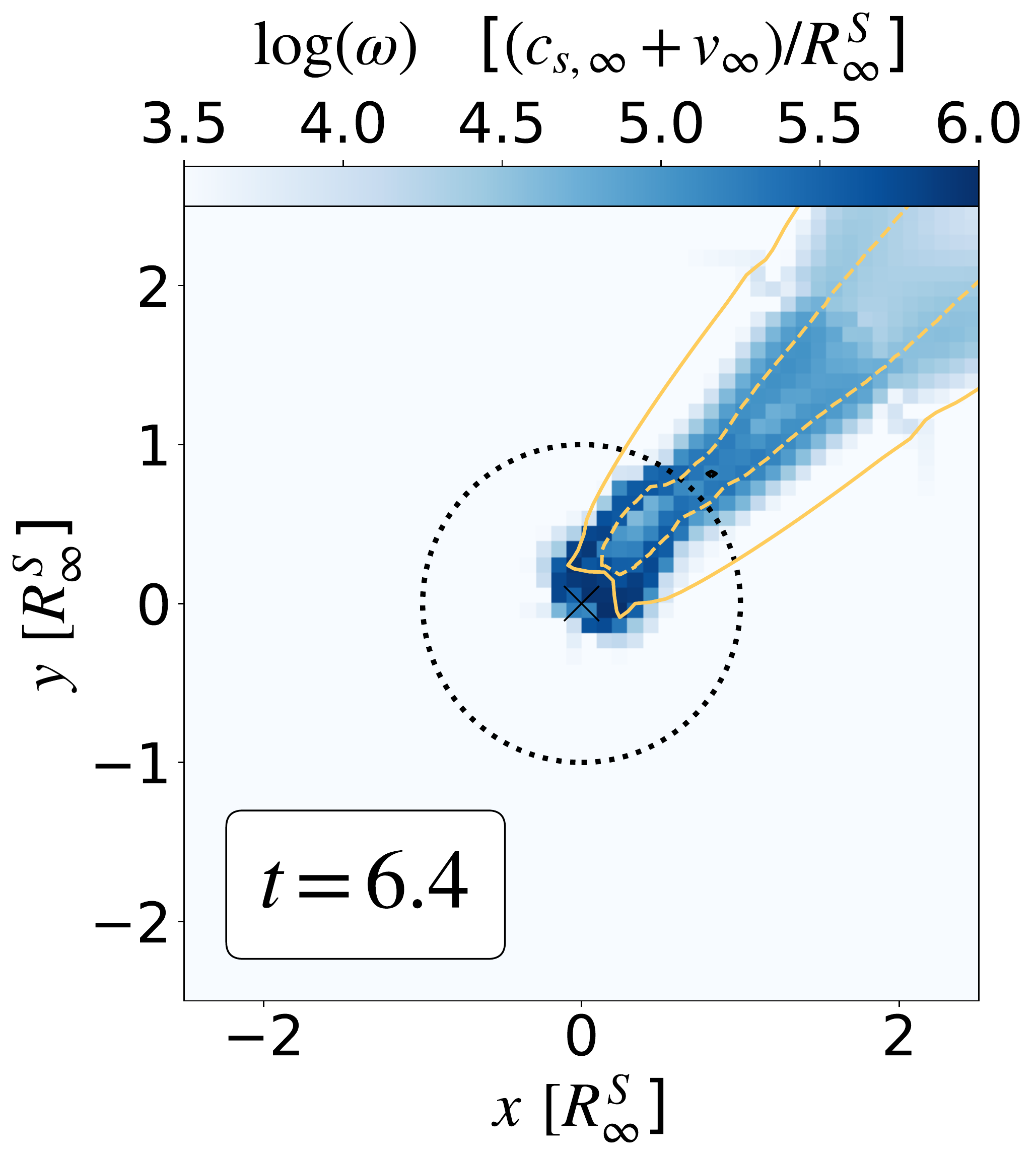}&	
		\includegraphics[width=0.32\textwidth]{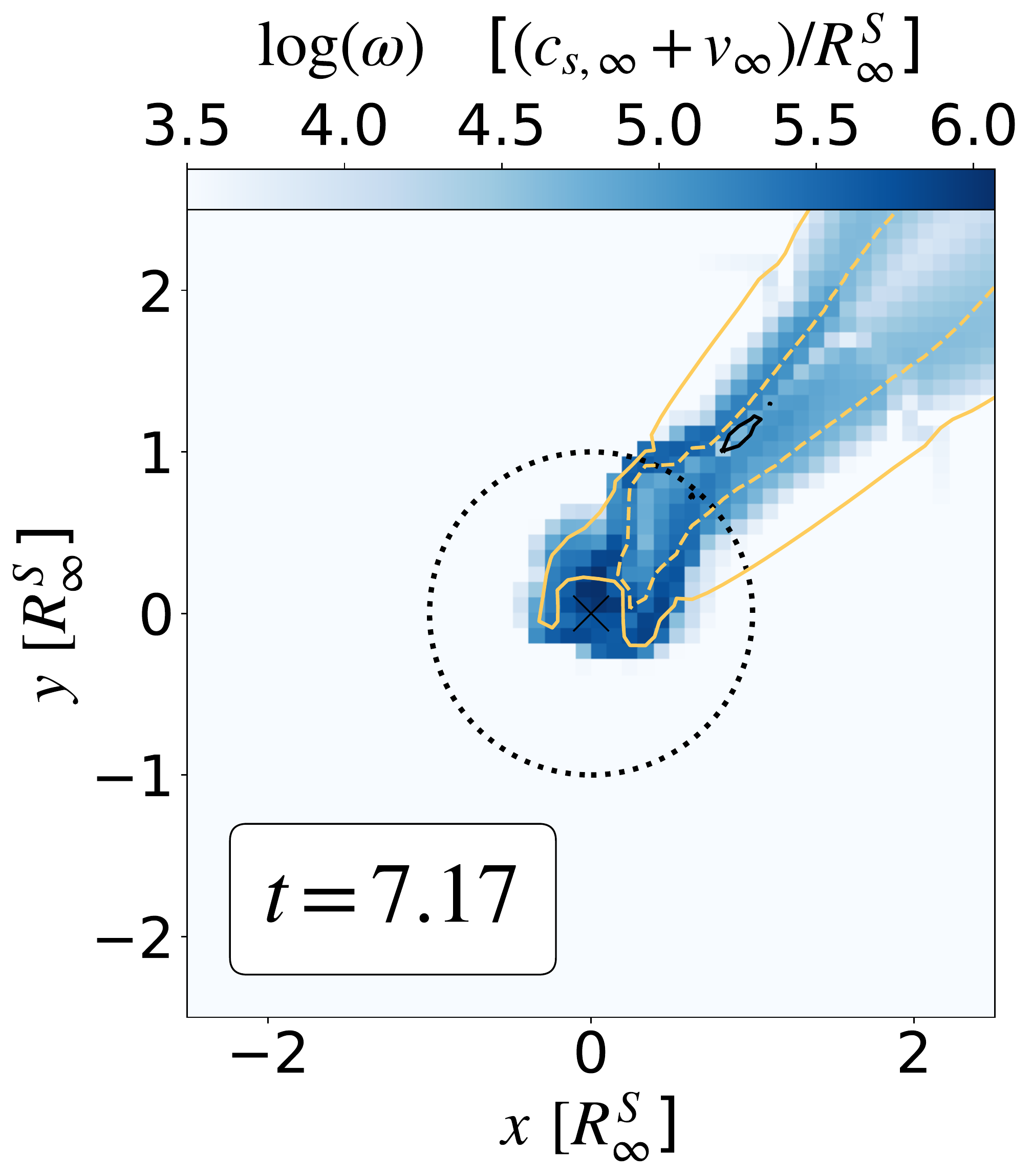}\\
	\end{tabular}
	\caption{{Vorticity slices for the simulation and times shown in Figure \ref{fig:eddies_gamma1.0}. Black contours show the sonic surfaces, whereas yellow contours denote entropy $K/ K_\infty = 0.5$ (solid line) and $K/ K_\infty = 0.1$ (dashed line). The location of the accretor is denoted by a black cross, and $R^S_\infty$ is shown as the dotted black circe. A minimum value of $\log(\omega) = 3.5$ was enforced on the plot to bring out the vorticity structure within the bowshock and wake. The circular features in the vorticity at radii of $\sim 2.5 \ R^S_\infty$ in the right hand two plots occurs due to a transitions in the refinement grid.}}
	\label{fig:vorticity_gamma1.0}
\end{figure*}

Indeed, repeating our experiment to explore the origin of the instability in the quasi-isothermal case, by shrinking the accretor for a $\mathcal{M}_\infty = 3$ simulation from $N=3.5$ (m3n3.5i), which leads to a steady state solution, to $N=7$ (m3n7i), which is unstable (see Figure \ref{fig:eddies_gamma1.0}) reveals that the instability originates downstream of the sink in the region of the stagnation point of the flow. {Instead of an increase in entropy in front of the accretor, like in the adiabatic case, the quasi-isothermal case shows a low entropy region within the entire bowshock (Figure \ref{fig:vorticity_gamma1.0}). This low entropy region is present for both stable and unstable simulations, with no signs of entropy perturbations, and we therefore conclude that this instability is not linked to the advective-acoustic cycle found in the adiabatic case.}

\begin{figure}
	\centering
	\includegraphics[width=\columnwidth]{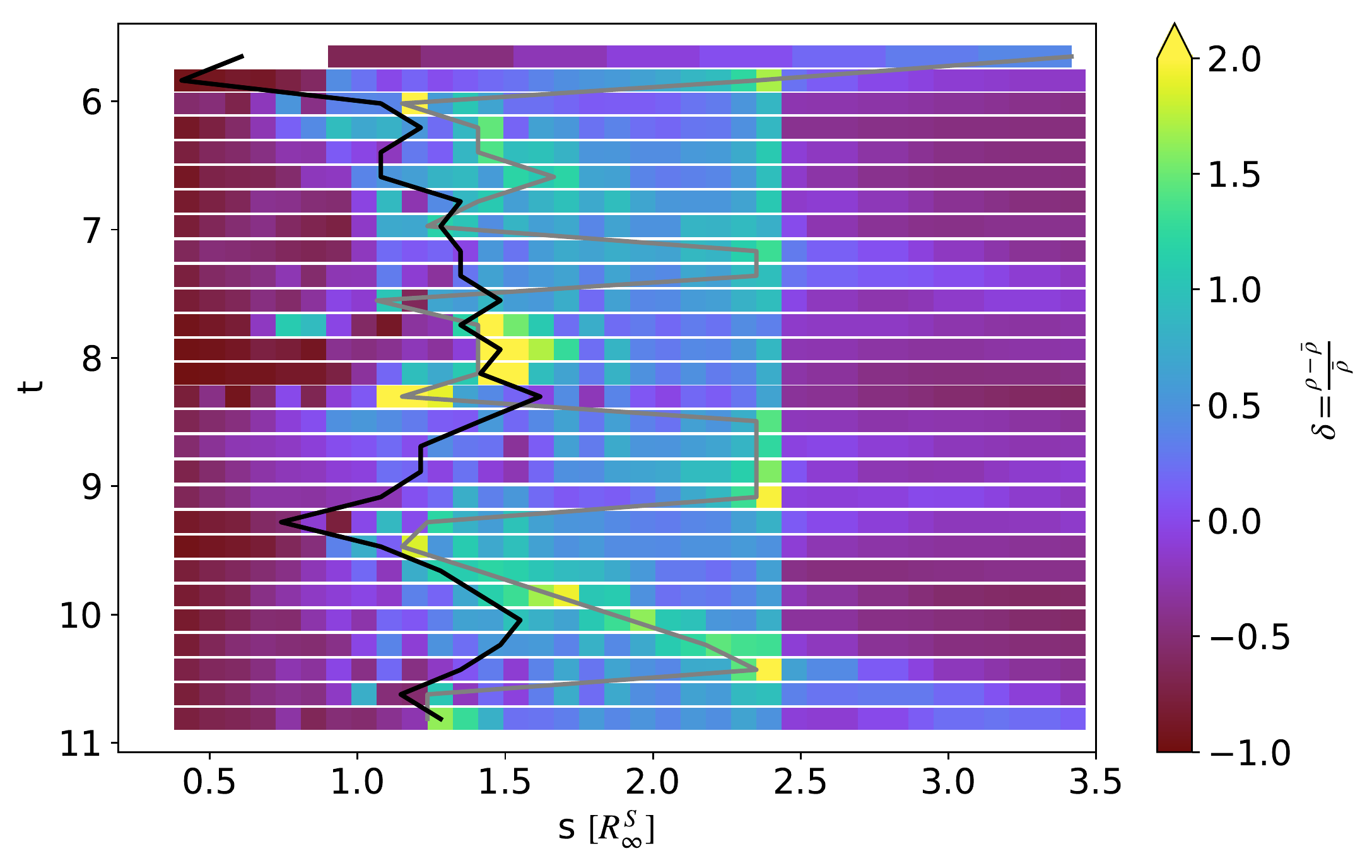}
	\caption{{Average density contrast $\delta$ of a slice of wake at a distance $s$ from the accretor, for the simulation shown in Figure \ref{fig:eddies_gamma1.0}. The black line denotes the position of the stagnation point, and the grey line highlights the position of the density contrast peak. Average densities are computed within thin cylinders of radius $R^S_\infty$ and height $ds$, aligned parallel to the axis of symmetry of the wake, and compared to the mass weighted density  $\bar{\rho}$ of a cylinder height $3.5 \ R^S_\infty$. The density contrast at $s \sim 2.5 \ R^S_\infty$ occurs due to a transition in the refinement grid.}}
	\label{fig:longitudinal}
\end{figure}

{Instead, this could be one of the two instabilities of the accretion column discussed in \citet{Foglizzo2005}: either the longitudinal instability  first described in \citet{Cowie1977}, where the authors consider the impact of a small density perturbation in the accretion column, or the transverse instability of the accretion column from \citet{Soker1990}. Investigating the longitudinal case first, Figure \ref{fig:longitudinal} shows the density contrast $\delta$ along the wake. As predicted by \citet{Cowie1977}, density perturbations form near the stagnation point but we do not see the amplification of waves travelling towards the accretor discussed by \citet{Foglizzo2005}. Instead, density perturbations propagate downstream from the accretion point, and have all but faded by $s=3 \ R^S_\infty$, with no evidence of a self-sustaining feedback loop. We therefore conclude that the longitudinal instability is not responsible for unsettling the wake.}

\begin{figure}
	\centering
	\includegraphics[width=\columnwidth]{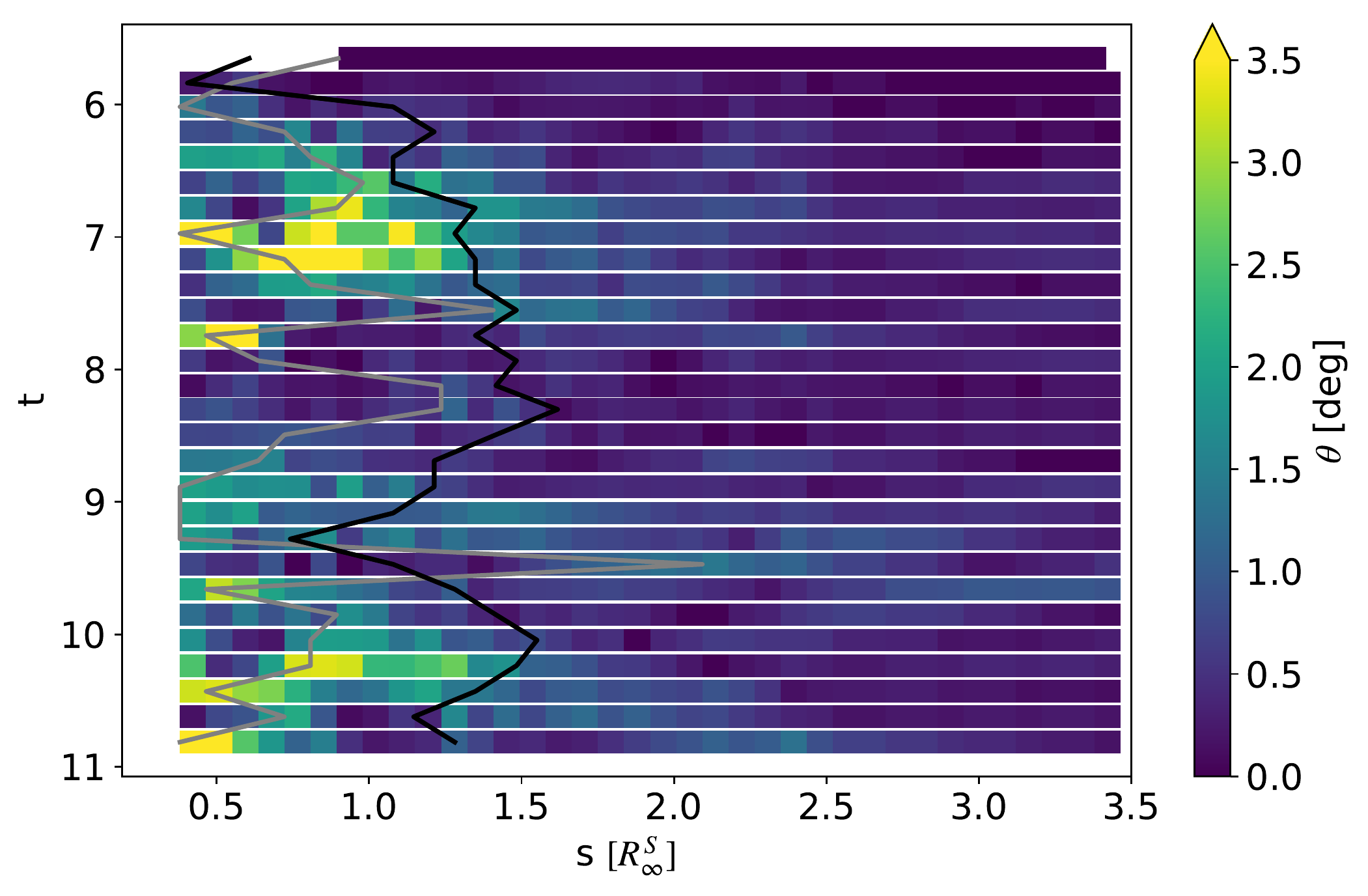}
	\caption{{Three-dimensional angle $\theta$ between the axis of symmetry of the wake and the centre of mass of a slice at distance $s$ along the wake. The black line denotes the position of the stagnation point, and the grey line highlights the position of the largest angle at a given point in time. The centre of mass at each distances $s$ is measured within a cylinder of radius $R^S_\infty$ and height $ds$. }}
	\label{fig:transverse}
\end{figure}

{Figure \ref{fig:transverse} shows that the stagnation point also plays an important role in the transverse displacement of the wake. The three-dimensional angle $\theta$, measured between the axis of symmetry of the problem and  centre of mass of a given slice along the wake, is largest in the region bounded by the stagnation point and the accretor. This confirms the visual conclusions from Figure \ref{fig:gamma1_density}, where the largest perpendicular displacement of the wake occurs at or near the stagnation point for all three simulations with $N\geq7$. We see no evidence for the predicted increase of amplifications with increasing distance from the accretor \citep{Soker1990}, and in fact see little transverse displacement of the wake beyond the stagnation point at all. We therefore conclude that while a local transverse instability can be observed in the three-dimensional simulations presented here, its effect is constrained to the supersonic region between the stagnation point and the accretor. If this region is unresolved, such as in cases of very low resolution, the accretion column, and by extension the wake, remain stable.}\

{The accretion column also remains stable in the adiabatic case previously discussed, as both the longitudinal and the transverse instability rely on the assumption that the radius of the shock cone is small compared to other distances, i.e. that the wake is very narrow. While theoretically both instabilities should appear for sufficiently high Mach numbers in the adiabatic case, the advective-acoustic instability acts to increase the shock opening angle  for moderate to high Mach numbers, suppressing the instabilities of the accretion column.}

\begin{figure}
	\centering
	\includegraphics[width=\columnwidth]{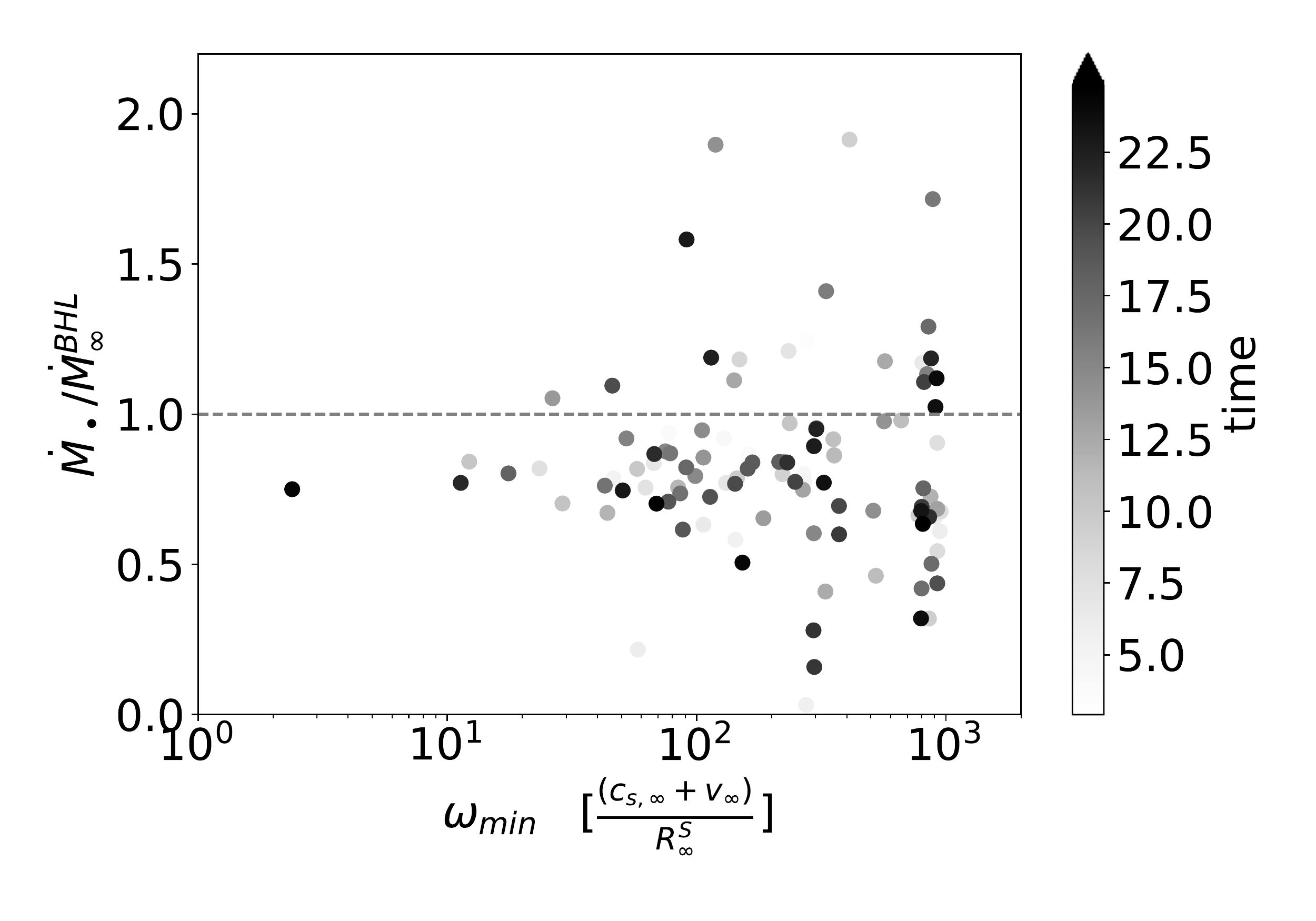}
	\caption{{Evolution of the instantaneous accretion rate $\dot{M}_\bullet$ onto the black hole versus the minimum vorticity, $\omega_{min}$ within the accretion column for simulation shown in Figure \ref{fig:eddies_gamma1.0}. The vorticity is measured within a hemisphere with a radius of $5 \Delta x$, centred on the accretor and oriented downstream.}}
	\label{fig:accretion_column_g1-0}
\end{figure}

{Contrary to the adiabatic case, accretion onto the black hole remains effective even in the presence of turbulence. As can be seen in Figure \ref{fig:accretion_column_g1-0}, gas properties in the wake remain in the low vorticity - high accretion regime, where $\omega_{min} < 10^3 $, throughout the course of the simulation despite the fact that the isothermal case was run for an extra 10 dynamical times. The vorticity slices in Figure \ref{fig:vorticity_gamma1.0} confirm that a low vorticity region continues to exist within the narrow bow shock, allowing gas to continue feeding the black hole. }

A variety of other instabilities have been suggested to potentially arise from Hoyle-Lyttleton 'like' accretion flows, such as individual vortex rings in  \citet{Kim2009} for a non-accreting massive perturber embedded in an adiabatic 2D supersonic flow, or shock cone vibrations as reported in \citet{Lora-Clavijo2013} when investigating 2D adiabatic BHL accretion using general relativity instead of Newtonian gravity. However, neither of these phenomena are observed here.
Finally, we stress that the main numerical limitation of the simulations presented here is that the size of the accretor is intrinsically tied to the resolution, with $r^* \approx 2 \times \Delta x_\mathrm{min}$. To truly investigate the nature of the instabilities would require exploring the impact of resolution and accretor size independently, which is beyond the scope of this work. We expect that this shortcoming will only affect the details of the instability triggering and their exact intensity.


\section{Conclusions}
\label{sec:conclusion}

As numerical capabilities increase, we are able to cover a larger range of length scales in simulations. In this paper, we explored the impact of resolution on the accretion and drag force for a sink particle in grid simulations, using sub-grid algorithms based on the Bondi-Hoyle-Lyttleton (BHL) interpolation formula for accretion, and the linear analytic drag force estimate of \citet{Ostriker1998}. In all simulations presented here, the size of the accretion region shrunk with increasing resolution, and both analytic formulas were evaluated using mass-averaged, kernel-weighted quantities in the immediate vicinity of the accretor. We adopted an idealised BHL setup, where originally homogeneously distributed gas with a uniform Mach number was allowed to settle in the gravitational potential of the accreting particle, with the relevant scale radius, $R^S_\infty(\mathcal{M}_\infty)$, resolved by a fixed number of resolution elements, $N = R^S_\infty(\mathcal{M}_\infty)/\Delta x_\mathrm{min} $ throughout the simulation. We then investigated the impact of a wide range of resolutions, $0.01 < N < 500$, Mach numbers $0 \leq \mathcal{M}_\infty \leq 10$, and two values of the adiabatic index $\gamma= 1.3334, 1.0001$, on the accretion rate and drag force sub-grid algorithms onto our accretors, modelled as sink particles with radius of $r*/R^S_\infty \simeq 2/N$.

We found that, as expected, for very low resolutions, i.e. $N < 1$, where the local mass weighted quantities reflect the ones at infinity because the gravitational influence of the sink is small, the accretion sub-grid algorithm closely followed the analytic values. Accretion rates converged for $N>100$ and Mach numbers $\mathcal{M}_\infty < 1.5$, although not to the analytic value set by the accretion algorithm as accretion onto the black hole transitioned from BHL to supply limited accretion. However, in this situation, the accretion rates measured onto the accretor all converged to values within a factor 
$\simeq 2$ of the BHL analytic interpolation formula, with the largest discrepancies found for transsonic flows.

On the other hand, at higher Mach numbers, $\mathcal{M}_\infty \geq 1.5$, and higher resolution, instabilities appeared and started to dominate the flow to such an extent that the accretion rate dropped to $\dot{M}/\dot{M}^\mathrm{BHL}_\infty \approx 0.1$ in the adiabatic runs with $\gamma = 1.3334$. In the quasi isothermal case ($\gamma = 1.0001$), accretion rates remained within a factor $\simeq 2$ of the BHL rate over the whole range of Mach numbers probed here, despite the presence of instabilities in the accretion column at high Mach numbers ($\mathcal{M}_\infty \geq 1.5$).
 
These findings are in agreement with previous work by \citet{Ruffert1996,Foglizzo2005} as to the origin of these instabilities, at least in the adiabatic case when $\gamma  \simeq 4/3$. They are caused by an acoustic-advective feedback loop which develops in the sub-sonic region between the shock front and the accretor, as described analytically in \citet{Foglizzo2005,Foglizzo2009}. {In the quasi-isothermal case, where $\gamma=1.0001$, we instead observe a transverse instability of the accretion column in the region between the stagnation point and the accretor, which locally displaces this column perpendicularly to the axis of symmetry of the problem, reminiscent of the transverse instability discussed in \citet{Soker1990}.}

The drag force due to the wake, calculated from the cell density in the box, converged quickly with resolution, again in agreement with results by \citet{Ruffert1995,Ruffert1996}. However, as the bow shock and accretion column began to be resolved by the simulation, the local mass weighted quantities used in the sub-grid algorithm to evaluate the analytic drag force according to the formula of \citet{Ostriker1998} started to poorly reflect the values of these quantities ``at infinity''. Indeed, not only did the magnitude of the drag force estimated in this way fluctuate on extremely short time scales, but its direction actually flipped, so that even after adding the contribution from the resolved wake, the net force on the accretor remained an \textit{acceleration} with respect to the gas flow, rather than a drag. This clearly unphysical behaviour solely occurs because the local relative velocity becomes dominated by gas flowing down the accretion column. Therefore, sub-grid algorithms for the drag force should be avoided as soon as the characteristic radius $R^S_\infty$ becomes larger than the size of the accretor.

The main limitation of the conclusions we reach about the sub-grid algorithms presented in this work, is that quantities ``at infinity'' are poorly defined in more realistic galaxy evolution or cosmological simulations. For example, we have measured the scale radius for quantities ``at infinity'', taken to be the boundary of the box. However, embedded in a galaxy, the flow far from the accretor is not uniform or isothermal, and the gravitational field due to the accreting black hole not necessarily dominant. We also caution that it is important to resolve $R^S_\infty$, which can be much smaller than the Bondi radius for highly supersonic flows which are quite frequent in galaxy simulations.

Moreover, while we measure fairly similar trends in both the adiabatic and quasi-isothermal cases probed here, our simulations do not include radiative cooling explicitly, which could significantly influence the sound speed in the vicinity of the accretor, and through it both the accretion rate and drag force. In addition, while we calculate the drag force, we do not allow the sink to move under its influence. This is particularly important for the instability dominated runs, where we would not expect the accretor to remain located on the axis of symmetry of the bow shock at all times. In less idealised simulations than those presented in this work, we also expect the sink to encounter non-homogeneous gas moving non-uniformly, other compact objects as well as gradients in the local gravitational potential. 

However, it is clear that the sub-grid models we have described in this work go beyond a basic implementation of an analytic BHL prescription. They are more versatile and reliable, capable of handling a wide range of resolutions and complex flow configurations. In particular, the smooth transition from an algorithm based on the analytic BHL accretion formula to SLA when local density and velocity features are resolved ensures that the accretion rate onto the black hole makes the best use of the local information available in the simulation, and should quite naturally converge to the correct solution once the size of the accretion region approaches the physical size of the black hole. By contrast, the drag force sub-grid algorithm becomes unphysical as resolution increases and extra caution regarding its implementation in less idealised simulations is needed, a question which we plan to address in a follow-up work. 

\section*{Acknowledgements}
The authors thank Will Potter for helpful suggestions, and the reviewer for a constructive referee report that significantly improved the depth and quality of this paper. The research of RSB is supported by Science and Technologies Facilities Council (STFC) and by the Centre National de la Recherche Scientifique (CNRS) on grant ANR-16-CE31-0011, and the research of AS and JD at Oxford is supported by the Oxford Martin School and Adrian Beecroft. This work is part of the Horizon-UK project, which used the DiRAC Complexity system, operated by the University of Leicester IT Services, which forms part of the STFC DiRAC HPC Facility (www.dirac.ac.uk ). This equipment is funded by BIS National E-Infrastructure capital grant ST/K000373/1 and  STFC DiRAC Operations grant ST/K0003259/1. DiRAC is part of the National E-Infrastructure. All visualisations were produced using the yt-project \citep{Turk2010}. 




\bibliographystyle{mnras}
\bibliography{references} 



\appendix

\section{Uniform initial conditions and small accretors}
\label{sec:initial_conditions}

\begin{figure*}
	\begin{tabular}{ccc}
		\includegraphics[width=0.32\textwidth]{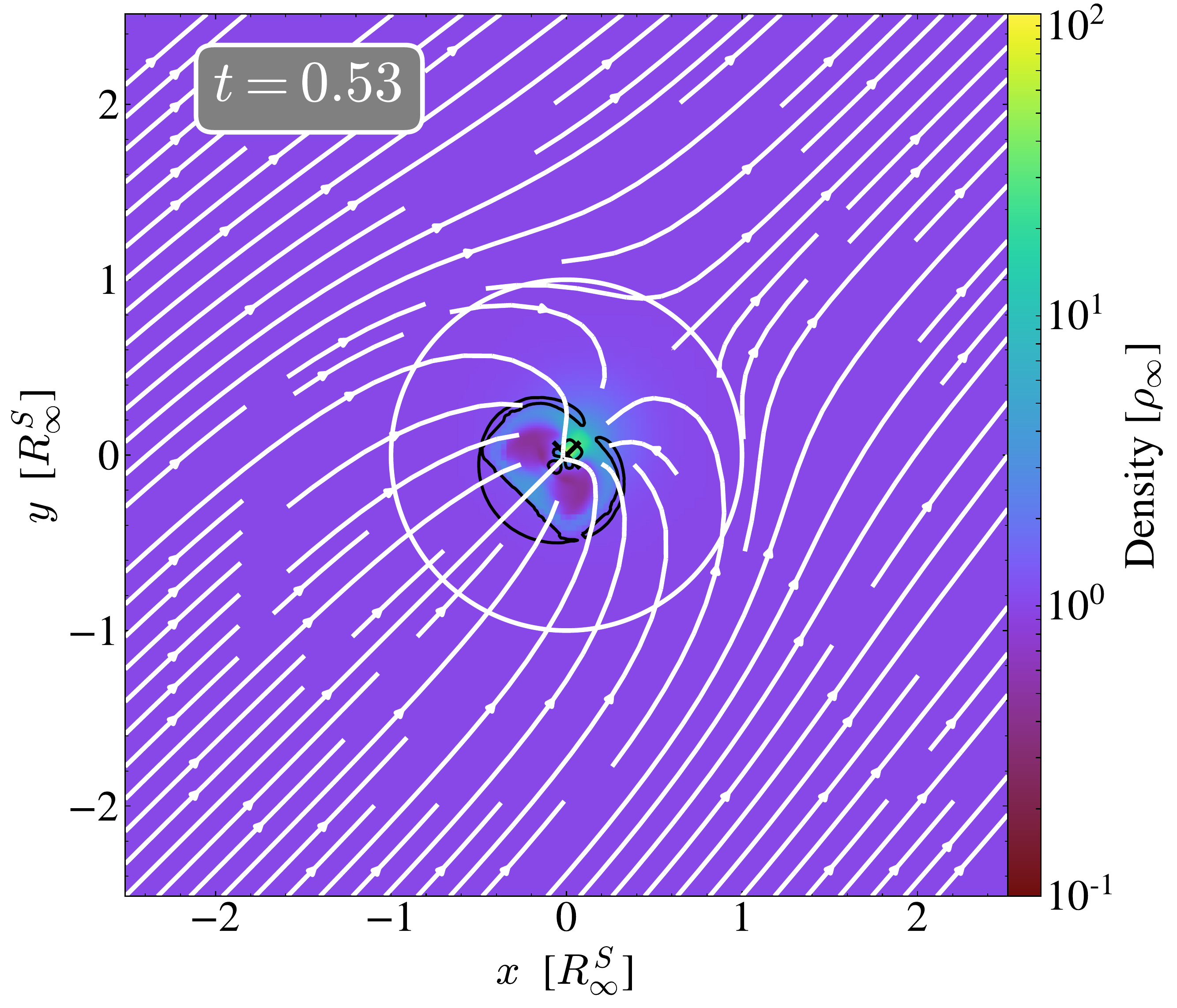} &
		\includegraphics[width=0.32\textwidth]{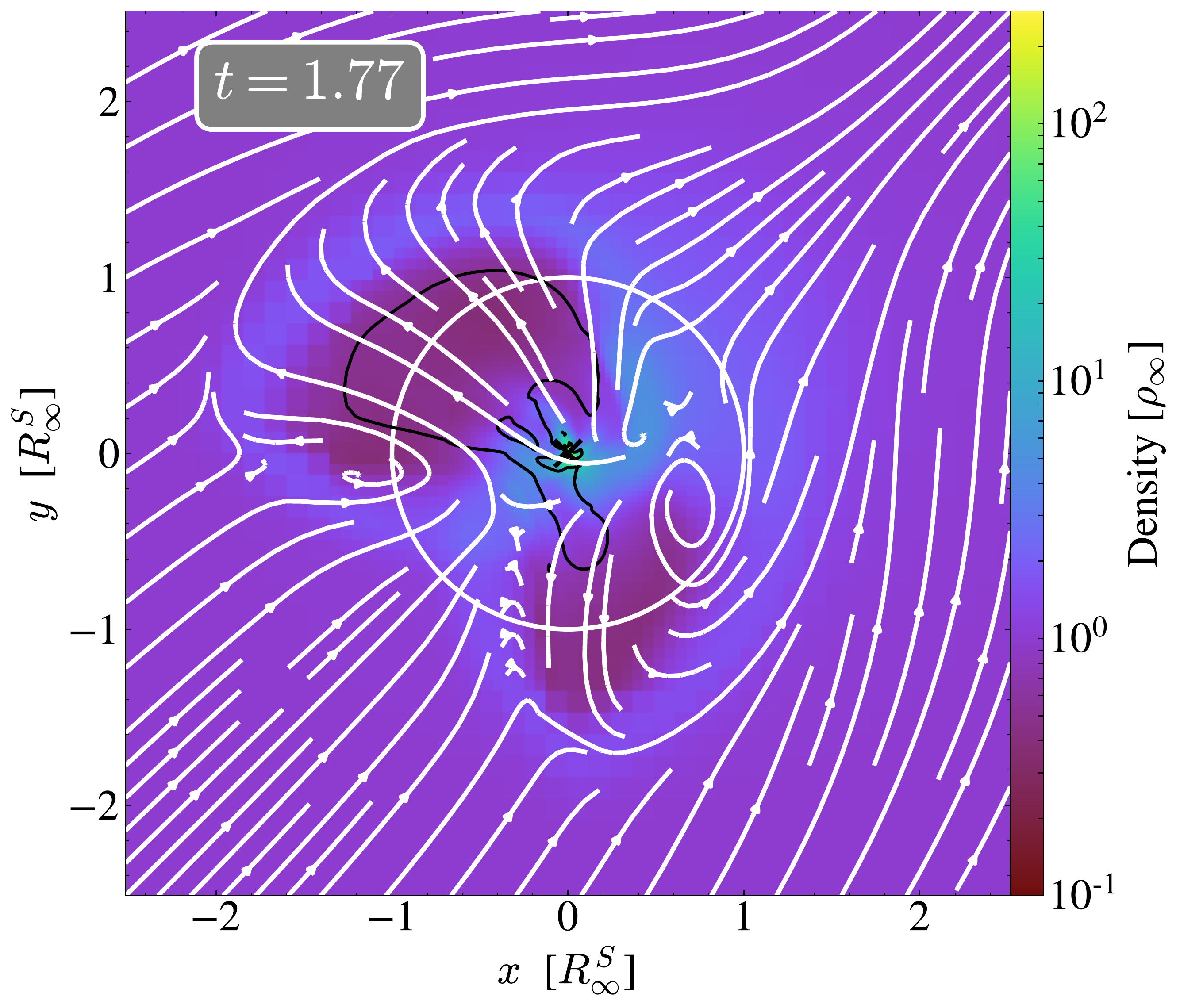} &
		\includegraphics[width=0.32\textwidth]{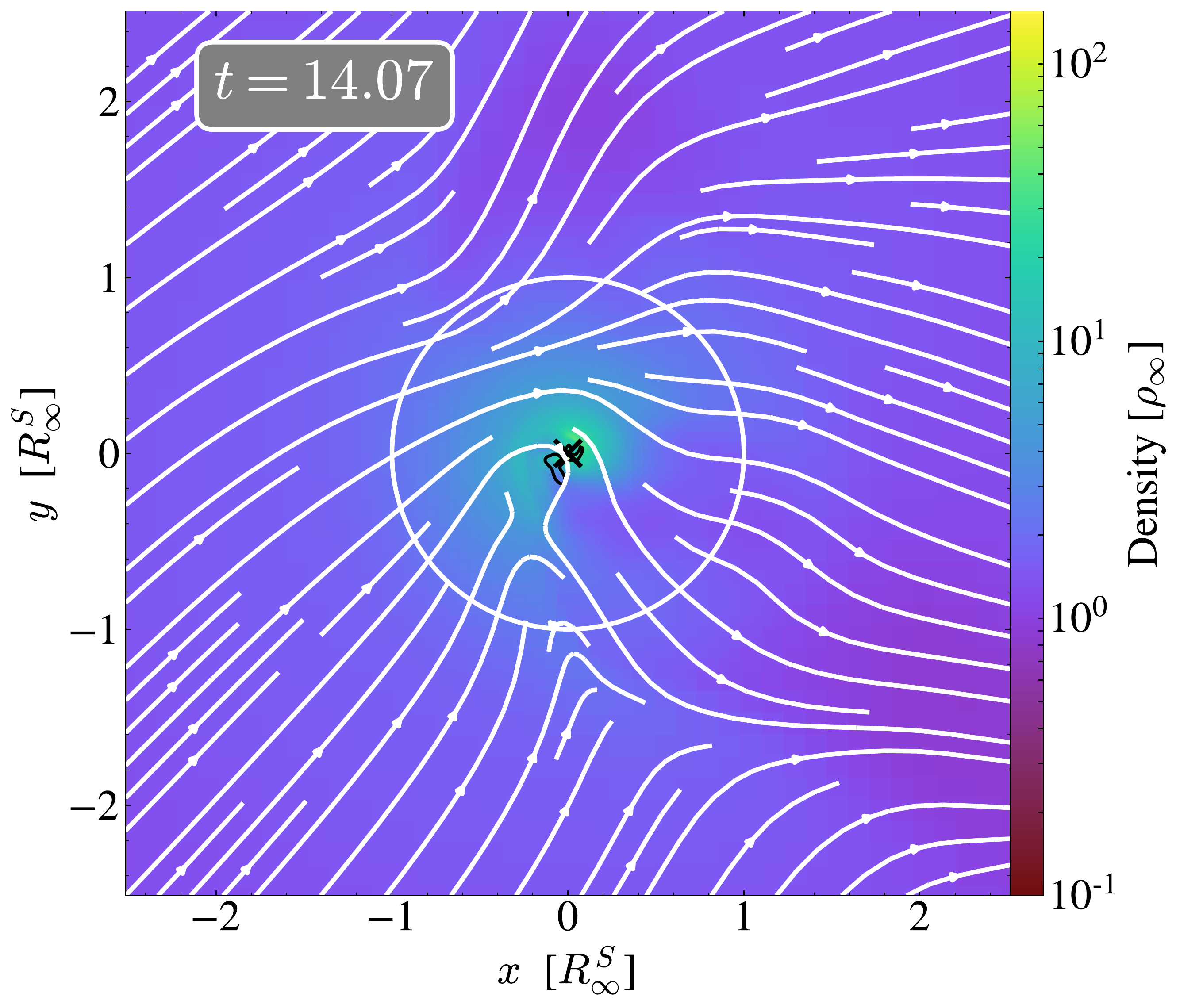} \\
		\includegraphics[width=0.32\textwidth]{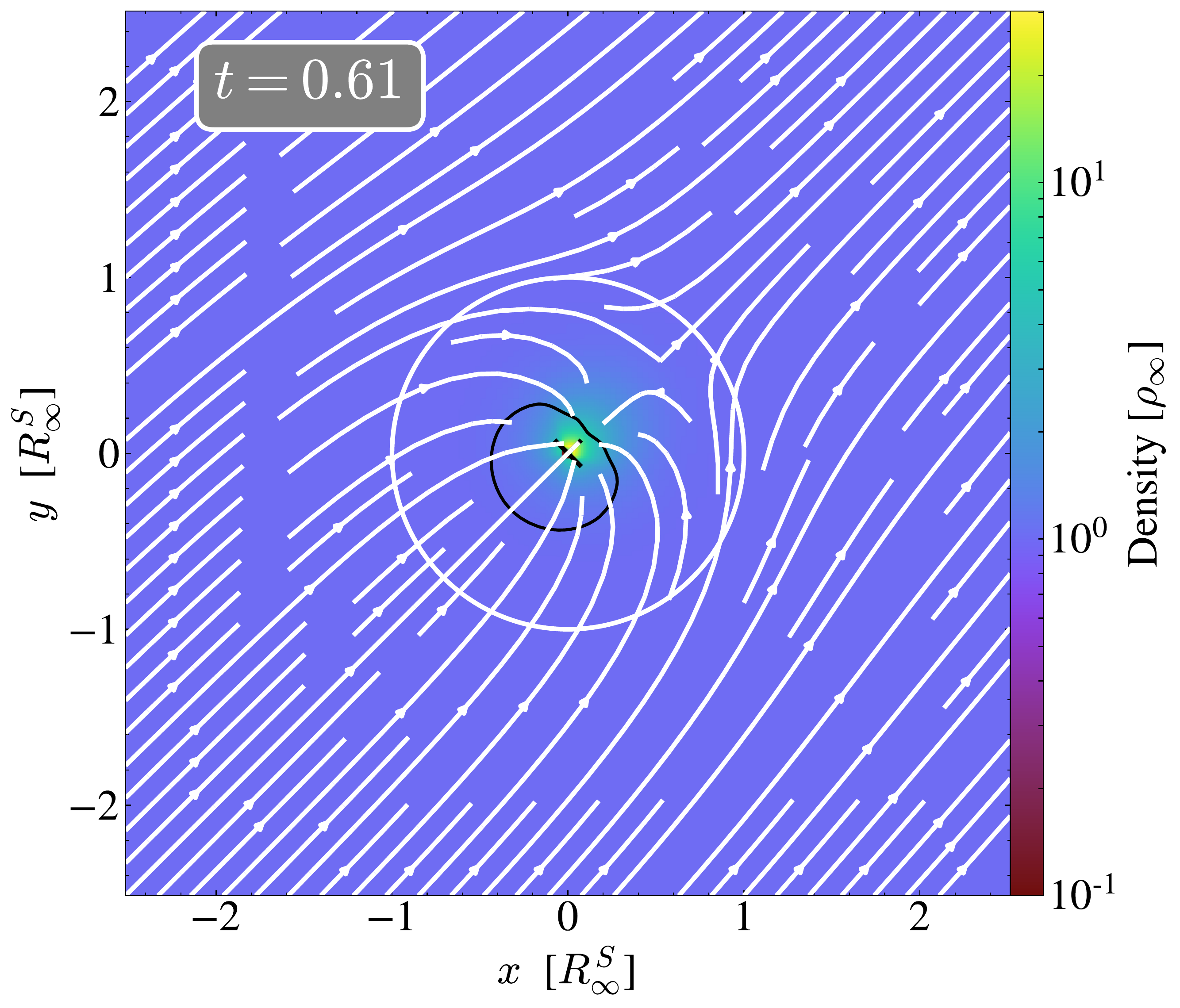} &
		\includegraphics[width=0.32\textwidth]{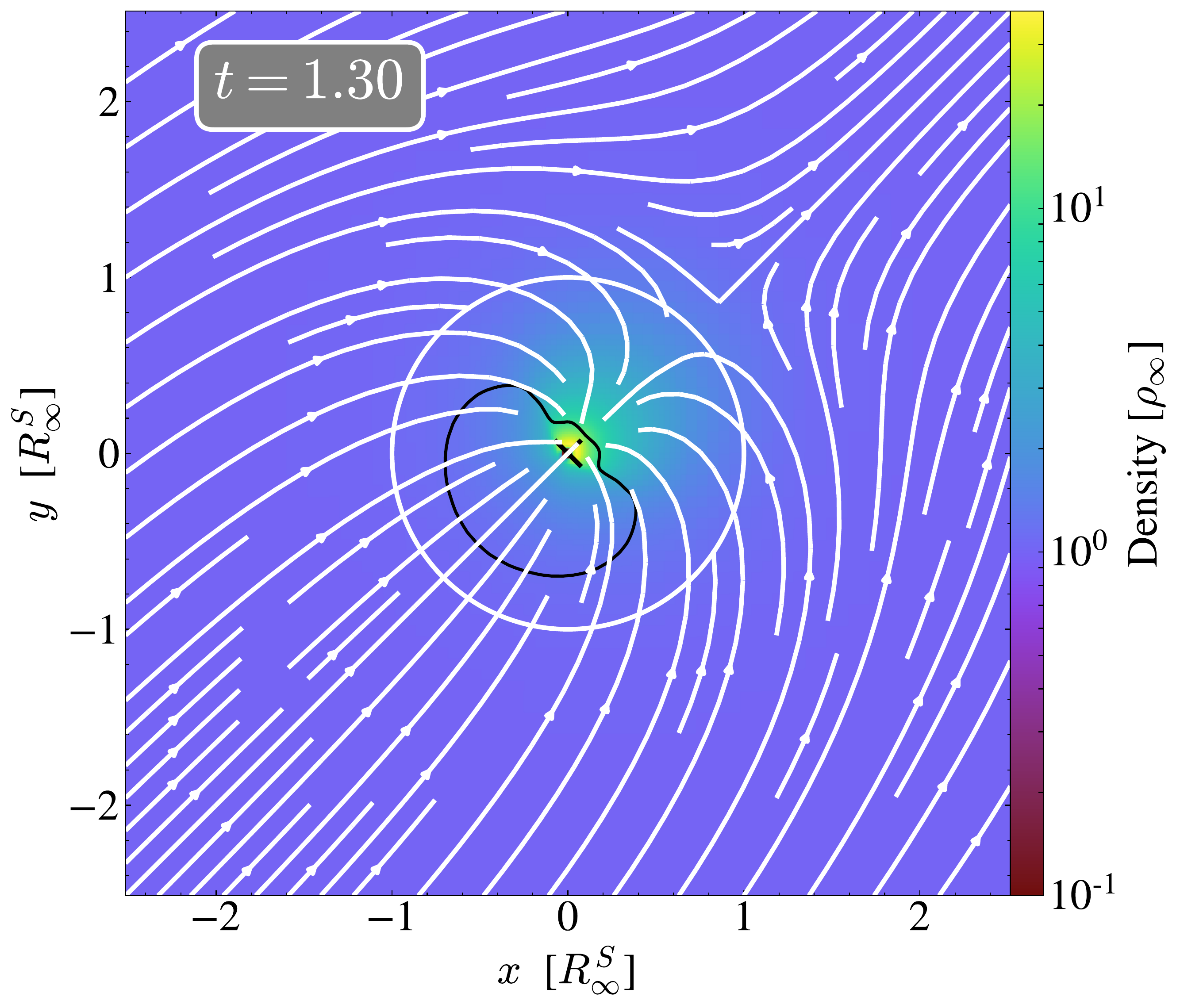} &
		\includegraphics[width=0.32\textwidth]{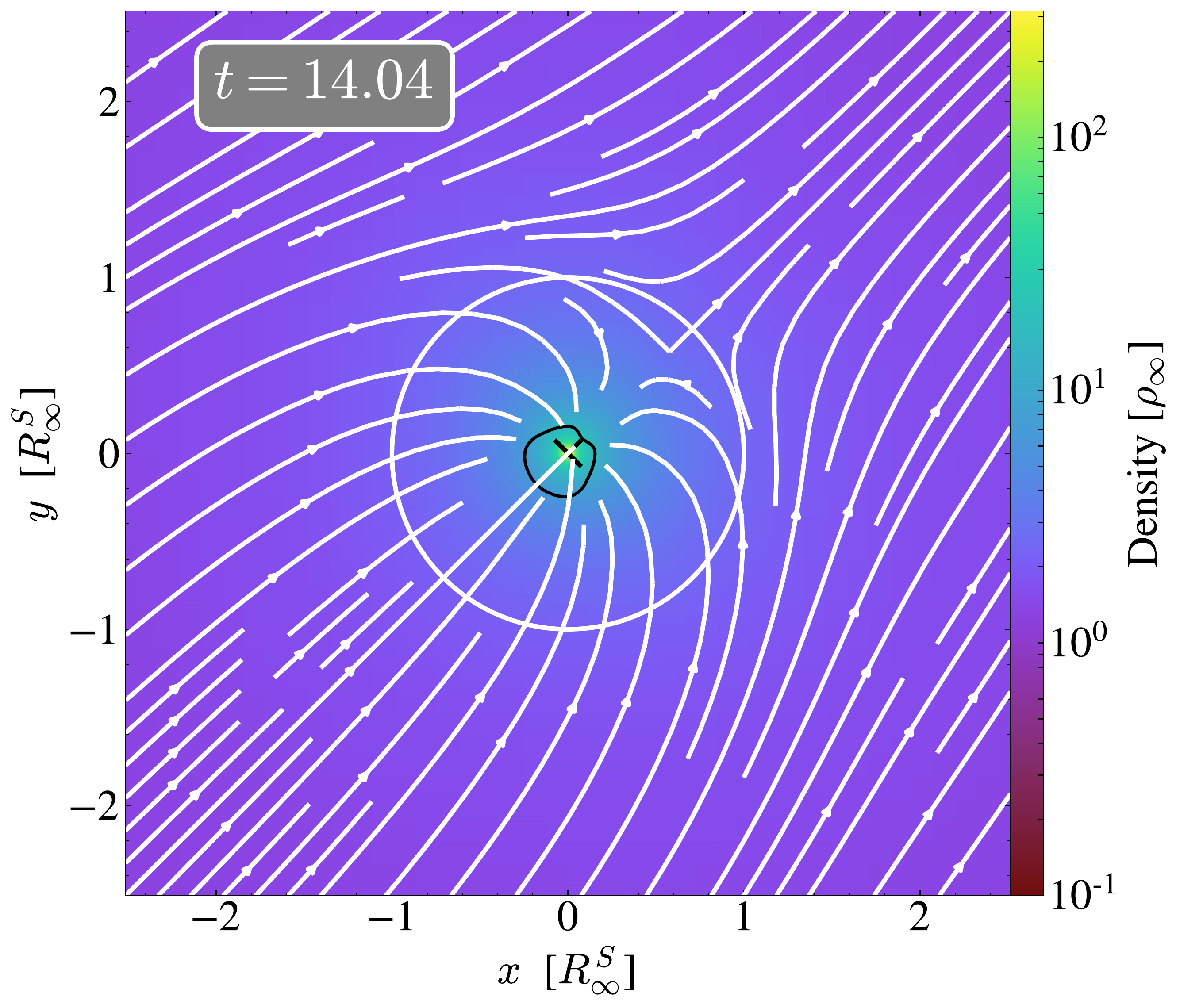} \\
	\end{tabular}
	\caption{Flow patterns for m0.5n262a\_A with all refinement levels available from the beginning (top row) and refinement levels triggered gradually (bottom row). The sink is marked by a black cross, and $R^S_\infty$ is shown as a white circle. Streamlines are annotated in white. Both simulations have identical final grid configurations, but the shock travelling upstream is avoided with gradual level releasing.}
	\label{fig:settling_density}
\end{figure*}

The simulations presented in this work use uniform initial conditions to simplify and homogeneise the setup of different flow configurations. After an initial period, during which the flow settles, the simulations  are assumed to have erased the memory of their initial conditions and evolved to their natural quasi-steady state solution. As can be seen in the top row of Figure \ref{fig:settling_density}, small accretors can cause shocks during the initial settling phase which generate density perturbations that remain present even after many dynamical time scales. In this Appendix, we present the method used here to minimise this spurious effect.

\begin{figure}
	\includegraphics[width=\columnwidth]{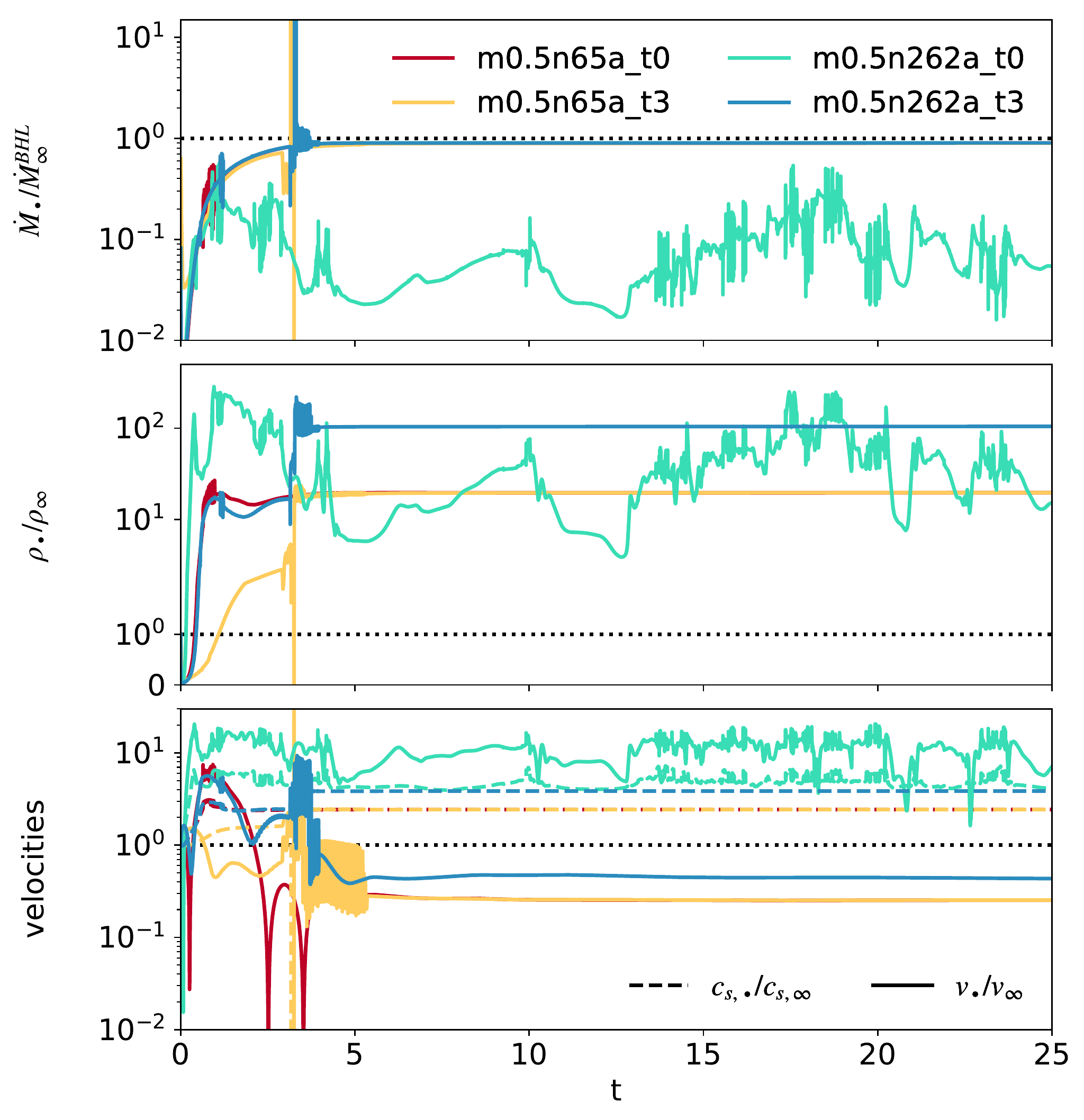}
    	\caption{Accretion rates and mass averaged sink properties for simulations with all refinement levels available from the beginning (m0.5n65a\_t0 and m0.5n262a\_t0), and simulations where the last two levels of refinement are added at $t=3$ (m0.5n65a\_t3 and m0.5n262a\_t3).}
    \label{fig:settling_profiles}
\end{figure}

The first precaution we take is to introduce the analytic gravity field gradually, with $F_{G} = G M_\mathrm{sink} / r^2 \times (t/t_{\rm full})$ while $t<t_{\rm full}$, where $t_{\rm full} = 2$.  For large accretors ($N = 65$), the simulations show some instability for $t<5$, visible in the profile plots of Figure \ref{fig:settling_profiles}, particularly when the bowshock detaches, but then settle into stable and converged solutions. However, for sufficiently small accretors, i.e. when $N>100$,  simulations of any Mach number show a shock forming in front of the sink and travelling upstream, driven and supported by eddies (see top row of Figure \ref{fig:settling_density} for an example). This 'circularisation' of the gas flow breaks the symmetry of the problem, which develops perturbations that fail to settle during more than 10 dynamical times. We only observed this phenomenon for sufficiently small accretors, where the density profile around the sink is more strongly peaked, causing the reverse shock. 

In order to minimise the effect of the initial conditions for small accretors, all simulations in this paper with $N>100$ were run using a reduced maximum level of refinement while $t < 3$, which is equivalent to an effective accretor size $N_{\rm eff}(t<3) \approx 90$. This avoids shocking the gas and allows the flow to settle. Around $t \simeq 3$, the extra levels of refinement are added in a staggered fashion to minimise instabilities caused by changing the grid structure: we first decrease the size of the accretor and then add an extra level of refinement per timestep until the target resolution is reached. 

Figure \ref{fig:settling_profiles} shows that some instabilities do occur during and just after adding refinement levels, as evidenced by the high frequency fluctuations in the gas properties $\rho_\bullet$, $v_\bullet$ and $c_{s,\bullet}$ caused by the gas ``sloshing'' at the bottom of the gravitational well. After a few dynamical times, the solution settles into the stable flow pattern observed for larger accretors (bottom row, Figure \ref{fig:settling_density}). A test run using $N=65$ (m0.5n65a), which is naturally stable, shows that both the run with the full resolution available from the beginning (n65\_t0 in Figure \ref{fig:settling_profiles}), and the run where the full resolution is reached at $t=3$ (n65\_t3 in Figure \ref{fig:settling_profiles}) converge to $\left<\dot{M}/\dot{M}_\infty^\mathrm{BHL}\right> = 0.89$, which is also recovered for the small accretor with staggered refinement, n263\_t3. By contrast, the simulation in which the small accretor is fully resolved from the start, n262\_t0, displays a lack of convergence, high frequency variations in the accretion rate and a much lower time averaged value of $\left<\dot{M} /\dot{M}_\infty^\mathrm{BHL}\right> = 0.28$. To ensure that simulations have sufficient time to settle, we measure all time averaged values in the paper for $t>10$.


\bsp	
\label{lastpage}
\end{document}